\documentclass[acmtog,nonacm]{acmart}

\AtBeginDocument{%
  }

\usepackage{amsmath}
\usepackage{bm}

\usepackage[ruled]{algorithm2e} 

\SetKwComment{Comment}{$//$\ }{}

\usepackage{empheq} 
\SetKwComment{Comment}{$//$\ }{}
\usepackage{tikz,pgfplots} 
\pgfplotsset{width=\linewidth, compat=1.9}
\usepackage{subcaption}
\usepackage{mathtools}
\usepackage{multirow}
\usepackage{dblfloatfix}

\usepackage[export]{adjustbox}
\usepackage{afterpage}
\usepackage{placeins}
\usepackage{makecell}

\definecolor{neworange}{rgb}{0.85,0.45,0.0}

\newcommand{\zeros}{\mathbf{0}} 
\newcommand{\ids}{\mathbf{I}} 

\newcommand{\pos}{\mathbf{q}} 
\newcommand{\vel}{\mathbf{v}} 
\newcommand{\bpos}{\bar{\mathbf{q}}} 
\newcommand{\bvel}{\bar{\mathbf{v}}} 
\newcommand{\ppos}{\tilde{\mathbf{q}}} 
\newcommand{\ext}{\mathbf{f}_{\mathrm{ext}}} 
\newcommand{\internal}{\mathbf{f}_{\mathrm{int}}} 

\newcommand{\force}{\boldsymbol{\lambda}}
\newcommand{\pene}{\boldsymbol{\delta}}

\newcommand{\rest}{\mathbf{d}}
\newcommand{\proj}{\mathbf{p}}
\newcommand{\weight}{w_i}

\newcommand{\mass}{\mathbf{M}} 
\newcommand{\map}{\mathbf{G}_i} 
\newcommand{\sys}{\mathbf{A}} 
\newcommand{\jac}{\mathbf{H}} 
\newcommand{\ortho}{\mathbf{R}}

\newcommand{\lset}{\mathcal{L}} 
\newcommand{\bset}{\mathcal{B}} 
\newcommand{\cset}{\mathcal{C}} 

\newcommand{\lhs}{\mathcal{A}}
\newcommand{\sparseinverse}{\mathbf{S}}
\newcommand{\rhs}{\mathcal{K}}
\newcommand{\adjoint}{\mathbf{z}}
\newcommand{\kernal}{\mathbf{K}}

\newcommand{\real}{\mathbb{R}} 
\newcommand{\fb}{\phi_{\mathrm{FB}}}
\newcommand{\sfb}{\phi_{\mathrm{FB}}^{\epsilon}}

\newcommand{\deform}{\mathbf{F}}
\newcommand{\pdeform}{\mathbf{P}}
\newcommand{\svdU}{\mathbf{U}}
\newcommand{\svdV}{\mathbf{V}}
\newcommand{\svds}{\boldsymbol{\sigma}}
\newcommand{\svdss}{\boldsymbol{\theta}}
\newcommand{\svdS}{\boldsymbol{\Sigma}}
\newcommand{\svdSS}{\boldsymbol{\Theta}}
\newcommand{\rot}{\boldsymbol{\Omega}}
\newcommand{\svdW}{\mathbf{W}}
\newcommand{\vect}[1]{\mathrm{vec}(#1)}
\newcommand{\diag}[1]{\mathrm{diag}(#1)}
\newcommand{\Diag}[1]{\mathrm{Diag}(#1)}

\newcommand{\dd}{\mathrm{d}}
\newcommand{\pp}[2]{\frac{\partial#1}{\partial#2}}
\newcommand{\inv}[1]{#1^{-1}}
\newcommand{\trans}[1]{#1^{\top}}
\newcommand{\fd}[1]{\big(\pp{L}{#1}\big)_{\mathrm{FD}}}
\newcommand{\NA}{\textemdash}
\usepackage{pifont}
\newcommand{\cmark}{\ding{51}} 
\newcommand{\xmark}{\ding{55}} 
\newcommand{\metric}[2]{\makebox[3.2em][l]{#1}\,#2}

\newcommand{\loss}{L}
\newcommand{\grad}{g}

\begin{document}

\title{Fast and Reliable Gradients for Deformables Across Frictional Contact Regimes}

\author{Ziqiu Zeng}
\affiliation{%
  \institution{National University of Singapore}
  \country{Singapore}
}
\affiliation{%
  \institution{Prana Lab}
  \country{USA}
}
\email{zzeng@nus.edu.sg}

\author{Gang Yang}
\affiliation{%
  \institution{National University of Singapore}
  \country{Singapore}
}
\email{ygang@u.nus.edu}

\author{Zhenhao Huang}
\affiliation{%
  \institution{National University of Singapore}
  \country{Singapore}
}
\email{huangzhenhao@u.nus.edu}

\author{Bingyang Zhou}
\affiliation{%
  \institution{National University of Singapore}
  \country{Singapore}
}
\email{bingyang.zhou@u.nus.edu}

\author{Yulin Li}
\affiliation{%
  \institution{National University of Singapore}
  \country{Singapore}
}
\email{yline@nus.edu.sg}

\author{Jason Pho}
\affiliation{%
  \institution{Prana Lab}
  \country{USA}
}
\email{pho@pranalab.ai}

\author{Siyuan Luo}
\authornote{Corresponding author.}
\affiliation{%
  \institution{National University of Singapore}
  \country{Singapore}
}
\email{sy.luo@nus.edu.sg}

\author{Fan Shi}
\authornote{Corresponding author.}
\affiliation{%
  \institution{National University of Singapore}
  \country{Singapore}
}
\email{fan.shi@nus.edu.sg}

\begin{abstract}
Differentiable physics engines enable gradient-based system identification and inverse-dynamics control, but reliable gradients remain elusive in the regimes that motivate inverse problems: frictional contact and large-deformation elasticity. We present an analytical adjoint framework that addresses this gap for deformable bodies with Signorini--Coulomb friction, derived by implicit differentiation of the converged coupled residual of an NCP-based forward solver. Three components make this framework practical: (i) a smoothed Fischer--Burmeister formulation with matched forward/backward smoothing for continuous cross-regime gradients; (ii) a closed-form SVD-based Jacobian for local projections that supports isotropic hyperelastic models such as Neo-Hookean; and (iii) the first Woodbury-preconditioned Krylov adjoint solver, jointly exploiting the elastic and contact structure of the linearized adjoint matrix and uniformly outperforming simpler preconditioners across all contact regimes. The resulting GPU pipeline delivers accurate, efficient gradients across benchmarks, baseline comparisons, material identification, and inverse-dynamics control on locomotion and manipulation.
\end{abstract}



\begin{teaserfigure}
  \includegraphics[width=1.0\textwidth]{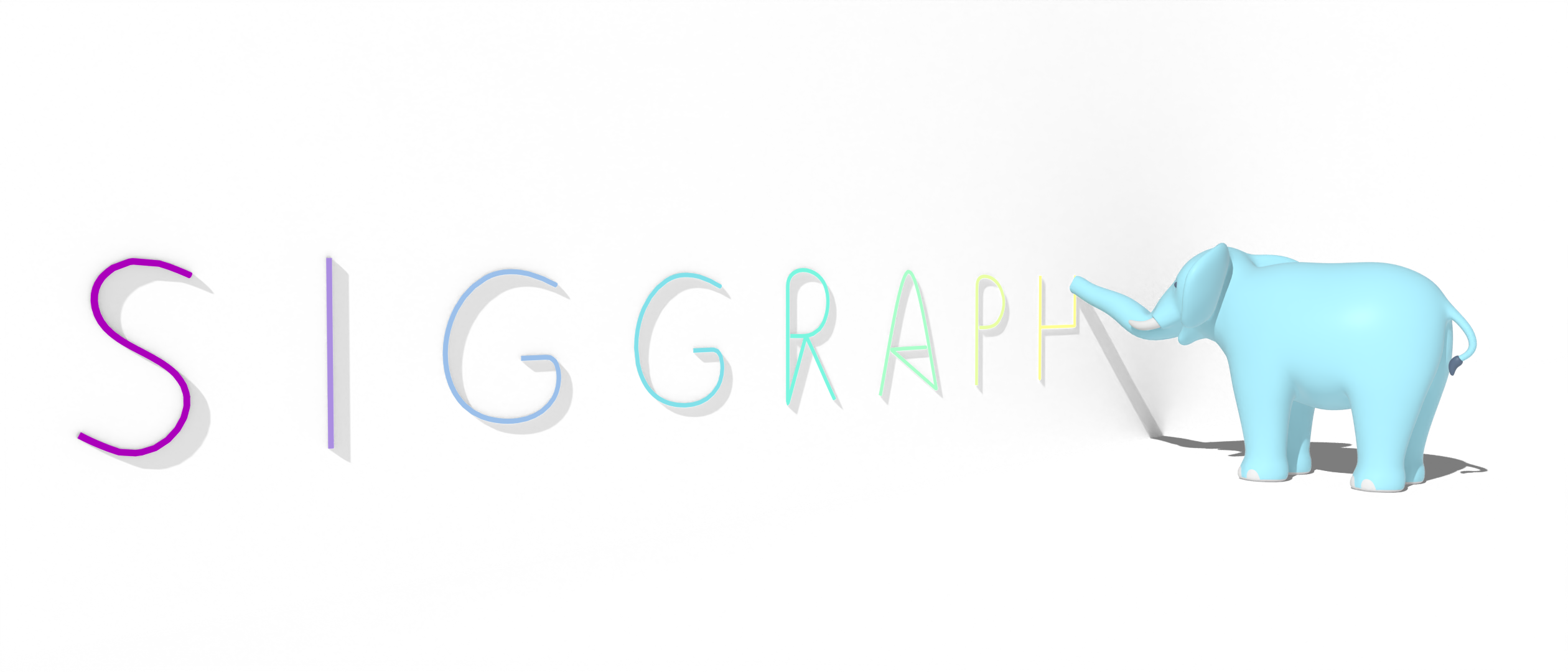}
  \caption{\textbf{Gradient-based inverse control of a soft elephant trunk.} We optimize a cable-driven elastic trunk to write the eight letters of ``SIGGRAPH'' on a wall. Our analytical adjoint drives a $5{,}220$-step writing trajectory by sequentially propagating gradients through a sliding optimization window, mapping the desired tip trajectory to the four cables’ per-step target lengths and tracing each letter to within $\approx\!1.4\%$ of its size. Each letter is optimized in about a minute, and the full eight-letter word completes in $\approx\!18$\,min end-to-end.}  \Description{A 3D render of a soft elephant trunk writing the word SIGGRAPH on a wall.}
  \label{fig:teaser}
\end{teaserfigure}

\maketitle

\section{Introduction}
\label{sec:intro}

The growing convergence of graphics, robotics, and machine learning is shifting attention from forward simulation---generating motion from known parameters---toward inverse problems: inferring material properties from observed behavior, or optimizing control inputs to achieve desired interactions \cite{newbury2024review}.
Differentiable simulation enables this shift by formulating the physics engine as a differentiable computational graph, allowing gradients of task-level objectives to flow back to physical parameters and controls.
The central difficulty is that the phenomena that make inverse problems interesting---frictional contact, large deformations, and long interaction horizons---are precisely those that make gradients difficult to compute reliably.

Existing approaches to differentiable simulation can be broadly classified by how they obtain gradients through the forward solver.
\emph{Automatic differentiation} and unrolling-based methods \cite{hu2019difftaichi, macklin2024warp, newton2025} propagate derivatives through every solver operation, offering generality at the cost of memory and computation that scale with the number of time steps and solver iterations.
\emph{Analytical adjoint} methods derive a linear adjoint system by implicitly differentiating the converged forward solution via the implicit function theorem, decoupling gradient derivation from the iteration path of the forward solver; they differ chiefly in how this adjoint linear system is then solved.
One family reuses the forward solver's iteration structure as a stationary fixed-point (preconditioned Richardson) scheme on the adjoint linear system \cite{du2021diffpd, li2022diffcloth}; convergence is then governed by the spectral radius of the underlying splitting, and gradient quality can become entangled with backward-solver tolerances and iteration counts \cite{holl2020learning}.
Another family solves the adjoint linear system with direct or Krylov methods \cite{geilinger2020add, huang2024differentiable, shen2025elastic}, treating the backward pass as a linear-algebra problem distinct from the forward nonlinear iteration.
We adopt this latter route.
However, the specific challenges depend on the forward formulation: its contact model, constitutive law, and solver structure each introduce distinct sub-problems.

In this work, we develop an analytical adjoint framework for a coupled position-velocity system with NCP-based frictional contact and projective elasticity.
Three parallel challenges emerge from this formulation.
(1)~\textbf{Contact non-smoothness.}
The complementarity conditions that enforce Signorini--Coulomb contact are inherently non-differentiable at regime transitions (contact/separation, static/kinetic friction).
Under a hard NCP formulation, this non-smoothness produces exactly zero gradients in the contact regime; penalty and barrier alternatives provide nonzero gradients but couple gradient quality to barrier or stiffness settings (Section~\ref{sec:ncp}).
We introduce a smoothed Fischer--Burmeister function applied consistently in both the forward and backward passes, providing nonzero, continuous gradients across all contact regimes while preserving the physical fidelity of complementarity-based contact.
(2)~\textbf{Elastic projection generality.}
Existing PD-based differentiable simulators rely on PD-compatible approximations of hyperelasticity (e.g., ARAP-based) whose projections admit closed-form Jacobians.
We derive an SVD-based Jacobian for the local projection operator that supports isotropic hyperelastic models such as Neo-Hookean.
(3)~\textbf{Adjoint system conditioning.}
The adjoint system is a fully determined linear system whose coefficient matrix combines a sparse elastic block and a low-rank contact block, assembled from the converged forward solution.
Prior work reuses the forward solver's nonlinear iteration for the backward pass, a natural but suboptimal choice for a linear problem; we instead employ a GPU-parallel Krylov solver paired with a Woodbury preconditioner that exploits both blocks, outperforming alternatives across all contact regimes (Section~\ref{sec:solver}).

Our framework builds on the forward solver of \citet{zeng2025fast}, which formulates deformable multi-body dynamics with Coulomb friction as a coupled residual system solved by a non-smooth Newton method within a GPU-parallel local-global iteration scheme.
Our differentiation relies on three structural properties of this formulation:
the solver converges to a well-defined fixed point, enabling implicit differentiation via the implicit function theorem;
all physics are assembled into a unified residual, from which analytical Jacobians can be derived;
and frictional contact is modeled via NCP complementarity, encoding exact Signorini--Coulomb conditions as a complement to penalty- and barrier-based formulations.

In summary, our contributions are:
\begin{itemize}
    \item \textbf{Framework.} An analytical adjoint framework for NCP-based deformable contact dynamics, integrating smoothed-FB contact gradients, SVD-based projection gradients, and a Woodbury-preconditioned adjoint solver into a single GPU pipeline whose modules share a common adjoint matrix and reuse the forward NCP caches (Delassus operator and block-diagonal contact kernel) inside the preconditioner.
    \item \textbf{Analytical Differentiation.} Implicit-function-theorem differentiation of the converged coupled residual; a smoothed Fischer--Burmeister formulation with matched forward/backward smoothing for cross-regime contact gradients; and a closed-form SVD-based Jacobian for local projections that supports standard isotropic hyperelastic models (e.g., Neo-Hookean).
    \item \textbf{Adjoint Solver.} The first Woodbury-preconditioned Krylov adjoint solver for differentiable simulation, exploiting the cached Delassus operator from the forward NCP solve and the block-diagonal contact kernel to deliver convergence that outperforms alternatives across SPD, ill-conditioned SPD, and asymmetric contact regimes.
    \item \textbf{Experiments.} Gradient-accuracy benchmarks against finite differences, comparisons against DiffPD, Polyfem, and Newton, material identification tasks, and inverse-dynamics control on locomotion and manipulation tasks.
\end{itemize}

\begin{figure*}[!t]
    \centering
    \includegraphics[width=\linewidth]{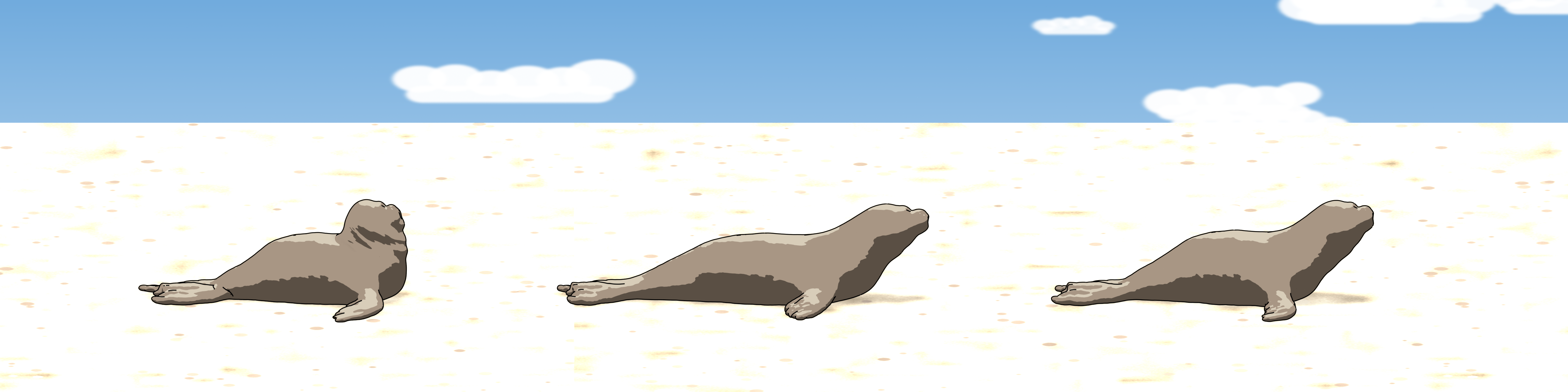}
    \caption{\textbf{Gradient-optimized seal locomotion gait (preview).}
    An elastic seal with $14$ muscle groups, optimized from a sinusoidal initialization (which alone drifts \emph{backward}) to a forward-crawl gait at $\sim 1.1$\,m/s, propelled entirely by frictional ground contact.
    See Section~\ref{sec:applications} for details.}
    \Description{Filmstrip showing successive frames of an elastic seal crawling forward on a ground plane through coordinated muscle contraction.}
    \label{fig:seal_stride}
\end{figure*}

Figure~\ref{fig:framework_overview} (Section~\ref{sec:framework}) shows our framework at a glance, and Figure~\ref{fig:seal_stride} previews a representative application detailed in Section~\ref{sec:applications}.

\section{Related Work}
\label{sec:related}

\subsection{Gradient Computation for Physics Simulation}
Computing gradients through physical simulation is the foundation of differentiable simulation, enabling gradient-based optimization for system identification, control, and design \cite{newbury2024review}.
Two principal strategies exist, complemented by a learning-based alternative.

\emph{Automatic differentiation} (AD) propagates derivatives at the elementary-operation level \cite{baydin2018automatic} and is supported by frameworks such as PyTorch \cite{paszke2019pytorch}, JAX, Warp \cite{macklin2024warp} / Newton \cite{newton2025}, DiffTaichi \cite{hu2019difftaichi}, and the earlier MLS-MPM-based ChainQueen \cite{hu_chainqueen_2019}.
While highly general, AD incurs memory and computational overhead that scales with simulation length and solver complexity---each time step and each inner-solver iteration of an implicit step adds a layer to the unrolled graph---which can be prohibitive for long-horizon or implicitly-solved systems.

\emph{Analytical adjoint} methods derive a linear adjoint system by implicit differentiation of the converged forward solver, decoupling gradient derivation from the iteration path of the forward pass \cite{geilinger2020add, qiao2021efficient, hahn_real2sim_2019, du2021diffpd, li2022diffcloth, huang2024differentiable, shen2025elastic}.
Methods within this family differ chiefly in how the resulting linear system is solved.
DiffPD \cite{du2021diffpd} and DiffCloth \cite{li2022diffcloth, yu2023diffclothai} reuse the forward PD local-global structure as a semi-implicit fixed-point (preconditioned Richardson) iteration on the adjoint linear system; this is efficient when the spectral radius of the underlying splitting is small, but can stagnate or diverge when it is not.
ADD~\cite{geilinger2020add}, Polyfem~\cite{huang2024differentiable}, and \citet{shen2025elastic} instead solve the adjoint linear system with direct linear solvers, treating the backward pass as a linear-algebra problem distinct from the forward nonlinear iteration.
Our work also falls in this latter family: we implicitly differentiate the converged coupled residual of \cite{zeng2025fast} and solve the resulting adjoint linear system with a GPU-parallel Krylov method paired with a Woodbury preconditioner that exploits its sparse-plus-low-rank structure (Section~\ref{sec:solver}).
As an alternative to direct differentiation, \emph{neural surrogates}~\cite{li_learning_2019, sanchezgonzalez_learning_2020, ma_learning_2023} learn dynamics or constitutive laws from observations, trading physical fidelity for differentiability by construction.

\begin{table*}[t]
\centering
\caption{\textbf{Qualitative comparison of representative differentiable simulation frameworks.}
``Local-global'' refers to the family of implicit solvers that alternate between a parallelizable local step and a global linear solve, including PD~\cite{bouaziz2023projective} and ADMM-PD variants.}
\label{tab:comparison}
\footnotesize
\setlength{\tabcolsep}{4pt}
\begin{tabular*}{\textwidth}{@{\extracolsep{\fill}}lcccccc@{}}
\toprule
Method & Discretization & Time Integ. & Contact / Friction & Differentiation & Dynamic & GPU \\
\midrule
ADD \cite{geilinger2020add} & Rigid/artic. & Implicit & Penalty / approx.\ Coulomb & Adjoint & \cmark & \xmark \\
DiffPD/DiffCloth \cite{du2021diffpd, li2022diffcloth} & FEM / Mass-spring & Implicit (PD) & Penalty / compl. & Adjoint & \cmark & \xmark \\
DiffXPBD \cite{stuyck2023diffxpbd} & PBD & Implicit (XPBD) & Constraint proj. & Analytical & \cmark & \xmark \\
DiffTaichi \cite{hu2019difftaichi, huang2021plasticinelab} & MPM & Explicit & Grid-based & AD (source) & \cmark & \cmark \\
JAX-FEM \cite{xue2023jaxfem} & FEM & Quasi-static & \NA & AD (JAX) & \xmark & \cmark \\
tatva \cite{pundir2026tatva} & FEM & Implicit & Penalty & AD (global) & \cmark & \cmark \\
Newton \cite{newton2025} & FEM / MPM & Semi-implicit & Penalty & AD (tape) & \cmark & \cmark \\
Polyfem \cite{huang2024differentiable} & FEM & Implicit (Newton) & IPC barrier & Adjoint & \cmark & \xmark \\
\citet{shen2025elastic} & FEM & Implicit & IPC barrier & Mixed 2nd-order & \cmark & \xmark \\
\textbf{Ours} & \textbf{FEM} & \textbf{Implicit (local-global)} & \textbf{NCP / Coulomb} & \textbf{Adjoint} & \cmark & \cmark \\
\bottomrule
\end{tabular*}
\end{table*}

\subsection{Differentiable Contact Models}

The choice of contact model in the forward simulation shapes the quality and reliability of the resulting gradients.
We organize existing approaches by their contact formulation and discuss the associated differentiation strategies.

\paragraph*{Penalty and compliant models.}
Penalty methods, going back to classical cloth-contact treatments \cite{bridson_robust_2002}, model contact forces as smooth functions of penetration depth and are naturally amenable to differentiation \cite{geilinger2020add, heiden2019interactive, jatavallabhula2021gradsim}.
ADD \cite{geilinger2020add} is a representative framework in this family: it employs smooth penalty forces and derives analytical adjoint gradients for rigid and articulated multi-body systems with friction.
However, simple penalty models (e.g., piecewise linear or ReLU-based) yield zero gradients in inactive contact regions, blocking the optimizer from discovering new contact modes \cite{turpin2022grasp}.
More critically, \citet{shi2023unified} demonstrate that accurate gradient and Hessian information from penalty-based collision energies requires careful treatment of stiff potentials, as naive penalty formulations can produce severely ill-conditioned derivatives.

\paragraph*{Barrier methods and IPC}
Incremental Potential Contact (IPC) \cite{li2020incremental} formulates contact as a barrier energy that guarantees intersection-free trajectories under large deformation, and has been extended to a variety of body types and discretizations \cite{li2021codimensional, lan2021medial, ferguson2021intersection, lan2022affine, he2026m, chen2022a, huang2025stiffgipc}.
Its log-barrier potential is smooth and differentiable wherever the barrier is active, motivating several differentiable IPC pipelines for elastodynamics with friction~\cite{huang2024differentiable}, rotation-rich rigid-body dynamics~\cite{romanya2025painless}, garment recovery with diffusion priors~\cite{li2025dress}, and discrete-geometry artefact mitigation~\cite{du2024robust, huang2025geometric}; interior-point hyperelasticity can be further accelerated via preconditioned nonlinear CG~\cite{shen_preconditioned_2024}.
While barrier methods excel at ensuring intersection-free configurations, their gradients exhibit characteristic tradeoffs: the barrier force and its derivatives decay exponentially away from the contact surface and grow unboundedly as the gap approaches zero, which can challenge gradient-based optimization near mode transitions.
Additionally, IPC treats normal and frictional contact sequentially rather than as a coupled system: the friction force at each time step is computed from the barrier-derived normal force with the tangential contact frame evaluated at the previous configuration~\cite{li2020incremental}, rather than solving normal and frictional responses simultaneously as in complementarity-based formulations.
The two families thus occupy complementary design points: IPC's barrier favors intersection-free guarantees, while complementarity-based formulations preserve exact Signorini--Coulomb conditions. We adopt the latter (Section~\ref{sec:ncp}).

\paragraph*{Complementarity-based methods.}
Complementarity formulations model contact as a Nonlinear Complementarity Problem (NCP) or its linearized variant (LCP), rigorously enforcing Signorini--Coulomb conditions \cite{degrave2019differentiable}; a related primal/dual descent template~\cite{macklin_primaldual_2020} unifies penalty- and constraint-based dynamics under a common optimization view.
Forward solvers in this family---such as Projected Gauss-Seidel (PGS) and non-smooth Newton methods \cite{macklin2019non}---deliver high physical fidelity but produce non-smooth solution mappings.
To obtain gradients, one may implicitly differentiate the KKT conditions, but truncation of iterative solves can leave complementarity constraints unsatisfied, degrading gradient accuracy.
The reverse PGS method \cite{qiao2021efficient} mitigates this via adjoint-based gradient computation.
Impulse-based formulations are easier to differentiate but can yield inaccurate gradients at collision instants due to time discretization (e.g., the rigid-body impulse subset of DiffTaichi~\cite{hu2019difftaichi}); time-of-impact methods \cite{zhong2022differentiable, zhong2023improving} address this by resolving the precise collision time.
Our work operates within the NCP family: we adopt the non-smooth Newton solver of \cite{zeng2025fast} as our forward model and introduce a smoothed Fischer--Burmeister residual specifically to enable well-defined differentiation of the converged solution.
Related forward simulators integrate complementarity-based contact into projective dynamics \cite{ly2020projective} and peridynamics \cite{lu_projective_2024}, though without addressing differentiability.

\paragraph*{Optimization-based contact.}
A parallel line of work formulates the forward contact step as a QP or QCQP and extracts gradients by implicitly differentiating the optimality (KKT) conditions.
\citet{liang2019differentiable} apply this approach to differentiable cloth simulation, and \citet{le2021differentiable} and \citet{menager2025differentiable} extend it to soft and articulated multi-body dynamics with friction.

\paragraph*{Position-based and projective methods.}
Projective Dynamics (PD) \cite{bouaziz2023projective} and XPBD \cite{macklin2016xpbd} achieve robust forward integration through iterative position-level projections, accelerated by various forward-solver strategies~\cite{liu_quasinewton_2017, wang_chebyshev_2015, wang_descent_2016, peng_anderson_2018}.
DiffPD \cite{du2021diffpd} extends PD into a differentiable form using a semi-implicit fixed-point backward solve with complementarity-based static friction, and DiffCloth \cite{li2022diffcloth, yu2023diffclothai} further incorporates Coulomb dry friction via the complementarity and branch-switching formulation of \citet{ly2020projective}.
DiffPD has been further adapted to downstream control and manipulation tasks~\cite{nava_fast_2022, li_contact_2022}.
However, the branch-switching formulation introduces gradient discontinuities at regime transitions (contact/separation, static/kinetic friction), which can impede gradient-based optimization when the solution trajectory crosses branch boundaries.
DiffXPBD \cite{stuyck2023diffxpbd} derives analytical gradients for XPBD-based constrained dynamics.
DiffQN~\cite{cai_diffqn_2025} recasts differentiable elastodynamics as a quasi-Newton method with low-rank Hessian updates, an alternative to PD's constant-Hessian assumption.
A further limitation shared by DiffPD and DiffCloth is that their local projection derivatives are derived for PD-compatible approximations (e.g., ARAP with volume-preserving constraints) whose projections admit closed-form Jacobians; differentiating directly through standard isotropic hyperelastic energies such as Neo-Hookean requires an SVD-based projection Jacobian, which these frameworks do not provide and which we develop in Section~\ref{sec:pd}.
A common limitation of stationary fixed-point solvers applied to the adjoint linear system is that gradient quality depends on the iteration's convergence, entangling gradient information with backward-solver tolerances and iteration counts \cite{holl2020learning}.
The semi-implicit fixed-point iteration used in DiffPD and DiffCloth is tailored to the structure of PD-based adjoint systems; Section~\ref{sec:backward_comparison} shows it does not readily extend to the adjoint systems arising from NCP-based contact formulations, which exhibit stronger coupling and ill-conditioning.

\paragraph*{Second-order and hybrid methods.}
\citet{shen2025elastic} propose a hybrid second-order differentiable simulation scheme that combines analytical first-order gradients with Hessian information computed via complex-step finite differences, demonstrating convergent sims for elastic locomotion with contact.
Their approach offers richer curvature information that can accelerate quasi-Newton optimization, complementing our first-order adjoint (linear in parameter count); similar mixed-order ideas could extend to NCP-based settings, with our smoothed FB serving as the first-order component.

\paragraph*{GPU-accelerated differentiable deformable simulation.}
Several frameworks target GPU-based differentiable simulation of deformable bodies, each occupying a distinct point in the design space.
DiffTaichi~\cite{hu2019difftaichi} and its ecosystem (e.g., PlasticineLab~\cite{huang2021plasticinelab}) employ source-code-transformation AD on GPU-compiled kernels to differentiate through MLS-MPM solvers, achieving high throughput for soft-body manipulation; however, MPM's particle-based discretization differs fundamentally from mesh-based FEM, and contact is resolved implicitly through the background grid rather than via explicit contact models.
JAX-FEM~\cite{xue2023jaxfem} provides a GPU-accelerated differentiable FEM solver via JAX's AD infrastructure, supporting hyperelastic materials at scale ($>$7M DOFs), but targets quasi-static inverse design and lacks dynamic contact handling.
\citet{pundir2026tatva} recently propose a GPU-scalable FEM framework using globally-applied AD with JVP-based matrix-free solvers for implicit time integration, supporting deformable-body contact via penalty forces; as a concurrent preprint, its differentiable contact capabilities have not yet been independently validated.
Newton~\cite{newton2025} (formerly Warp \texttt{warp.sim}~\cite{macklin2024warp}) is a GPU-accelerated physics engine with multiple solver backends; however, only its semi-implicit integrator provides differentiability for deformable bodies, using tape-based AD with penalty contact; the more advanced XPBD and VBD solvers do not support backward passes at the time of this writing.

\paragraph*{Summary.}
Table~\ref{tab:comparison} provides a qualitative comparison of representative differentiable simulation frameworks.
Existing analytical adjoint methods either simplify contact physics (e.g., penalty mollification in ADD) or yield adjoint structures outside our sparse-plus-low-rank decomposition (Section~\ref{sec:solver}); the PD-based instances of this family (DiffPD, DiffCloth) further solve the adjoint linear system with a stationary fixed-point iteration whose convergence is governed by the spectral radius of the splitting, and their local projection derivatives are derived for PD-compatible approximations rather than standard isotropic hyperelastic energies.
GPU-accelerated AD frameworks offer hardware efficiency but either inherit the $O(T)$ memory scaling of tape-based differentiation (Newton, DiffTaichi), operate on particle-based discretizations without explicit contact models (DiffTaichi), or lack dynamic contact altogether (JAX-FEM).
Our framework combines NCP-based Coulomb friction, a fully analytical adjoint formulation with $O(1)$ per-step backward memory, and GPU-parallel solvers, occupying a point in this design space---mesh-based FEM with rigorous frictional contact and efficient GPU gradient computation---that is not addressed by existing work.

\section{Background}
\label{sec:background}

This section reviews the forward simulation framework of~\cite{zeng2025fast}, which serves as the foundation for the differentiable pipeline developed in Section~\ref{sec:method}.
Three structural properties of this formulation make it particularly amenable to implicit differentiation:
(i)~all physics---elasticity, binding constraints, and frictional contact---are assembled into a single coupled residual (Equation~\eqref{eq:dynamics}), from which analytical Jacobians can be derived;
(ii)~frictional contact is modeled via NCP complementarity, encoding exact Signorini--Coulomb conditions as a complement to penalty- and barrier-based formulations;
and (iii)~the forward solver drives this residual to convergence via a Newton-type method, providing the well-defined fixed point required by the implicit function theorem.

\subsection{Deformable Multi-body Dynamics}
Following~\cite{martin2011example}, given positions $\bpos$ and velocities $\bvel$ at time $t$, implicit Euler time integration solves for new positions $\pos$ at time $t+h$:
\begin{equation}
    \pos = \arg\min_{\pos'} \Big( \frac{1}{2h^2} \big\|\mass^{\frac{1}{2}}(\pos'-\ppos)\big\|^2_F + \sum\nolimits_i\psi_i(\pos')  \Big),
    \label{eq:implicit}
\end{equation}
where $\mass$ is the mass matrix, $\ppos = \bpos+h\bvel+h^2\inv{\mass}\ext$ is the inertial prediction, and $\psi_i$ denotes the $i$-th elastic energy.
Incorporating contact via Lagrange multipliers $\force$~\cite{macklin2019non} yields:
\begin{subequations}
\begin{align}
    \mass(\pos-\ppos) - h^2\internal(\pos) - h^2\sum_{j\in\lset}\trans{\jac}_j\force &= \zeros \\
    \forall j\in\lset, \quad \boldsymbol{\phi}_j(\pos, \force) &= \zeros,
\end{align}
\label{eq:lagrange}
\end{subequations}
where $\internal$ denotes the internal elastic forces, $\lset$ indexes the Lagrangian constraints---binding pairs $\bset\subseteq\lset$ and contact pairs $\cset\subseteq\lset$---$\jac$ is the contact Jacobian mapping the state to the constraint space, and $\boldsymbol{\phi}$ encodes the bilateral condition for $\bset$ and the Signorini--Coulomb condition for $\cset$.

\subsection{Contact Constraints}

\subsubsection{Bilateral Binding} Such constraints arise in joints and prescribed attachments (e.g., needle~\cite{Adagolodjo_robotic_2019} and cable~\cite{coevoet_software_2017} constraints).
For each $b\in\bset$, the residual is defined via a linear spring with stiffness $E_b$ and rest displacement $\rest_b$:
\begin{subequations}
\begin{align}
    \pene_b &\coloneqq \jac_b\pos - \rest_b \\
    \boldsymbol{\phi}_b &\coloneqq \pene_b + E_b\force_b.
\end{align}
\label{eq:binding}
\end{subequations}

\subsubsection{Non-interpenetration and Frictional Contact}
Given a contact normal $\mathrm{n}_c\in\real^3$, the constraint displacement $\pene_c\in\real^3$ decomposes into a normal gap $\delta_n\in\real$ and a tangential component $\pene_f\in\real^2$, with $\trans{\jac}_c=[\trans{\jac}_n,\trans{\jac}_f]$.
The complementarity conditions are expressed as residuals using the Fischer--Burmeister function $\fb(x,y)\coloneqq x+y-\sqrt{x^2+y^2}$:
\begin{subequations}
\begin{align}
    \pene_c &= \begin{bmatrix} \delta_n \\ \pene_f \end{bmatrix} \coloneqq
    \begin{bmatrix} \jac_n\pos - d_n \\ \jac_f(\pos-\bpos) - \rest_f \end{bmatrix} \\
    \boldsymbol{\phi}_c &= \begin{bmatrix} \phi_n \\ \phi_f \\ \boldsymbol{\phi}_{\parallel} \end{bmatrix} \coloneqq
    \begin{bmatrix} \fb(\delta_n, \lambda_n) \\ \fb(\|\pene_f\|, \mu\lambda_n - \|\force_f\|) \\
    \|\force_f\|\pene_f + \|\pene_f\|\force_f \end{bmatrix},
\end{align}
\label{eq:contact}
\end{subequations}
Note that normal and tangential components depend on the state in different ways (via $\pos$ and $\vel$ respectively), which enables a structured treatment of frictional contact in the unified residual.

\subsection{Projective Internal Force}
We model elastic internal forces via local-global solvers~\cite{bouaziz2023projective, overby_admm_2017}.
For each element, a projective state $\proj_i$ is obtained by minimizing a proximal objective that balances a quadratic penalty toward the current configuration against a constitutive potential $\zeta_i(\cdot,\xi_i)$:
\begin{subequations}
\begin{align}
    \proj_i &= \arg\min_{\proj'_i} \big( \frac{\weight}{2} \|\proj'_i-\map\pos\|_F^2 + \zeta_{i}(\proj'_i, \xi_i) \big) \\
    \internal &= -\textstyle\sum_i \weight\trans{\map}\big(\map\pos-\proj_i\big)
\end{align}
\label{eq:internal}
\end{subequations}
where $\weight$ is the stiffness weight and $\map$ is the discrete differential operator for element $i$.

Integrating Equations~\eqref{eq:binding}--\eqref{eq:internal} into Equation~\eqref{eq:lagrange} and concatenating all constraints, the complete dynamics becomes:
\begin{subequations}
\begin{align}
    \begin{aligned}
    \overbrace{\big(\mass + h^2\textstyle\sum_i \weight\trans{\map}\map\big)}^{\sys}\pos -
    \overbrace{\big(\mass\ppos + h^2\textstyle\sum_i \weight\trans{\map}\proj_i\big)}^{\mathbf{b}}& \\
    - h^2\trans{\jac}_b\force_b - h^2\trans{\jac}_c\force_c &= \zeros
    \end{aligned}& \\
    \jac_b\pos - \rest_b + E_b\force_b = \zeros& \\
    \label{eq:constraint_ncp_n}\fb(\delta_n, \lambda_n) = \zeros& \\
    \label{eq:constraint_ncp_f}\fb\big(\|\pene_f\|, \mu\lambda_n - \|\force_f\|\big) = \zeros& \\
    \label{eq:constraint_direction}\|\force_f\|\pene_f + \|\pene_f\|\force_f = \zeros& \\
    \label{eq:constraint_pd}\weight(\proj_i-\map\pos) + \nabla_{\proj_i}\zeta_{i}(\proj_i, \xi_i) = \zeros&.
\end{align}
\label{eq:dynamics}
\end{subequations}
Throughout the paper, $\xi_i$ denotes local material parameters (e.g., Young's modulus, Poisson's ratio), $\weight$ is the stiffness weight, $E_b$ controls binding compliance, and $\mu$ is the friction coefficient.
Section~\ref{sec:method} shows how to differentiate this coupled system implicitly to obtain gradients with respect to all states and parameters.

\paragraph*{GPU acceleration via sparse-inverse factorization.}
The forward solver of~\cite{zeng2025fast} accelerates the global solves on GPU by precomputing the Cholesky factor $L$ of $\sys = L L^\top$ and storing $\sparseinverse = L^{-1}$ as a sparse matrix, so that $\sys^{-1} = \sparseinverse^\top \sparseinverse$ is recovered \emph{exactly} via two sparse matrix--vector products. Both the global linear solve $\sys^{-1}\mathbf{b}$ and the assembly of the Schur-complement (Delassus) operator $\jac \sys^{-1} \trans{\jac}$ thereby reduce to SpMVs amenable to massive parallelism.
The remainder of the paper inherits this convention: all applications of $\sys^{-1}$ in both the forward and backward passes use this exact sparse-inverse factorization for GPU acceleration unless otherwise stated.

\section{Framework}
\label{sec:framework}

\begin{figure*}[!t]
    \centering
    \includegraphics[width=\linewidth]{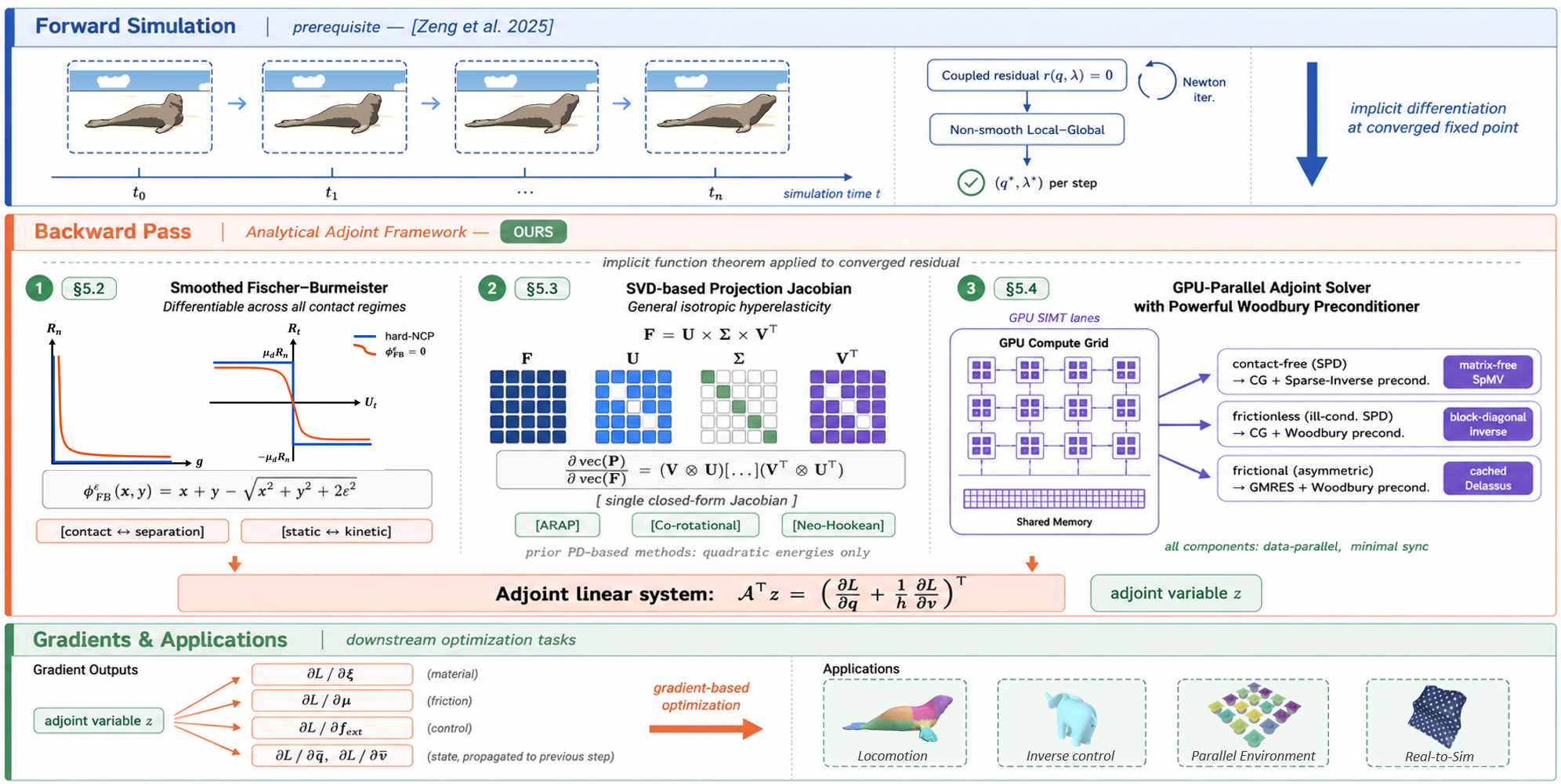}
    \caption{\textbf{Framework overview.}
    \emph{Top (blue):} the forward simulator of~\citet{zeng2025fast} converges each time step to a fixed point.
    \emph{Middle (orange):} three analytical components---smoothed Fischer--Burmeister contact (Section~\ref{sec:ncp}), SVD-based projection Jacobian (Section~\ref{sec:pd}), and Woodbury-preconditioned adjoint solver (Section~\ref{sec:solver})---assemble into a single adjoint linear system.
    \emph{Bottom (green):} the resulting gradients drive downstream optimization across material identification, friction estimation, locomotion, and inverse control (Section~\ref{sec:applications}).}
    \Description{Three horizontal blocks. Top: forward simulation showing a frame sequence and the coupled residual solver. Middle: backward pass showing the three analytical components (smoothed Fischer-Burmeister plot, SVD diagram, GPU compute grid with Woodbury-preconditioned adjoint solver). Bottom: gradient outputs feeding into application icons (material ID, friction estimation, soft-body locomotion, inverse control).}
    \label{fig:framework_overview}
\end{figure*}

Our differentiable simulation framework consists of three coupled phases (Figure~\ref{fig:framework_overview}): a forward simulator that converges each time step to a fixed point, an analytical adjoint that differentiates this fixed point in closed form, and a gradient stream that feeds downstream optimization tasks.

\paragraph*{Forward simulation.}
At each time step the forward simulator of \citet{zeng2025fast} solves a coupled residual $r(q,\lambda) = 0$ that integrates elastic projections, bilateral binding constraints, and complementarity-based frictional contact within a non-smooth local--global iteration; we treat its converged state $(q^*, \lambda^*)$ as a prerequisite (Section~\ref{sec:background}).
The solver further caches three structural quantities---the elastic inverse $\sys^{-1}$ in its sparse-inverse form $\sparseinverse^\top\sparseinverse$, the Delassus operator $\jac\sys^{-1}\trans{\jac}$, and the block-diagonal contact kernel $\kernal$---which the backward pass reuses without recomputation.

\paragraph*{Backward (analytical adjoint) pass.}
Applying the implicit function theorem to the converged residual yields a linear adjoint system whose coefficient matrix decomposes into a sparse elastic block and a low-rank contact correction.
We make this differentiation well-defined and practically tractable through: a smoothed Fischer--Burmeister contact formulation that provides continuous gradients across contact regimes (Section~\ref{sec:ncp}); a closed-form SVD-based Jacobian for the local elastic projection that supports standard isotropic hyperelastic energies (e.g., Neo-Hookean), going beyond the PD-compatible approximations of prior PD-based differentiable work (Section~\ref{sec:pd}); and a GPU-parallel Krylov solver with a Woodbury preconditioner that jointly exploits the elastic and contact structure of the linearized adjoint matrix (Section~\ref{sec:solver}).

\paragraph*{End-to-end GPU pipeline.}
Executing both the forward fixed-point solve and the backward adjoint solve entirely on the GPU further avoids the host--device transfers that frequently dominate wall-clock cost in mixed-architecture differentiable simulators, allowing per-stage speedups to compound across the pipeline rather than be amortized against transfer overhead.

\paragraph*{Gradient outputs and applications.}
The adjoint variable $z$ contracts against per-step structural matrices to produce gradients with respect to material parameters $\partial L/\partial \xi$, friction $\partial L/\partial \mu$, external forces or muscle activations $\partial L/\partial f_{\mathrm{ext}}$, and the previous state $\partial L/\partial \bar{q}, \partial L/\partial \bar{v}$ (propagated across time steps).
These gradient streams feed gradient-based optimization in Section~\ref{sec:applications}, covering material identification, friction estimation, soft-body locomotion, and inverse-dynamics control.

\section{Method}
\label{sec:method}

This section details the three analytical components of the backward pass introduced in Section~\ref{sec:framework}: the smoothed Fischer--Burmeister contact gradient, the SVD-based projection Jacobian, and the GPU-parallel adjoint solver.

\subsection{Adjoint Form}
\label{sec:adjoint}

Within one implicit time step, the kinematic relation $\vel=\frac{1}{h}(\pos-\bpos)$ makes $\vel$ an explicit function of $\pos$ and $\bpos$, so that $\pp{\vel}{\pos}=\frac{1}{h}$ and $\pp{\vel}{\bpos}=-\frac{1}{h}$.
In contrast, the next-step position $\pos$ is defined only implicitly through the converged forward dynamics in Equation~\eqref{eq:dynamics}, which couples $\pos$ to several auxiliary variables (e.g., $\force_b$, $\force_c$, and $\proj_i$).
To derive partial derivatives of $\pos$ with respect to the previous state and other inputs, we take the total differential of Equation~\eqref{eq:dynamics} together with the auxiliary optimality/constraint relations. 
This yields the coupled differential system:
\begin{subequations}
\begin{align}
    \label{eq:diff_main}\begin{aligned}
        \sys\dd\pos - \mass\dd\bpos - h\mass\dd\bvel - h^2\dd\ext - h^2\sum\nolimits_i \trans{\map}\proj_i\dd\weight& \\
        - h^2\sum\nolimits_i\weight\trans{\map} \dd\proj_i - h^2\trans{\jac_b}\dd\force_b - h^2\trans{\jac_c}\dd\force_c &= \zeros
    \end{aligned}& \\
    \dd\force_b + \inv{E_b}\jac_b\dd\pos - \inv{E_b}\dd\rest_b = \zeros& \\
    \label{eq:diff_ncp}\dd\force_c - \pp{\force_c}{\pos}\dd\pos - \pp{\force_c}{\bpos}\dd\bpos - \pp{\force_c}{\mu}\dd\mu = \zeros& \\
    \label{eq:diff_pd}\dd\proj_i - \pp{\proj_i}{\pos}\dd\pos - \pp{\proj_i}{\weight}\dd\weight - \pp{\proj_i}{\xi_i}\dd\xi_i = \zeros&.
\end{align}
\label{eq:differential} 
\end{subequations}
Equations~\eqref{eq:diff_ncp}--\eqref{eq:diff_pd} provide the local differential relations for frictional contact and local projections, respectively. 
Substituting the resulting expressions into Equation~\eqref{eq:diff_main} yields a linear system that relates $\dd\pos$ to differentials of all inputs and parameters:
\begin{equation}
\begin{aligned}
    \lhs\dd\pos &= \rhs_{\bpos}\dd\bpos + \rhs_{\bvel}\dd\bvel + \rhs_{\ext}\dd\ext + \rhs_{\rest_b}\dd\rest_b \\
    &+ \rhs_{E_b}\dd E_b + \rhs_{\mu}\dd\mu + \rhs_{\weight}\dd\weight + \rhs_{\xi_i}\dd\xi_i,
\end{aligned}
\end{equation}
which implicitly represents the partial derivatives of $\pos$ with respect to each variable on the right-hand side. 
The matrix blocks are
\begin{subequations}
\begin{align}
    \lhs &= \sys - \overbrace{h^2\sum\nolimits_i\weight\trans{\map}\pp{\proj_i}{\pos}}^{\Delta\sys} + \overbrace{h^2\trans{\jac_b}\inv{E_b}\jac_b}^{\rhs_b} + \overbrace{h^2\trans{\jac}_c(-\pp{\force_c}{\pos})}^{\rhs_c} \\
    \rhs_{\bpos} &= \mass + h^2\trans{\jac}_c\pp{\force_c}{\bpos},\; \rhs_{\bvel} = h\mass,\; \rhs_{\ext} = h^2\ids \\
    \rhs_{\rest_b} &= h^2\trans{\jac_b}\inv{E_b},\; \rhs_{E_b} = -h^2\trans{\jac_b}\inv{E_b}\force_b,\; \rhs_{\mu} = h^2\trans{\jac}_c\pp{\force_c}{\mu} \\
    \rhs_{\weight} &= h^2\trans{\map}(\proj_i+\weight\pp{\proj_i}{\weight}),\; \rhs_{\xi_i} = h^2\weight\trans{\map}\pp{\proj_i}{\xi_i} 
\end{align}
\label{eq:differential_2} 
\end{subequations}

In backpropagation, these partial derivatives are combined with the loss gradients $\pp{L}{\pos}$ and $\pp{L}{\vel}$ at the end of the time step.
For example, assuming row-vector gradients, the gradient with respect to $\bpos$ follows from the chain rule:
\begin{equation}
    \pp{L}{\bpos} = \pp{L}{\pos}\pp{\pos}{\bpos} + \pp{L}{\vel}\pp{\vel}{\bpos} = \underbrace{\big(\pp{L}{\pos} + \frac{1}{h}\pp{L}{\vel}\big)\inv{\lhs}}_{\trans{\adjoint}}\rhs_{\bpos} - \frac{1}{h}\pp{L}{\vel}.
\end{equation}
Rather than explicitly forming $\inv{\lhs}$ or individual sensitivities $\pp{\pos}{x}$, we introduce an adjoint variable $\adjoint$ defined by
\begin{equation}
    \trans{\lhs}\adjoint=\trans{\big(\pp{L}{\pos} + \frac{1}{h}\pp{L}{\vel}\big)}.
\label{eq:adjoint}
\end{equation}
Once $\adjoint$ is computed, gradients with respect to all inputs and parameters follow from simple matrix--vector products with the corresponding right-hand-side terms in Equation~\eqref{eq:differential_2}.
For all other variables $x\neq\bpos$, since $\pp{\vel}{x}=\zeros$, the gradient reduces to
\begin{equation}
    \pp{L}{x}=\trans{\adjoint}\rhs_{x}.
    \label{eq:compute_gradient}
\end{equation}
As illustrated in Figure~\ref{fig:framework_overview}, the gradients of sequential inputs (e.g., $\pp{L}{\ext}$) are final results for the current step, the gradients of global parameters (e.g., $\pp{L}{E_b}$) are accumulated across time steps, and the gradients of state variables $\pp{L}{\bpos}$ and $\pp{L}{\bvel}$ are propagated to the previous step.
In Sections~\ref{sec:ncp} and~\ref{sec:pd}, we detail the differentiation and assembly of $\rhs_c, \rhs_{\mu}, \rhs_{\weight}, \rhs_{\xi_i}$ and $\Delta\sys$.

\subsection{NCP Contact Gradient}
\label{sec:ncp}

We first recall how $\pene_c$ depends on the current and previous states.
With the tangent projector $\mathbf{P}_f := \Diag{0,\mathbf{I}_{2\times 2}}$, the partial derivatives needed for Jacobian assembly are
\begin{equation}
\pp{\pene_c}{\pos} = \jac_c,\quad \pp{\pene_c}{\bpos} = - \mathbf{P}_f\jac_c,
\label{eq:diff_contact}
\end{equation}
which will be combined with $\pp{\force_c}{\pene_c}$ derived below to form the contact contribution to the global Jacobian blocks.
Note that Equation~\eqref{eq:diff_contact} treats the contact Jacobian $\jac_c$ (and hence the local contact frame) as fixed with respect to $\pos$ within one time step.
This frozen-frame linearization is standard in complementarity-based contact solvers~\cite{macklin2019non, erleben_methodology_2018} and introduces an error of $O(h^2)$, consistent with the accuracy of the implicit Euler time integration.

Differentiating Equations~\eqref{eq:constraint_ncp_n}--\eqref{eq:constraint_direction} is complicated by the non-smoothness induced by complementarity: the Fischer-Burmeister residual is not differentiable at $(x,y)=(0,0)$, the locus that contact--separation and static--kinetic regime transitions traverse.
To obtain well-defined and numerically stable derivatives across all contact regimes, we adopt a smoothed Fischer-Burmeister function:
\begin{equation}
    \sfb(x,y)\coloneqq x+y-\sqrt{x^2+y^2+2\epsilon^2}
\end{equation}
with a small $\epsilon>0$, and use the same residual consistently in both the forward solve and the backward pass.
In practice, $\epsilon$ is held constant throughout the simulation and optimization; we set $2\epsilon^2=10^{-12}$ in all experiments unless otherwise noted, a value chosen to preserve forward fidelity while keeping the backward subgradients well-conditioned across regime transitions (Section~\ref{sec:epsilon_sweep}).
Specifically, the zero-level set $\sfb(x,y)=0$ is equivalent to the positive branch of $xy=\epsilon^2$, which is smooth for $\epsilon>0$.
In the following, we will frequently simplify derivative expressions on the zero-level set, since this is exactly the regime encountered after the converged solution:
\begin{equation}
    \pp{\sfb}{x} = \frac{y}{x+y}, \quad \pp{\sfb}{y} = \frac{x}{x+y}.
\label{eq:diff_fb}
\end{equation}

\paragraph*{Forward-backward smoothing consistency.}
Equation~\eqref{eq:diff_fb} is valid only when the same parameter $\epsilon$ used in the forward smoothing also appears in the backward subgradient evaluation, since the simplified ratio form relies on the converged manifold $xy = \epsilon^2$; we provide the derivation and an empirical validation on a friction-active task in Appendix~\ref{sec:smoothness_consistency}.
Throughout the rest of the paper, $\epsilon$ denotes a single parameter shared by both passes.

The remaining nonlinearity in the NCP stems from the Euclidean norm $\|\cdot\|$, which intrinsically couples the two tangential components.
For $\epsilon > 0$, the smoothed Fischer-Burmeister formulation ensures that the denominators involving $\|\pene_f\|$ and $\|\force_f\|$ are well-defined for almost all iterates after convergence.
Observing from Equation~\eqref{eq:constraint_direction} that $\pene_f$ and $\force_f$ are collinear with opposite orientations, we introduce an orthogonal transformation $\ortho$ for each contact.
This transformation rotates the contact variables into a local contact-aligned frame, thereby decoupling the normal component, the tangential magnitude, and the tangential direction.
As a result, the frictional complementarity constraints admit a linearized representation with a simple, nearly block-diagonal structure.
For more details on the construction of $\ortho$, we refer the reader to Appendix~\ref{sec:apx_A1}.
After expressing $\pene$, $\force$, and their differentials in the contact-aligned basis:
\begin{subequations}
\begin{align}
    \ortho\pene_c = 
    \begin{bmatrix} 
        \delta_n \\ 
        \|\pene_f\|\\
        0
    \end{bmatrix},&\quad
    \ortho\dd\pene_c = 
    \begin{bmatrix} 
        \dd\delta_n \\ 
        \dd\|\pene_f\|\\
        \frac{(\pene_f\times\dd\pene_f)_n}{\|\pene_f\|}
    \end{bmatrix} \\
    \ortho\force_c = 
    \begin{bmatrix} 
        \lambda_n \\ 
        -\|\force_f\|\\
        0
    \end{bmatrix},&\quad
    \ortho\dd\force_c = 
    \begin{bmatrix} 
        \dd\lambda_n \\ 
        -\dd\|\force_f\|\\
        -\frac{(\force_f\times\dd\force_f)_n}{\|\force_f\|}
    \end{bmatrix},
\end{align}
\label{eq:ortho}
\end{subequations}
where $(\,\cdot\,)_n$ denotes the normal component which is the only nonzero component. 

Substitute Equation~\eqref{eq:diff_fb} into the differential of Equations~\eqref{eq:constraint_ncp_n}--\eqref{eq:constraint_direction}:
\begin{subequations}
\begin{align}
    \frac{\dd\delta_n}{\delta_n} + \frac{\dd\lambda_n}{\lambda_n} &= 0 \\
    \frac{\dd\|\pene_f\|}{\|\pene_f\|} + \frac{\dd\big(\mu\lambda_n - \|\force_f\|\big)}{\mu\lambda_n - \|\force_f\|} &= 0 \\
    \Big( \frac{\pene_f}{\|\pene_f\|}\times\frac{\dd\pene_f}{\|\pene_f\|}
    - \frac{\force_f}{\|\force_f\|}\times\frac{\dd\force_f}{\|\force_f\|} \Big)_n &= 0.
\end{align}
\label{eq:constraint_ortho}
\end{subequations}

Comparing Equations~\eqref{eq:ortho} and \eqref{eq:constraint_ortho} and substituting Equation~\eqref{eq:diff_contact}, we obtain:
\begin{subequations}
\begin{align}
    \label{eq:partial_contact}\pp{\force_c}{\pene_c} &= - \kernal_c = - \trans{\ortho}
    \begin{bmatrix} 
        \frac{\lambda_n}{\delta_n} & 0 & 0 \\ 
        -\mu\frac{\lambda_n}{\delta_n} & \frac{\mu\lambda_n - \|\force_f\|}{\|\pene_f\|} & 0 \\
        0 & 0 & \frac{\|\force_f\|}{\|\pene_f\|} 
    \end{bmatrix} \ortho \\
    \rhs_c &= h^2\trans{\jac_c}\kernal_c\jac_c,\quad \rhs_{\bpos} = \mass + h^2\trans{\jac_c}\kernal_c\mathbf{P}_f\jac_c  \\
    \pp{\force_c}{\mu} &= - \mathbf{k}_{\mu} = - \trans{\ortho}
    \begin{bmatrix} 
        0 \\ 
        \lambda_n \\
        0
    \end{bmatrix}, \quad
    \rhs_{\mu} = -h^2\trans{\jac_c}\mathbf{k}_{\mu}.
\end{align}
\label{eq:final_diff_contact}
\end{subequations}

With Equation~\eqref{eq:final_diff_contact}, we obtain a unified formulation for differentiating NCP-based contact constraints.
As illustrated in the smoothed FB panel of Figure~\ref{fig:framework_overview}, the original (non-smoothed) complementarity formulation yields uninformative gradients: both finite-difference and analytical derivatives collapse to zero almost everywhere.
In contrast, our smoothed Fischer--Burmeister formulation provides a continuous relaxation across the non-smooth regimes of contact and separation, as well as static and kinetic friction, enabling well-defined gradients and natural transitions between regimes.
Theoretically, the smoothing replaces the hard complementarity condition $xy=0$ with $xy=\epsilon^2$, introducing a small perturbation to the contact forces.
At our working $2\epsilon^2 = 10^{-12}$, this perturbation is empirically negligible: Section~\ref{sec:epsilon_sweep} demonstrates that $1\,\text{\textperthousand}$ $\mu$-precision is preserved on a slip/stick benchmark.

\paragraph*{Cross-regime gradients for long-horizon optimization.}
The smoothed FB residual encodes contact/separation and static/kinetic transitions within a single continuous expression, yielding nonzero, directionally accurate gradients in every regime: with $\lambda_n \delta_n = \epsilon^2 > 0$, even contact-bound configurations satisfy $\partial \delta_n / \partial F \neq 0$, in contrast to the hard NCP limit where this derivative vanishes.
Unlike the constitutive stiffness parameters $k$ (penalty) and $\kappa$ (IPC)---whose gradient signal is inseparable from the force model and either shrinks as $1/k$ or spans the barrier's unbounded dynamic range---$\epsilon$ is a constraint-level regularization that relaxes $\lambda_n \delta_n = 0$ to $\lambda_n \delta_n = \epsilon^2$ without altering the force law: $\lambda_n$ remains a Lagrange multiplier determined by force balance.
This separation matters most over long horizons: the adjoint chain rule multiplies one contact Jacobian per regime crossing, so branch-switching formulations~\cite{du2021diffpd, li2022diffcloth} discard cross-regime sensitivity at each switch and penalty Jacobians attenuate exponentially with $k$, whereas smoothed FB keeps every Jacobian nonzero and preserves an informative chain across arbitrarily many transitions (Table~\ref{tab:long_horizon}).

\subsection{Local Energy Gradient}

For the local projections, Equation~\eqref{eq:constraint_pd} may be differentiated into the form of Equation~\eqref{eq:pd_expanded}.
However, this expression involves the Hessian $\nabla^2_{\proj_i}\zeta_i(\proj_i,\xi_i)$, which is typically not available in closed form for general hyperelastic models such as Neo-Hookean.
Existing PD-based differentiable frameworks~\cite{du2021diffpd, li2022diffcloth} derive the same analytical adjoint linear system but solve it via a semi-implicit fixed-point iteration that reuses the forward PD factorization; the stationary splitting requires the projection Jacobian to appear in closed form for the global step to remain factorization-free, and is therefore limited to PD-compatible approximations of hyperelasticity (e.g., ARAP with volume-preserving constraints). Our local step instead uses a standard Neo-Hookean energy directly.
Classical projected-Newton schemes restore convexity at the element level via eigenvalue clamping~\cite{teran_robust_2005} or absolute-value filtering~\cite{chen_stabler_2024} (see~\cite{kim_dynamic_2020} for a broader overview); these target the forward elastic Hessian rather than the prox-map Hessian relevant to our backward pass.
We take a different approach: rather than relying on a closed-form projection Hessian, we differentiate the projection operator itself via its SVD structure, yielding a closed-form Jacobian that applies to any isotropic hyperelastic model (instantiated for Neo-Hookean in our experiments).

\label{sec:pd}
\begin{equation}
\begin{aligned}
    \zeros &= \big(\weight\ids + \nabla^2_{\proj_i}\zeta_i(\proj_i, \xi_i)\big)\dd\proj_i - \weight\map\dd\pos &\\ 
    &+ (\proj_i-\map\pos)\dd\weight+ \pp{}{\xi_i}\nabla_{\proj_i}\zeta_i(\proj_i, \xi_i)\dd\xi_i,
\end{aligned}
\label{eq:pd_expanded}
\end{equation}

As an illustrative example, ARAP-style projections are defined through an SVD-based mapping of the deformation gradient.
Although the projection is computed algorithmically, it remains a deterministic function of the deformation as long as the SVD is non-degenerate, which enables analytic differentiation.
Concretely, we assume a mild singular-value separation condition $|\sigma_i-\sigma_j|>\tau$ for $i\neq j$ with a small threshold $\tau$; the degenerate case is treated in Appendix~\ref{sec:apx_B4}.

We define the projected state vector $\proj = \vect{\pdeform}$, where $\pdeform=\svdU\svdSS\trans{\svdV}$ is the optimized deformation gradient after local projection, sharing the same singular basis as $\deform$ but with modified singular values $\svdss(\svds)$ determined by the constitutive model.
By differentiating the SVD of $\deform$ and propagating through the projection $\pdeform(\deform)$ (see Appendix~\ref{sec:apx_B0} for the full derivation), we obtain the vectorized Jacobian:
\begin{subequations}
\begin{align}
    &\begin{aligned}
    \pp{\vect{\pdeform}}{\vect{\deform}} &= (\svdV\otimes\svdU) \big[ \mathbf{D}\svdW\trans{\mathbf{D}} + \Diag{\vect{\mathbf{M}}} \\
    &+ \Diag{\vect{\mathbf{M}}} \mathbf{T} \big] (\trans{\svdV}\otimes\trans{\svdU})
    \end{aligned} \\
    &\quad\quad\Delta\sys = h^2\sum\nolimits_i\weight\trans{\map}\pp{\vect{\pdeform}}{\vect{\deform}}\map
\end{align}
\end{subequations}
Here $\otimes$ denotes the Kronecker product, $\svdW \coloneqq \pp{\svdss}{\svds}$ is the model-dependent singular-value derivative (Appendix~\ref{sec:apx_B2}), and $\mathbf{D}$, $\mathbf{T}$ are fixed permutation matrices that convert between matrix and vector representations (Appendix~\ref{sec:apx_B1}).
The matrices $\mathbf{M}$ encode the coupling between singular-vector rotations and singular-value changes under the projection; their entries are
\begin{equation}
    M_{ii}=0,\quad M_{ij} = \frac{\sigma_i\theta_i-\sigma_j\theta_j}{\sigma_i^2-\sigma_j^2},
\end{equation}
which remain bounded under the non-degeneracy assumption.
The differentiation of elastic parameters (Young's modulus, Poisson's ratio) follows by composing this Jacobian with $\svdW$ and is detailed in Appendix~\ref{sec:apx_B3}.

\subsection{Fast Adjoint Linear Solver}
\label{sec:solver}

In previous sections, we derive a fully differentiable formulation for contact-involved dynamics by relaxing the non-smooth NCP constraints and expressing their differentials in a unified, contact-aligned form.
This enables contact, friction, and elasticity to be consistently incorporated into the backward pass.
All gradient contributions assemble into a unified sparse linear system:
\begin{equation}
 \trans{(\sys - \Delta\sys + \rhs_b + \rhs_c)}\adjoint = \trans{\big(\pp{L}{\pos} + \frac{1}{h}\pp{L}{\vel}\big)}
\end{equation}

This adjoint system is a \emph{fully determined linear system}: the coefficient matrix $\lhs$ is assembled from quantities evaluated at the converged forward solution and is therefore a known constant matrix at the time of the backward pass.
This stands in contrast to the forward problem, which is nonlinear and requires iterative solvers that interleave local projections with global linear solves.
Prior PD-based differentiable frameworks~\cite{du2021diffpd, li2022diffcloth} derive the same linear adjoint system but solve it by reusing the forward solver's local-global iteration, yielding a stationary splitting of the form
\begin{equation}\label{eq:semi-implicit}
    \sys\,\mathbf{z}^{k+1} = (\sys - \lhs)\,\mathbf{z}^k + \mathbf{b},
\end{equation}
which treats $\sys$ implicitly and the remainder $\lhs - \sys$ explicitly.
This is a natural design; it mirrors the forward solver and reuses its precomputed factorization of $\sys$.
However, when viewed through the lens of numerical linear algebra, Equation~\eqref{eq:semi-implicit} is a stationary splitting iteration (preconditioned Richardson) on the linear system $\lhs\,\mathbf{z} = \mathbf{b}$, with convergence rate governed by the spectral radius $\rho(\sys^{-1}(\sys - \lhs))$.
Since the adjoint system is linear, it admits more efficient solvers: Krylov methods (CG for SPD, GMRES for non-symmetric systems) achieve faster convergence under the same per-iteration cost---each iteration requires one application of $\sys^{-1}$ and one matrix--vector product with $\lhs$---while providing rigorous convergence guarantees that the stationary iteration lacks.
Moreover, the precomputed factorization of $\sys$ serves directly as a preconditioner for these Krylov solvers, preserving the computational reuse that motivates the splitting approach.

Efficient adjoint solves nonetheless face challenges: $\lhs$ is large-scale and sparse, and is in general neither symmetric positive definite nor well-conditioned.
We analyze its structure under three contact settings---contact-free, frictionless contact, and frictional contact---and introduce corresponding acceleration strategies below.

\subsubsection{Contact-free}
We begin with the simplest setting, where no contact is involved and $\lhs = \sys - \Delta\sys$.
As shown in Equation~\eqref{eq:pd_expanded},
\begin{equation}
    \Delta\sys = h^2 \sum_i \weight\, \trans{\map}
    \underbrace{\big(\ids + \inv{\weight}\nabla^2_{\proj_i}\zeta_i\big)^{-1}}_{\mathbf{J}}
    \map.
\end{equation}
This expression reveals a structure analogous to $h^2 \sum_i \weight\, \trans{\map}\map$, as both arise from the aggregation of local constraints.
By construction, $\lhs$ is symmetric.
To establish positive definiteness, note that when each $\zeta_i$ corresponds to a projection onto a constitutive manifold, its on-manifold Hessian satisfies $\nabla^2_{\proj_i}\zeta_i \succeq 0$.
Since $\weight > 0$, we have $\inv{\mathbf{J}} = \ids + \inv{\weight}\nabla^2_{\proj_i}\zeta_i \succeq \ids$, and therefore $\mathbf{J} \preceq \ids$, i.e., $\ids - \mathbf{J} \succeq 0$.
Consequently, $\lhs = \sys - \Delta\sys = \mass + h^2 \sum_i \weight\, \trans{\map}(\ids - \mathbf{J})\map \succeq \mass \succ 0$.
The resulting system is symmetric positive definite and admits a unique solution, which we solve efficiently using conjugate gradient (CG) preconditioned by $\sys^{-1}$, applied via the sparse-inverse factorization $\sparseinverse^\top \sparseinverse$ from Section~\ref{sec:background}.

\paragraph*{Matrix Assembly} The matrix pattern of $\Delta\sys$ is invariant since the local projective constraints are predefined, and the contributing elements likewise remain fixed; we therefore precompute a mapping from elements to pattern entries and reuse it to assemble $\sys - \Delta\sys$, yielding efficient assembly in both memory and time.

\subsubsection{Frictionless Contact}
When contacts or bilateral constraints exist but without friction ($\mu=0$), we have $\ortho\kernal_c\trans{\ortho}=\Diag{\frac{\lambda_n}{\delta_n}, 0, 0}\succeq 0$ in Equation~\eqref{eq:partial_contact}.
We denote
\begin{equation}
    \rhs
    = h^2 \trans{\jac}\kernal\jac
    = h^2
    [\trans{\jac_b}, \trans{\jac_c}]
    \Diag{\inv{E_b}\ids_b, \kernal_c}
    \begin{bmatrix}
        \jac_b \\
        \jac_c
    \end{bmatrix}.
\end{equation}
This assembled contact contribution is symmetric positive semi-definite, and therefore
$\lhs = \sys - \Delta\sys + \rhs$ remains symmetric and, under the same assumptions as in the contact-free case, is typically positive definite.

In practice, however, $\rhs$ can be severely ill-conditioned, which significantly degrades the effectiveness of using $\sys$ alone as a preconditioner.
This ill-conditioning primarily arises from two sources.
First, binding constraints are often enforced with very small compliance (i.e., $E_b \ll 1$) to fix or actuate points, introducing stiff directions into the system.
Second, Equation~\eqref{eq:final_diff_contact} implies that the contact block may develop extremely large eigenvalues under large normal contact forces $\force_n$.
Although smoothing ensures that $\delta_n$ remains strictly positive, it does not fully resolve the resulting conditioning issues.
We therefore introduce several complementary strategies to address this challenge.

\paragraph*{Jacobi Preconditioner} From Equation~\eqref{eq:final_diff_contact}, frictionless contact contributes only to the diagonal of $\lhs$.
In this setting, a Jacobi preconditioner based on the inverse diagonal, $(\lhs_{\mathrm{diag}})^{-1}$, provides a simple and inexpensive means of accelerating convergence via Jacobi scaling.

\paragraph*{Woodbury Preconditioner}
To our knowledge this is the first time a Woodbury-form preconditioner has been deployed for the adjoint linear system of a differentiable contact simulator.
Our principal preconditioner for the contact regime is constructed to jointly exploit the two structural blocks of the linearized adjoint matrix: the sparse elastic part $\sys - \Delta\sys$ and the low-rank contact correction $\rhs$.
We split the inverse via the Woodbury identity, which lets us apply $\sys^{-1}$ (already factorized in the forward pass) on the elastic side and a small dense solve on the contact side, never forming or factorizing the augmented matrix $\sys - \Delta\sys + \rhs$ itself.
This dual exploitation is what lets Woodbury outperform the simpler alternatives in Section~\ref{sec:backward_comparison}, even in regimes (contact-free, low-stiffness) where Jacobi's diagonal scaling alone would seem to be enough.
Concretely, the identity reads:
\begin{equation}
    \begin{aligned}
    \inv{(\sys + \rhs)}
    &= \inv{(\sys + h^2 \trans{\jac}\kernal\jac)} \\
    &= \inv{\sys}
    - h^2 \inv{\sys}\trans{\jac}
    \Big( \inv{\kernal} + h^2
    \jac \inv{\sys}\trans{\jac}
    \Big)^{-1}
    \jac \inv{\sys},
    \end{aligned}
    \label{eq:woodbury_formula}
\end{equation}
Two structural reuses are what make this preconditioner cheap enough to apply once per Krylov iteration; without them, Woodbury would itself become the bottleneck.
First, the Delassus operator $\jac \inv{\sys}\trans{\jac}$ that appears inside the contact-space inverse is precisely the operator already assembled and cached during the forward NCP solve, so we sidestep an explicit Schur-complement construction whose cost alone would rival a full forward step.
Second, $\kernal$ is block-diagonal with $1\times1$ (frictionless) or $3\times3$ (frictional) blocks, so $\inv{\kernal}$ is obtained pointwise in parallel rather than via any global factorization.
Equation~\eqref{eq:woodbury_formula} therefore reduces the inverse of a large sparse system to solves with $\sys^{-1}$ and a small dense solve in the \emph{contact space}, whose dimension scales only with the number of active contact constraints.
The inner matrix is SPD in the frictionless case and asymmetric under friction.
We use a hybrid inner solver: a direct factorization (LLT for SPD, LU for asymmetric) at small active-contact counts ($n_c < 1000$), switching to an iterative method (CG for SPD, GMRES for asymmetric) above this threshold to avoid direct-factorization fill-in at scale.

\paragraph*{Matrix-free Multiplication} Unlike $\Delta\sys$, the contact contribution $\rhs$ is dynamic. 
Explicitly assembling $\rhs$ leads to non-negligible overhead.
We propose using the matrix-free approach that transforms every SpMV in iterative solvers like CG as factorized multiplications:
\begin{equation}
    \rhs \mathbf{x} = h^2 \trans{\jac}\kernal\jac \mathbf{x} = h^2 \trans{\jac} ( \kernal (\jac \mathbf{x}))
\end{equation}
without explicitly assembling the matrix.
Consequently, the full SpMV for $\lhs$ becomes a hybrid form that includes $\sys - \Delta\sys$ with assembled matrix and $\rhs$ with matrix-free operations.

\subsubsection{Frictional Contact}
When friction is present ($\mu > 0$), the assembled contact contribution in Equation~\eqref{eq:final_diff_contact} is generally non-symmetric, and the resulting adjoint system is no longer guaranteed to be SPD.
Accordingly, we switch from CG to Generalized minimal residual method (GMRES), while retaining the same block-structured assembly and preconditioning strategy.

\paragraph*{Summary}
The proposed adjoint solvers and preconditioning strategies are designed to be highly amenable to parallel execution: matrix assembly, sparse matrix--vector multiplications, matrix-free contact operators, and block-diagonal inverses all exhibit strong data parallelism and minimal synchronization, and the reduction of global solves to contact-space systems enables efficient GPU implementations.
Section~\ref{sec:backward_comparison} shows that this Woodbury-centric design outperforms simpler preconditioners (Jacobi, sparse-inverse) and the semi-implicit reuse of DiffPD/DiffCloth across contact-free, frictionless, and frictional regimes, both in iteration count and in wall-clock time.
The gap is largest in the frictional regime, where the semi-implicit baseline diverges outright and Jacobi lags by an order of magnitude.

\section{Results}
\label{sec:results}

All experiments are conducted on a workstation with an Intel\textregistered\ Ultra 9 285K CPU and an NVIDIA GeForce RTX 5090 GPU (32\,GB).
Section~\ref{sec:gradient_accuracy} validates gradient correctness against finite-difference references.
Section~\ref{sec:backward_comparison} analyzes the convergence of our adjoint solver across contact regimes, with the DiffPD/DiffCloth semi-implicit iteration included as a baseline.
Section~\ref{sec:epsilon_sweep} ablates the smoothing parameter $\epsilon$ that governs the cross-regime gradient signal.
Section~\ref{sec:baseline_comparison} provides end-to-end comparisons against DiffPD, Polyfem, and Newton, representing the PD-adjoint, IPC, and AD differentiation routes, with Newton additionally serving as the GPU-based comparison point.
Section~\ref{sec:applications} demonstrates practical applications on contact-rich control tasks.
Table~\ref{tab:overview} summarizes mesh statistics, peak contact load, and GPU wall times---resolved at the per-iteration and per-time-step granularity---across all experiments; gradient accuracy and optimization-stage statistics are reported separately in Table~\ref{tab:analysis}.

\begin{table*}[ht!]
\centering
\caption{\textbf{GPU performance summary across all experiments.}
For each task we list a pointer to its corresponding figure and the scene setup: time-step size $\Delta t$, number of time steps $T$, mesh size as vertex count (Vert.) and element count (Elem.; triangles for shells, tetrahedra for volumes), constitutive model, density $\rho$ (kg/m$^3$), Young's modulus $E$ (Pa, or stretch stiffness $k_s$ for ARAP/Co-rotation), Poisson's ratio $\nu$, and friction coefficient $\mu$ ($-$ for contact-free or frictionless settings).
\textbf{Bold values in the $\{E,\nu,\mu\}$ columns mark the parameter being identified by the optimization (the value shown is its ground-truth target).}
For both \emph{Fwd} and \emph{Bwd}, ``Per opt iter (s)'' is the total wall time of a single optimization iteration (one full rollout plus one adjoint sweep), and ``Per step (ms)'' is the corresponding amortized cost per simulation step.
``Memory'' is the peak GPU memory consumed during the full optimization, and ``Total'' is the end-to-end wall time, estimated as ``Per opt iter'' multiplied by the optimization iteration count.}
\label{tab:overview}
\footnotesize
\setlength{\tabcolsep}{4pt}
\renewcommand{\arraystretch}{0.9}
\begin{tabular*}{\textwidth}{@{\extracolsep{\fill}}lrrccccrrrrrrr@{}}
\toprule
\multirow{2}{*}{Example} & \multicolumn{7}{c}{Setup} & \multicolumn{2}{c}{Per opt iter (s)} & \multicolumn{2}{c}{Per step (ms)} & \multirow{2}{*}{\makecell{Memory\\(MB)}} & \multirow{2}{*}{\makecell{Total\\(min)}} \\
\cmidrule(lr){2-8} \cmidrule(lr){9-10} \cmidrule(lr){11-12}
& $\Delta t$ (s) & $T$ & Vert. & Elem. & Constitutive
     & $\rho$ (kg/m$^3$) / $E$ (Pa) / $\nu$ & $\mu$
     & Fwd & Bwd
     & Fwd & Bwd & & \\
\midrule
\multirow{3}{*}{Fig.\,\ref{fig:demo_nh_stiffness}} & \multirow{3}{*}{$0.05$} & \multirow{3}{*}{$300$} & $1.2K$ & $3.2K$ & \multirow{3}{*}{ARAP} & \multirow{3}{*}{$1000$ / $\mathbf{10^{4}}$ / $0$} & \multirow{3}{*}{$-$} & $0.500$ & $2.160$ & $1.67$ & $7.20$ & $922$ & $4.4$ \\
& & & $5.0K$ & $22.9K$ & & & & $1.246$ & $5.713$ & $4.15$ & $19.04$ & $2256$ & $11.6$ \\
& & & $9.9K$ & $48.4K$ & & & & $2.638$ & $11.854$ & $8.79$ & $39.51$ & $4138$ & $24.2$ \\
\cmidrule(lr){1-14}
\multirow{3}{*}{Fig.\,\ref{fig:demo_nh_stiffness}} & \multirow{3}{*}{$0.05$} & \multirow{3}{*}{$300$} & $1.2K$ & $3.2K$ & \multirow{3}{*}{Co-rotation} & \multirow{3}{*}{$1000$ / $\mathbf{10^{4}}$ / $0.3$} & \multirow{3}{*}{$-$} & $0.805$ & $2.364$ & $2.68$ & $7.88$ & $922$ & $5.3$ \\
& & & $5.0K$ & $22.9K$ & & & & $1.494$ & $5.814$ & $4.98$ & $19.38$ & $2256$ & $12.2$ \\
& & & $9.9K$ & $48.4K$ & & & & $3.363$ & $11.234$ & $11.21$ & $37.45$ & $4138$ & $24.3$ \\
\cmidrule(lr){1-14}
\multirow{3}{*}{Fig.\,\ref{fig:demo_nh_stiffness}} & \multirow{3}{*}{$0.05$} & \multirow{3}{*}{$300$} & $1.2K$ & $3.2K$ & \multirow{3}{*}{Neo-Hookean} & \multirow{3}{*}{$1000$ / $\mathbf{10^{4}}$ / $0.3$} & \multirow{3}{*}{$-$} & $1.113$ & $2.220$ & $3.71$ & $7.40$ & $922$ & $5.6$ \\
& & & $5.0K$ & $22.9K$ & & & & $1.931$ & $4.801$ & $6.44$ & $16.00$ & $2256$ & $11.2$ \\
& & & $9.9K$ & $48.4K$ & & & & $3.481$ & $10.927$ & $11.60$ & $36.42$ & $4138$ & $24.0$ \\
\cmidrule(lr){1-14}
\multirow{3}{*}{Fig.\,\ref{fig:demo_wind}} & \multirow{3}{*}{$0.05$} & \multirow{3}{*}{$300$} & $1.2K$ & $3.2K$ & \multirow{3}{*}{ARAP} & \multirow{3}{*}{$1000$ / $2{\times}10^{4}$ / $0$} & \multirow{3}{*}{$-$} & $0.497$ & $1.305$ & $1.66$ & $4.35$ & $902$ & $3.0$ \\
& & & $5.0K$ & $22.9K$ & & & & $1.185$ & $3.427$ & $3.95$ & $11.42$ & $2184$ & $7.7$ \\
& & & $9.9K$ & $48.4K$ & & & & $2.565$ & $6.473$ & $8.55$ & $21.58$ & $3998$ & $15.1$ \\
\cmidrule(lr){1-14}
\multirow{3}{*}{Fig.\,\ref{fig:demo_bending}} & \multirow{3}{*}{$0.01$} & \multirow{3}{*}{$500$} & $2.1K$ & $4.1K$ & \multirow{3}{*}{\makecell{ARAP\\\& Bending}} & \multirow{3}{*}{$1000$ / $10^{5}$ / $0$} & \multirow{3}{*}{$-$} & $0.673$ & $0.949$ & $1.35$ & $1.90$ & $1224$ & $2.7$ \\
& & & $5.1K$ & $10.0K$ & & & & $0.935$ & $1.415$ & $1.87$ & $2.83$ & $2218$ & $3.9$ \\
& & & $10.2K$ & $20.2K$ & & & & $1.853$ & $2.232$ & $3.71$ & $4.46$ & $3994$ & $6.8$ \\
\cmidrule(lr){1-14}
\multirow{3}{*}{Fig.\,\ref{fig:demo_poisson}} & \multirow{3}{*}{$0.01$} & \multirow{3}{*}{$200$} & $0.9K$ & $3.0K$ & \multirow{3}{*}{Neo-Hookean} & \multirow{3}{*}{$1000$ / $10^{6}$ / $\mathbf{0.4}$} & \multirow{3}{*}{$-$} & $0.800$ & $0.576$ & $4.00$ & $2.88$ & $782$ & $2.3$ \\
& & & $2.0K$ & $7.9K$ & & & & $0.839$ & $0.794$ & $4.19$ & $3.97$ & $1040$ & $2.7$ \\
& & & $4.9K$ & $21.2K$ & & & & $1.268$ & $1.263$ & $6.34$ & $6.32$ & $1704$ & $4.2$ \\
\cmidrule(lr){1-14}
\multirow{3}{*}{Fig.\,\ref{fig:demo_curtain}} & \multirow{3}{*}{$0.05$} & \multirow{3}{*}{$300$} & $2.1K$ & $4.1K$ & \multirow{3}{*}{\makecell{Neo-Hookean\\\& Bending}} & \multirow{3}{*}{$1000$ / $\mathbf{10^{4}}$ / $\mathbf{0.3}$} & \multirow{3}{*}{$-$} & $0.925$ & $1.189$ & $3.08$ & $3.96$ & $994$ & $3.5$ \\
& & & $5.1K$ & $10.0K$ & & & & $1.222$ & $1.876$ & $4.07$ & $6.25$ & $1654$ & $5.2$ \\
& & & $10.2K$ & $20.2K$ & & & & $1.666$ & $3.027$ & $5.55$ & $10.09$ & $2864$ & $7.8$ \\
\cmidrule(lr){1-14}
\multirow{3}{*}{Fig.\,\ref{fig:demo_bunny}} & \multirow{3}{*}{$0.01$} & \multirow{3}{*}{$100$} & $1.5K$ & $4.7K$ & \multirow{3}{*}{Neo-Hookean} & \multirow{3}{*}{$1000$ / $10^{7}$ / $0.3$} & \multirow{3}{*}{$\mathbf{0.1}$} & $2.696$ & $2.353$ & $26.96$ & $23.53$ & $830$ & $8.4$ \\
& & & $4.1K$ & $14.6K$ & & & & $2.832$ & $3.366$ & $28.32$ & $33.66$ & $1190$ & $10.3$ \\
& & & $9.8K$ & $50.7K$ & & & & $4.655$ & $9.152$ & $46.55$ & $91.52$ & $2396$ & $23.0$ \\
\cmidrule(lr){1-14}
\multirow{3}{*}{Fig.\,\ref{fig:demo_lifting}} & \multirow{3}{*}{$0.01$} & \multirow{3}{*}{$100$} & $2.0K$ & $8.9K$ & \multirow{3}{*}{Neo-Hookean} & \multirow{3}{*}{$1000$ / $5{\times}10^{5}$ / $0.3$} & \multirow{3}{*}{$0$} & $1.080$ & $1.030$ & $10.80$ & $10.30$ & $906$ & $3.5$ \\
& & & $4.9K$ & $24.2K$ & & & & $1.748$ & $2.019$ & $17.48$ & $20.19$ & $1356$ & $6.3$ \\
& & & $9.9K$ & $50.8K$ & & & & $3.024$ & $4.786$ & $30.24$ & $47.86$ & $2278$ & $13.0$ \\
\cmidrule(lr){1-14}
\multirow{3}{*}{Fig.\,\ref{fig:demo_tux}} & \multirow{3}{*}{$0.01$} & \multirow{3}{*}{$100$} & $2.0K$ & $9.4K$ & \multirow{3}{*}{Neo-Hookean} & \multirow{3}{*}{$1000$ / $10^{7}$ / $0.3$} & \multirow{3}{*}{$0.2$} & $0.957$ & $3.195$ & $9.57$ & $31.95$ & $926$ & $6.9$ \\
& & & $5.0K$ & $25.1K$ & & & & $1.574$ & $5.337$ & $15.74$ & $53.37$ & $1420$ & $11.5$ \\
& & & $9.8K$ & $51.4K$ & & & & $3.215$ & $8.330$ & $32.15$ & $83.30$ & $2444$ & $19.2$ \\
\cmidrule(lr){1-14}
\multirow{3}{*}{Fig.\,\ref{fig:demo_duck}} & \multirow{3}{*}{$0.05$} & \multirow{3}{*}{$600$} & $0.9K$ & $3.9K$ & \multirow{3}{*}{Neo-Hookean} & \multirow{3}{*}{$1000$ / $10^{5}$ / $0.3$} & \multirow{3}{*}{$0$} & $7.170$ & $1.888$ & $11.95$ & $3.15$ & $1168$ & $15.1$ \\
& & & $4.9K$ & $25.2K$ & & & & $12.952$ & $12.483$ & $21.59$ & $20.81$ & $3874$ & $42.4$ \\
& & & $10.0K$ & $52.9K$ & & & & $17.558$ & $24.183$ & $29.26$ & $40.30$ & $7638$ & $69.6$ \\
\cmidrule(lr){1-14}
\multirow{3}{*}{Fig.\,\ref{fig:demo_duck}} & \multirow{3}{*}{$0.05$} & \multirow{3}{*}{$600$} & $0.9K$ & $3.9K$ & \multirow{3}{*}{Neo-Hookean} & \multirow{3}{*}{$1000$ / $10^{5}$ / $0.3$} & \multirow{3}{*}{$0.1$} & $8.496$ & $4.443$ & $14.16$ & $7.41$ & $1176$ & $21.6$ \\
& & & $4.9K$ & $25.2K$ & & & & $14.374$ & $14.604$ & $23.96$ & $24.34$ & $3994$ & $48.3$ \\
& & & $10.0K$ & $52.9K$ & & & & $19.377$ & $31.690$ & $32.29$ & $52.82$ & $7982$ & $85.1$ \\
\cmidrule(lr){1-14}
\multirow{2}{*}{Fig.\,\ref{fig:seal_stride}} & \multirow{2}{*}{$0.01$} & \multirow{2}{*}{$550$} & $2.0K$ & $8.9K$ & \multirow{2}{*}{Neo-Hookean} & \multirow{2}{*}{$1000$ / $10^{6}$ / $0.3$} & \multirow{2}{*}{$0.5$} & $15.511$ & $6.192$ & $28.20$ & $11.26$ & $7664$ & $25.0$ \\
& & & $9.9K$ & $50.2K$ & & & & $26.753$ & $14.893$ & $48.64$ & $27.08$ & $16838$ & $47.9$ \\
\cmidrule(lr){1-14}
\multirow{3}{*}{Fig.\,\ref{fig:parallel_env_render}} & \multirow{3}{*}{$0.01$} & \multirow{3}{*}{$50$} & $4.1K$ & $7.7K$ & \multirow{3}{*}{\makecell{ARAP\\\& Bending}} & \multirow{3}{*}{$1000$ / $\mathbf{[10^{4},10^{6}]}$ / $0.4$} & \multirow{3}{*}{$0$} & $0.169$ & $0.593$ & $3.4$ & $11.9$ & $3070$ & $0.4$ \\
& & & $64.8K$ & $123.9K$ & & & & $3.227$ & $3.050$ & $64.5$ & $61.0$ & $6093$ & $3.8$ \\
& & & $518.7K$ & $991.2K$ & & & & $65.883$ & $14.715$ & $1317.7$ & $294.3$ & $29232$ & $45.2$ \\
\bottomrule
\end{tabular*}
\end{table*}

Rows in Table~\ref{tab:overview} sharing the same figure pointer but differing in Vert.\,/\,Elem. form a mesh-resolution ablation; the three Armadillo stiffness blocks form an elastic-model ablation; the final three-row block reports three representative scales ($N \in \{4, 64, 512\}$) from the full 8-point parallel-environment scaling sweep of Section~\ref{sec:parallel_sysid} (Total estimated at a representative $30$-iter optimization budget).
Optimization targets that lie outside the $\{E,\nu,\mu\}$ columns---body forces, bending coefficients, initial velocities---appear in Table~\ref{tab:analysis} with their initial and target values.

A consistent pattern in Table~\ref{tab:overview} is that the backward pass costs more per optimization iteration than the forward pass.
The adjoint Krylov solve uses a uniform residual tolerance of $10^{-8}$, much tighter than the forward physics requires; this tightness is essential because the gradient is propagated through every time step via the chain rule, so any per-step solver residual is amplified into vanishing or exploding gradients that corrupt the optimization signal.
The forward simulation, by contrast, tolerates partial residuals because per-step errors do not compound multiplicatively through the rollout.

\subsection{Gradient Accuracy}
\label{sec:gradient_accuracy}

We test our analytical gradients on the suite of 12 gradient-based system identification tasks summarized in Table~\ref{tab:analysis}.
The tasks span three differentiation challenges: (i) hyperelastic projection sensitivity, with seven elasticity demos covering ARAP, Co-rotation, and Neo-Hookean models, single- and joint-parameter targets, and volumetric and bending energies; (ii) cross-regime gradients through contact--separation transitions (Cow lifting); and (iii) friction-coupled gradients through static--kinetic transitions (Tux pulling, Bunny friction-coefficient identification) and repeated multi-mode contact events (Rubber-Duck frictionless and with friction).
Each task optimizes a single scalar parameter (or pair, for the Curtain joint $(E,\nu)$ task) toward a prescribed target using basic gradient descent with no quasi-Newton acceleration, so that convergence depends primarily on gradient quality.
The objective is a final-state matching loss,
\begin{equation}
    L = \|\pos_{\mathrm{final}}-\pos_{\mathrm{target}}\|^2,
\end{equation}
and gradients are validated against central finite-difference references,
\begin{equation}
    \fd{x}\coloneqq\frac{L(x+\eta)-L(x-\eta)}{2\eta}.
\end{equation}

\subsubsection{Elasticity}
We first isolate elasticity by disabling all contact interactions, so that the dynamics are governed solely by internal elastic forces.

\begin{figure}[!t]
    \centering
    \begin{subfigure}{\linewidth}
        \centering
        \includegraphics[width=\linewidth]{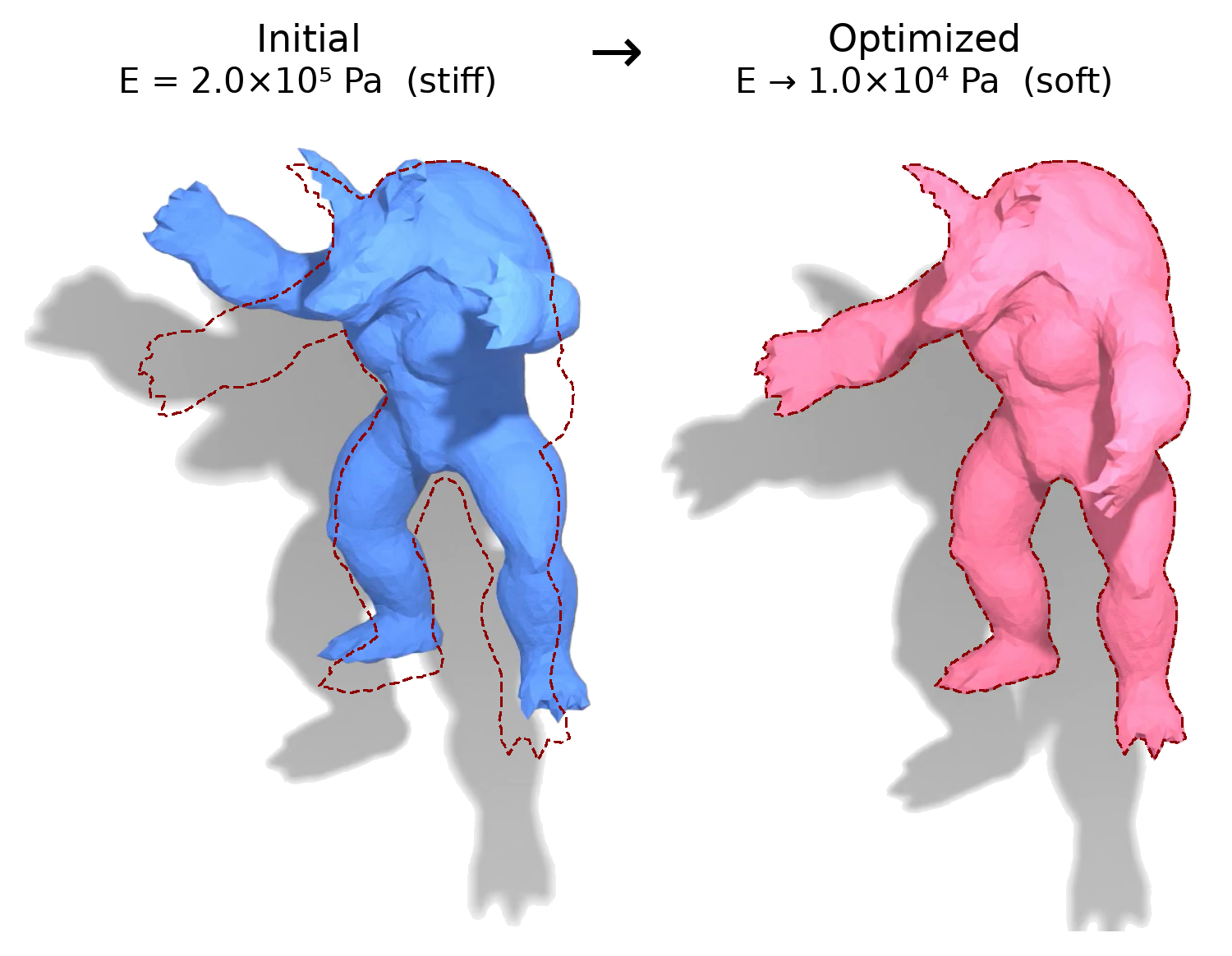}
        \caption{Configurations.}
    \end{subfigure}\\[2pt]
    \begin{subfigure}{\linewidth}
        \centering
        \includegraphics[width=\linewidth]{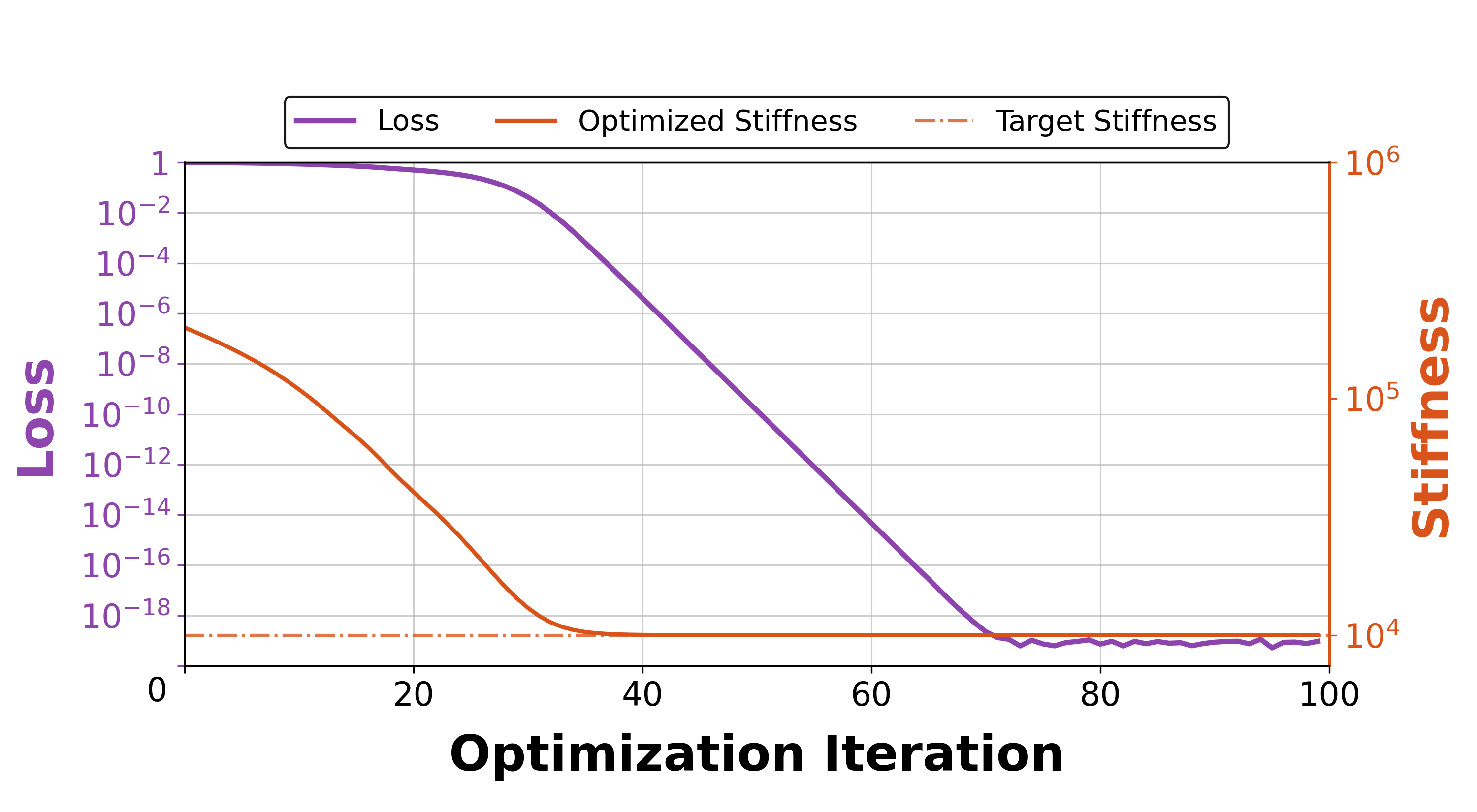}
        \caption{Loss and trajectory.}
    \end{subfigure}
    \caption{\textbf{Stiffness identification (Armadillo / NH).} Identifying $k_s$ from $2{\times}10^5$ toward $10^4$.}
    \Description{Armadillo Neo-Hookean stiffness identification: top row shows side-by-side renders of the initial and optimized armadillo against the target silhouette; bottom row plots loss decay and the stiffness trajectory across optimization iterations.}
    \label{fig:demo_nh_stiffness}
\end{figure}

As shown in Figure~\ref{fig:demo_nh_stiffness}, we perform stiffness identification on an Armadillo model, optimizing $k_s$ from $2\times 10^{5}$ toward $\tilde{k}_s = 10^{4}$.
To validate generality across constitutive laws, we run this task under three elasticity models---ARAP, co-rotational, and Neo-Hookean---using the same mesh ($1.2$K vertices) and target, changing only the SVD-based projection Jacobian $\svdW$.
Table~\ref{tab:analysis} reports MRE\textbar E of $2.10\times 10^{-2}$ for Co-rotation and $2.12\times 10^{-2}$ for Neo-Hookean (essentially identical) against $2.49\times 10^{-3}$ for ARAP, an order of magnitude lower.
This spread reflects the Poisson's-ratio setting of each run rather than the constitutive form: as recorded in Table~\ref{tab:overview}, ARAP uses $\nu = 0$ and therefore zero Lam\'e $\lambda$ (no volumetric coupling), whereas Co-rotation and Neo-Hookean both use $\nu = 0.3$ and share a comparable volumetric stress regime.
The agreement between Co-rotation and Neo-Hookean at matched $\nu$, despite their formally distinct $\svdW$ structures (a constant Sherman--Morrison expression vs.\ a $\theta$-dependent rational form), confirms that the analytical Jacobian is consistent across constitutive models; the only systematic predictor of MRE here is the physical regime, not the choice of $\svdW$.

\begin{figure}[!t]
    \centering
    \begin{subfigure}{\linewidth}
        \centering
        \includegraphics[width=\linewidth]{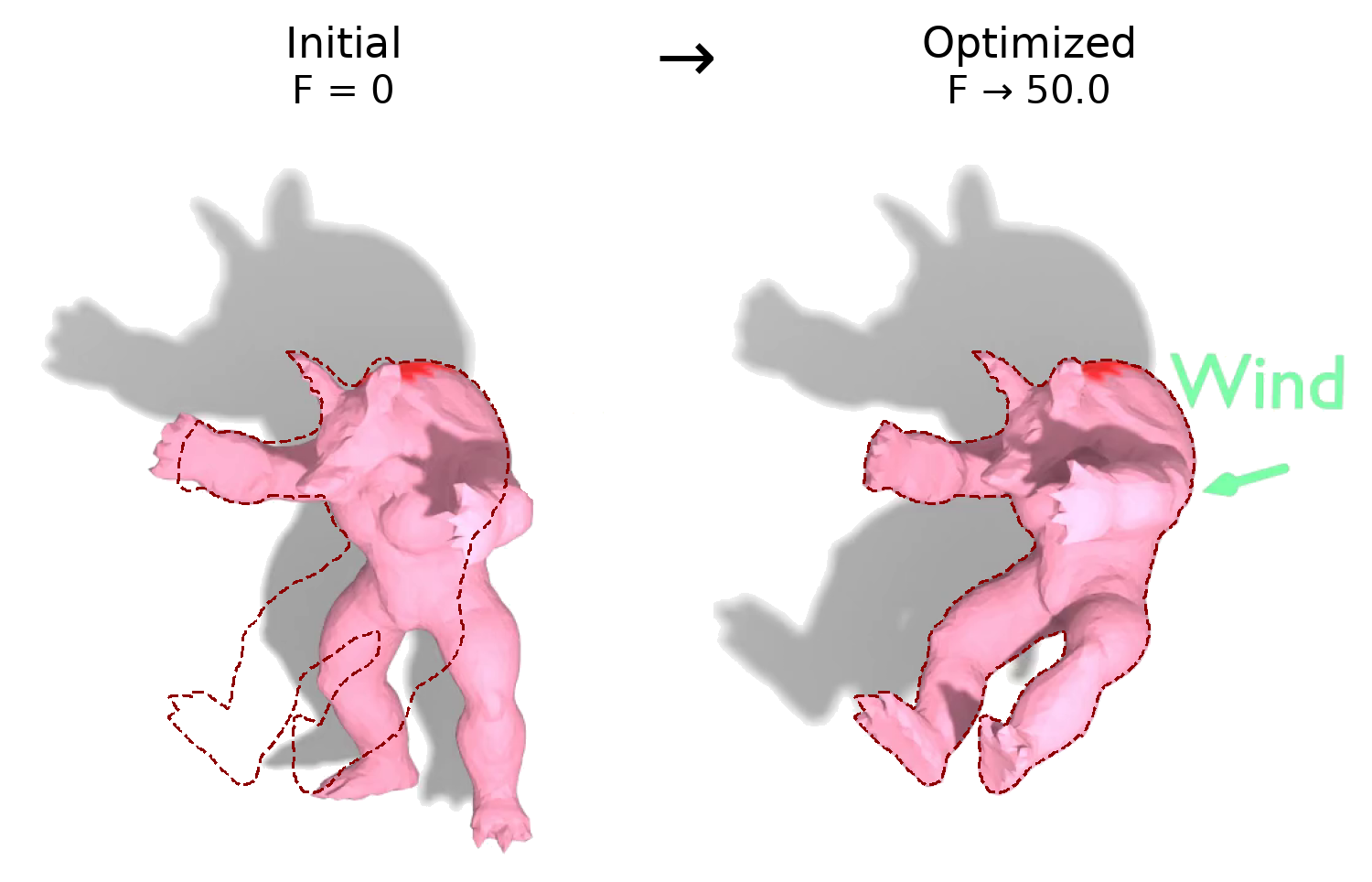}
        \caption{Configurations.}
    \end{subfigure}\\[2pt]
    \begin{subfigure}{\linewidth}
        \centering
        \includegraphics[width=\linewidth]{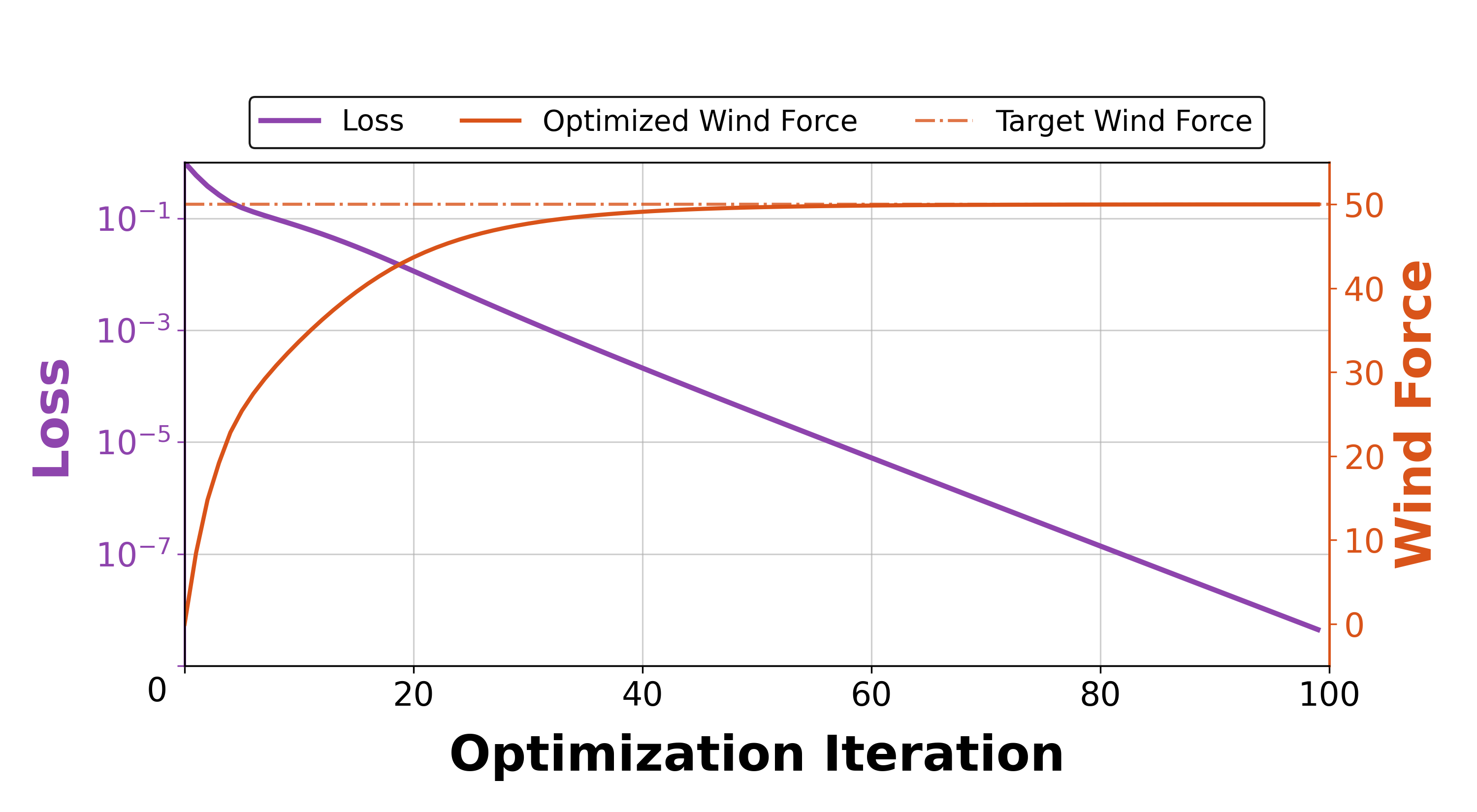}
        \caption{Loss and trajectory.}
    \end{subfigure}
    \caption{\textbf{Wind force identification (Armadillo).} Identifying $F_\mathrm{wind}$ from $0$ toward $50$.}
    \Description{Armadillo wind-force identification: top row shows the unforced and optimized armadillo deformations against the target silhouette; bottom row plots loss decay and the wind-force magnitude trajectory across optimization iterations.}
    \label{fig:demo_wind}
\end{figure}

\begin{figure}[!t]
    \centering
    \begin{subfigure}{\linewidth}
        \centering
        \includegraphics[width=\linewidth]{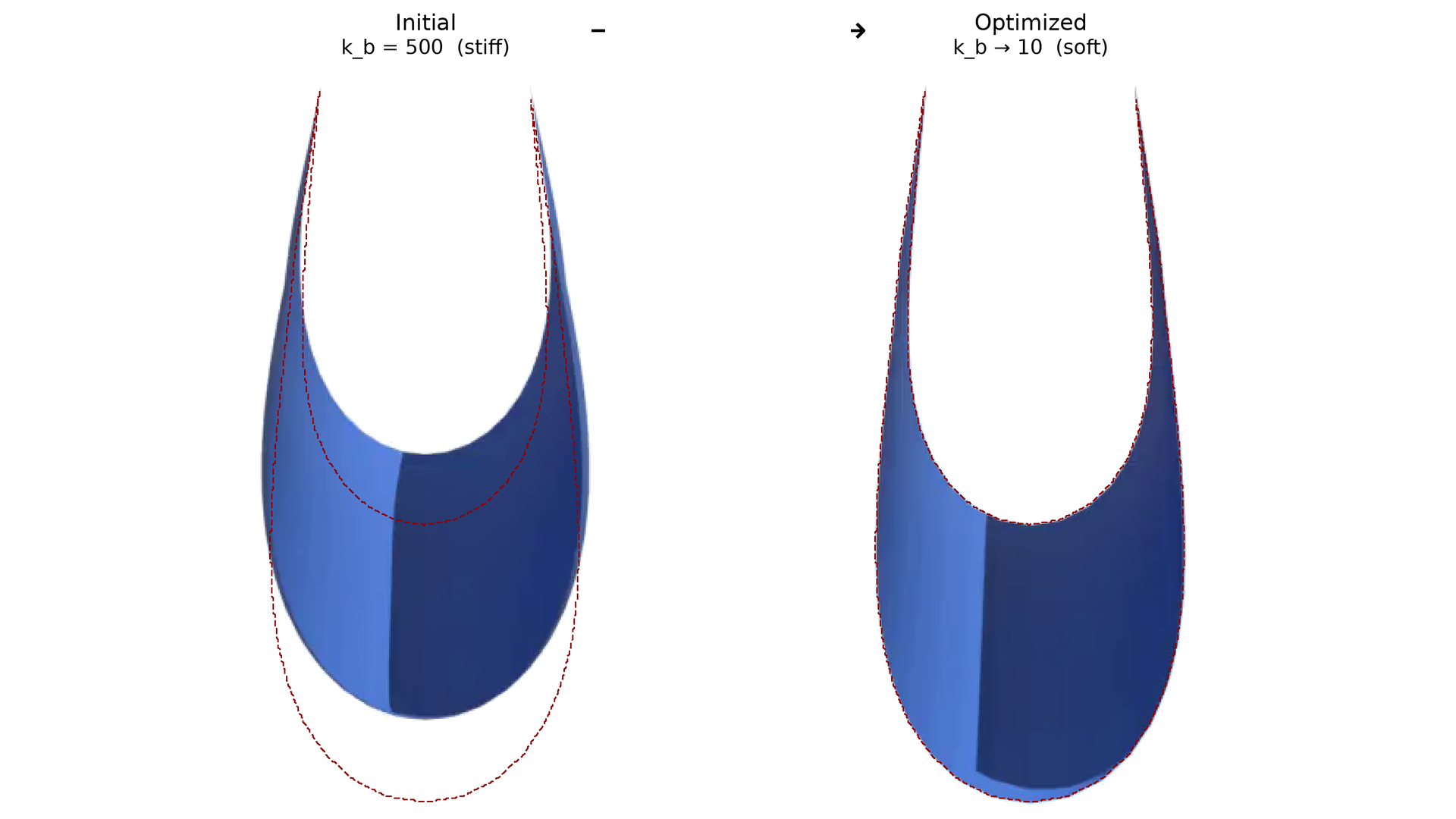}
        \caption{Configurations.}
    \end{subfigure}\\[2pt]
    \begin{subfigure}{\linewidth}
        \centering
        \includegraphics[width=\linewidth]{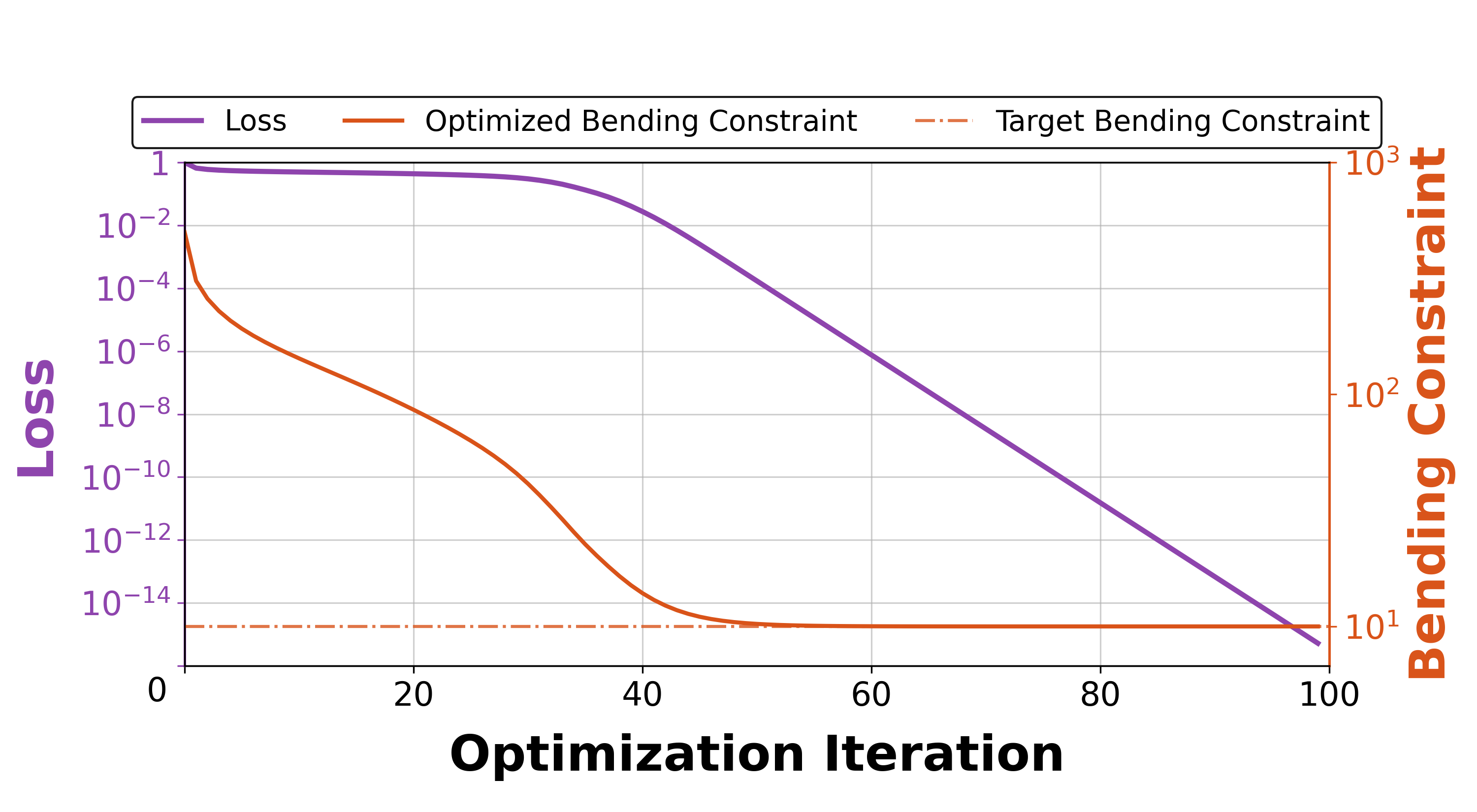}
        \caption{Loss and trajectory.}
    \end{subfigure}
    \caption{\textbf{Bending coefficient identification (Cloth).} Identifying $k_b$ from $5{\times}10^2$ toward $10$.}
    \Description{Cloth bending-coefficient identification: top row shows the stiff and optimized cloth shapes against the target; bottom row plots loss decay and the bending coefficient trajectory across optimization iterations.}
    \label{fig:demo_bending}
\end{figure}

We also identify wind force from zero toward $\tilde{f}_\mathrm{wind}=50$ (Figure~\ref{fig:demo_wind}), and bending coefficient $k_b$ of a rectangular cloth patch from $5\times10^{2}$ toward $\tilde{k}_b=10$ (Figure~\ref{fig:demo_bending}).

\begin{figure}[!t]
    \centering
    \begin{subfigure}{\linewidth}
        \centering
        \includegraphics[width=0.8\linewidth]{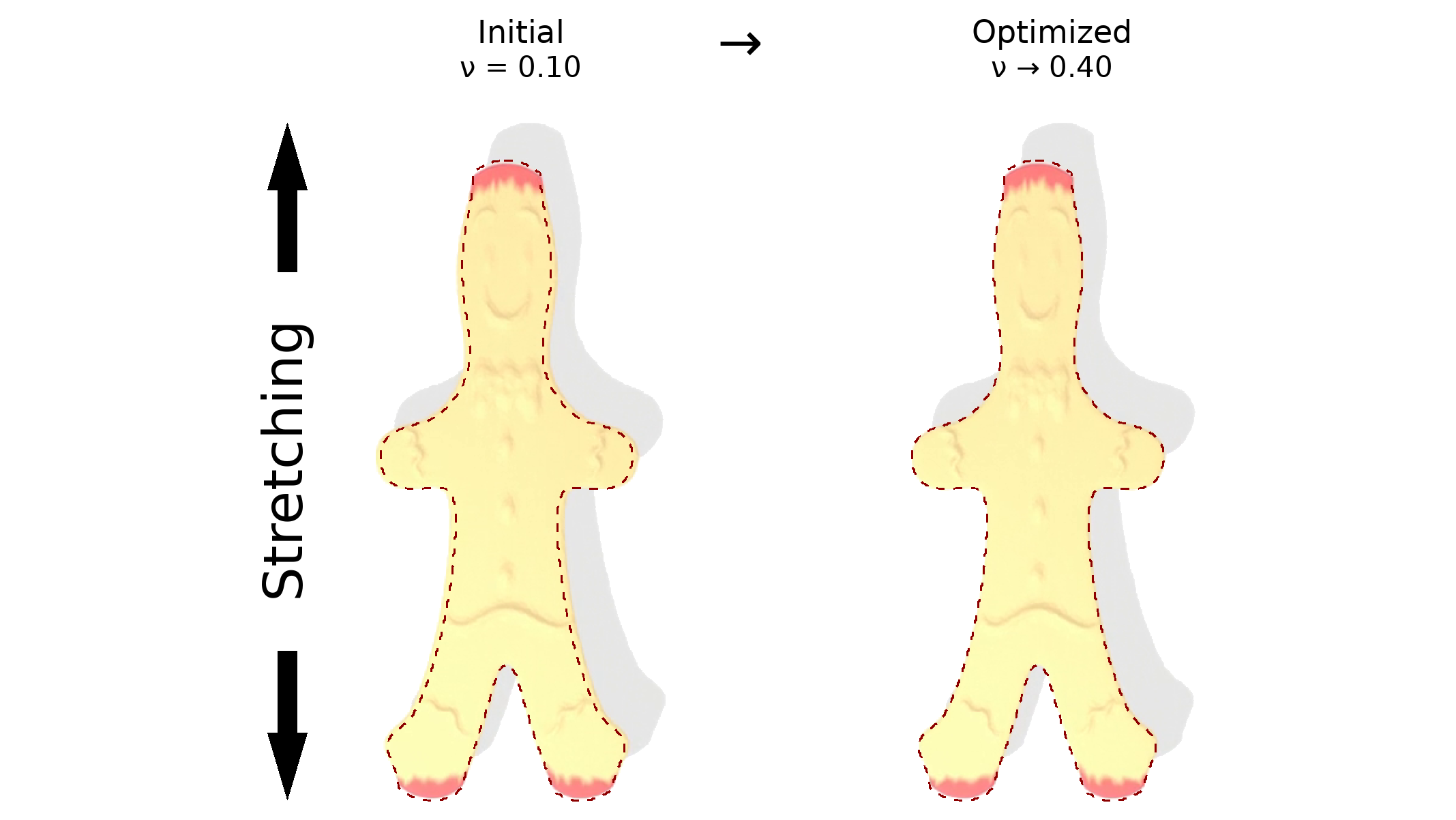}
        \caption{Configurations.}
    \end{subfigure}\\[2pt]
    \begin{subfigure}{\linewidth}
        \centering
        \includegraphics[width=\linewidth]{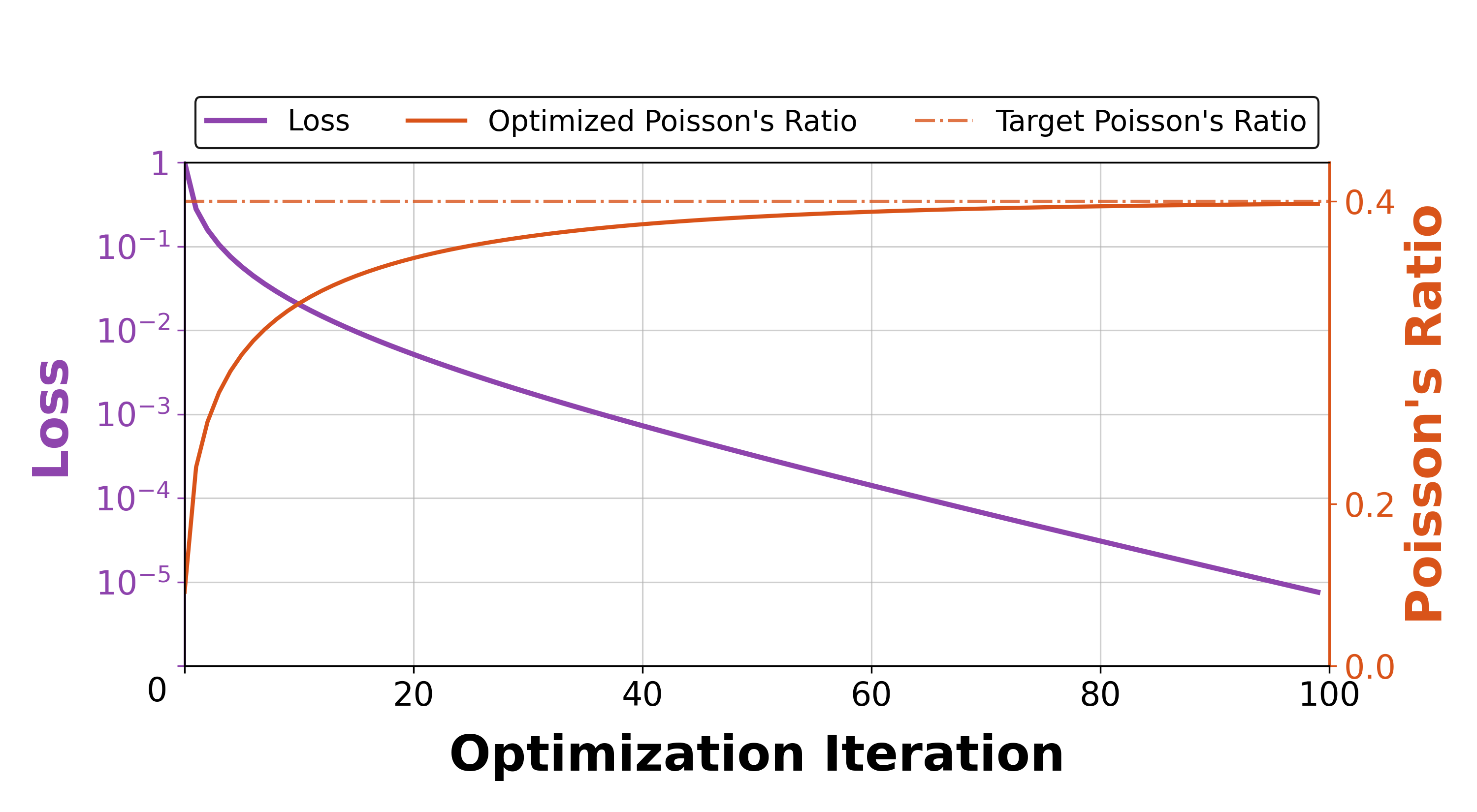}
        \caption{Loss and trajectory.}
    \end{subfigure}
    \caption{\textbf{Poisson's ratio identification (Gingerbread-Man / NH).} Identifying $\nu$ from $0.1$ toward $0.4$.}
    \Description{Gingerbread-Man Neo-Hookean Poisson's-ratio identification: top row shows side-by-side renders of the initial and optimized lateral stretching against the target shape; bottom row plots loss decay and the Poisson's ratio trajectory across optimization iterations.}
    \label{fig:demo_poisson}
\end{figure}

We next evaluate gradients under Neo-Hookean hyperelasticity, where stresses depend nonlinearly on the deformation gradient and parameter sensitivities are no longer simple scalings.
Poisson's ratio $\nu$ is particularly challenging because it governs transverse response and couples orthogonal directions; even small gradient errors manifest as visibly incorrect lateral deformation.
We identify $\nu$ from $0.1$ to $\tilde{\nu}=0.4$ on a Gingerbread-Man thin shell by matching the deformation under lateral stretching (Figure~\ref{fig:demo_poisson}).

\begin{figure}[!t]
    \centering
    \begin{subfigure}{\linewidth}
        \centering
        \includegraphics[width=\linewidth]{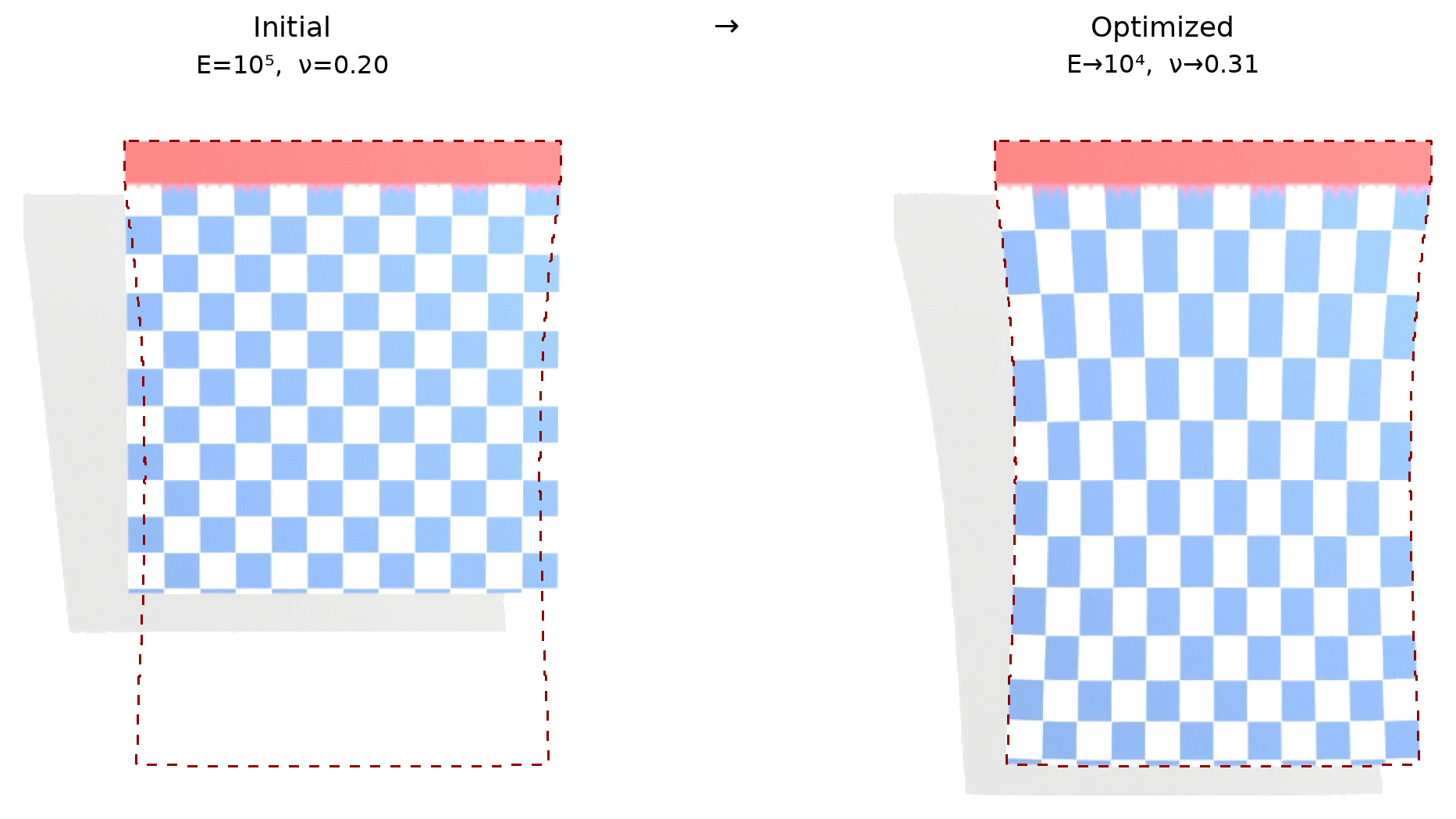}
        \caption{Configurations.}
    \end{subfigure}\\[2pt]
    \begin{subfigure}{\linewidth}
        \centering
        \includegraphics[width=\linewidth]{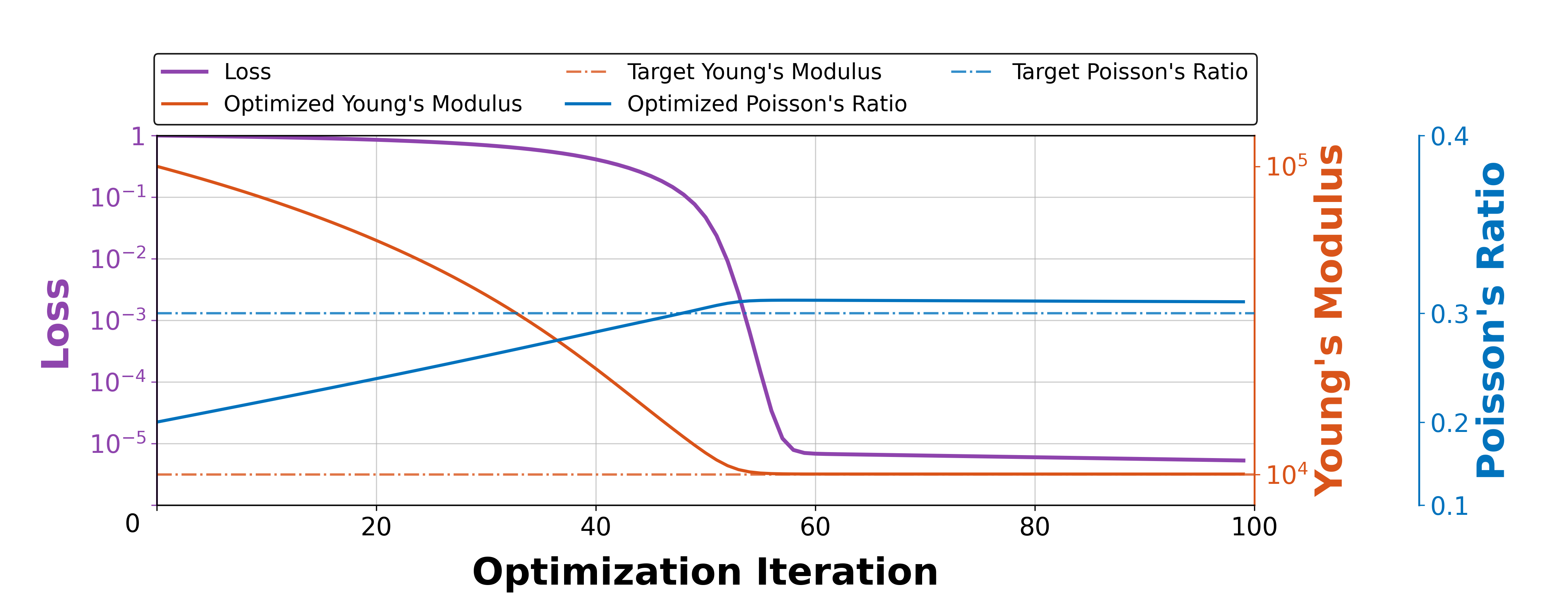}
        \caption{Loss and trajectories.}
    \end{subfigure}
    \caption{\textbf{Joint $(E, \nu)$ identification (Curtain / NH+B).} Jointly identifying $(E, \nu)$ from $(10^5, 0.2)$ toward $(10^4, 0.3)$.}
    \Description{Hanging-curtain joint Young-modulus and Poisson-ratio identification: top row shows the curtain under the initial and identified parameters next to the target draping; bottom row plots loss decay and the two parameter trajectories across optimization iterations.}
    \label{fig:demo_curtain}
\end{figure}

Finally, we perform joint identification of Young's modulus $E$ and Poisson's ratio $\nu$ on a Neo-Hookean curtain draping under gravity (Figure~\ref{fig:demo_curtain}).
This two-parameter setting is more demanding: $E$ and $\nu$ are coupled through the nonlinear constitutive response, and different combinations can produce similar deformations.
Our optimizer still converges smoothly, reaching $(\tilde{E},\tilde{\nu})=(10^{4},0.3)$ from $(10^{5},0.2)$ (Figure~\ref{fig:demo_curtain}).

Per-demo analytical-vs-finite-difference gradient curves for the unit-test cases are shown in Appendix Figure~\ref{fig:grad_grid}.

\subsubsection{Contact}
Differentiating through frictional contact is fundamentally harder than elasticity because the dynamics are mode-dependent: the normal direction switches between contact and separation, and the tangential direction between static and kinetic friction.
Our smoothed Fischer--Burmeister formulation avoids explicit branch enumeration by encoding all contact modes in a single differentiable residual.
The experiments below stress-test whether the gradients remain informative across contact regime transitions.

\begin{figure}[!t]
    \centering
    \begin{subfigure}{\linewidth}
        \centering
        \includegraphics[width=\linewidth]{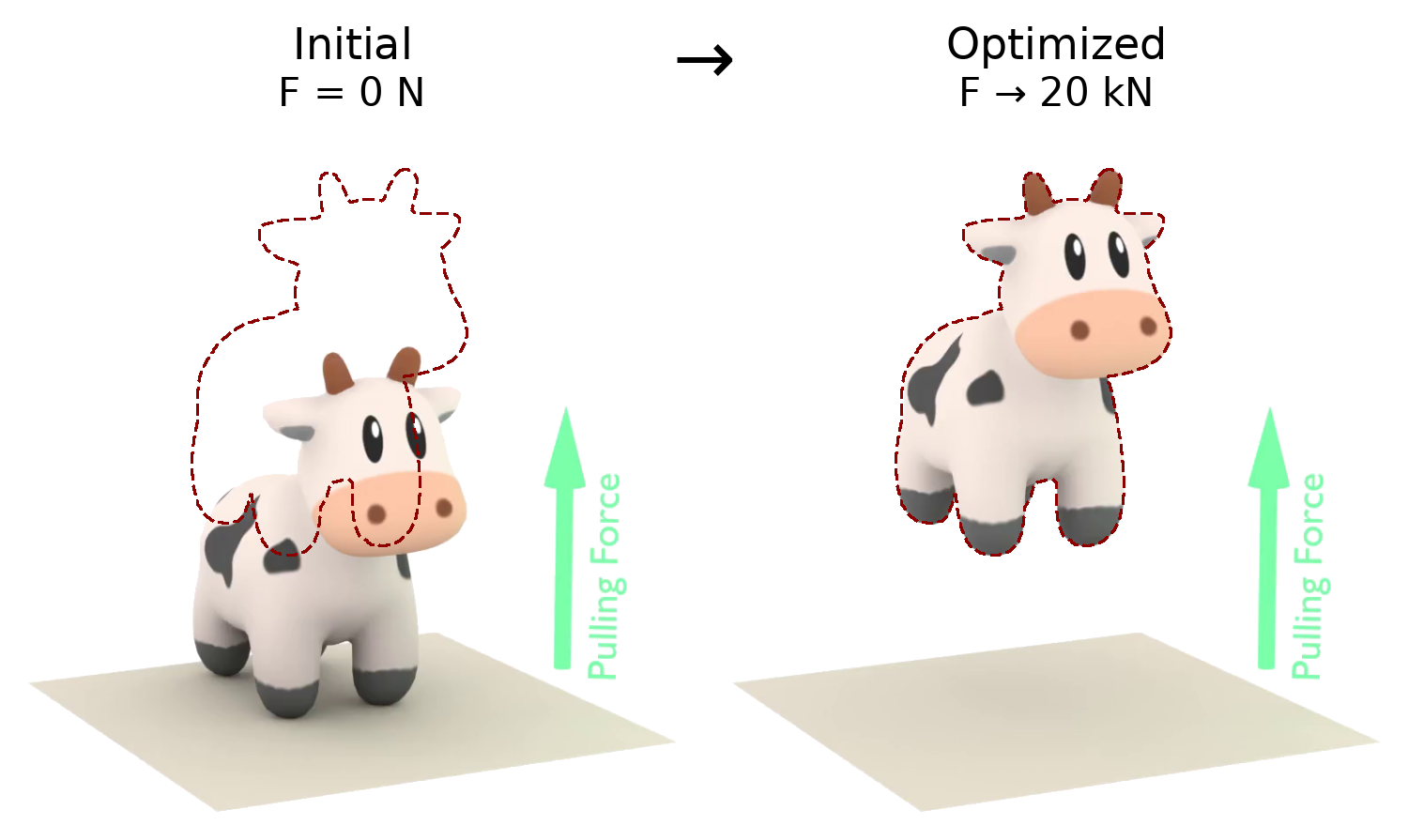}
        \caption{Configurations.}
    \end{subfigure}\\[2pt]
    \begin{subfigure}{\linewidth}
        \centering
        \includegraphics[width=\linewidth]{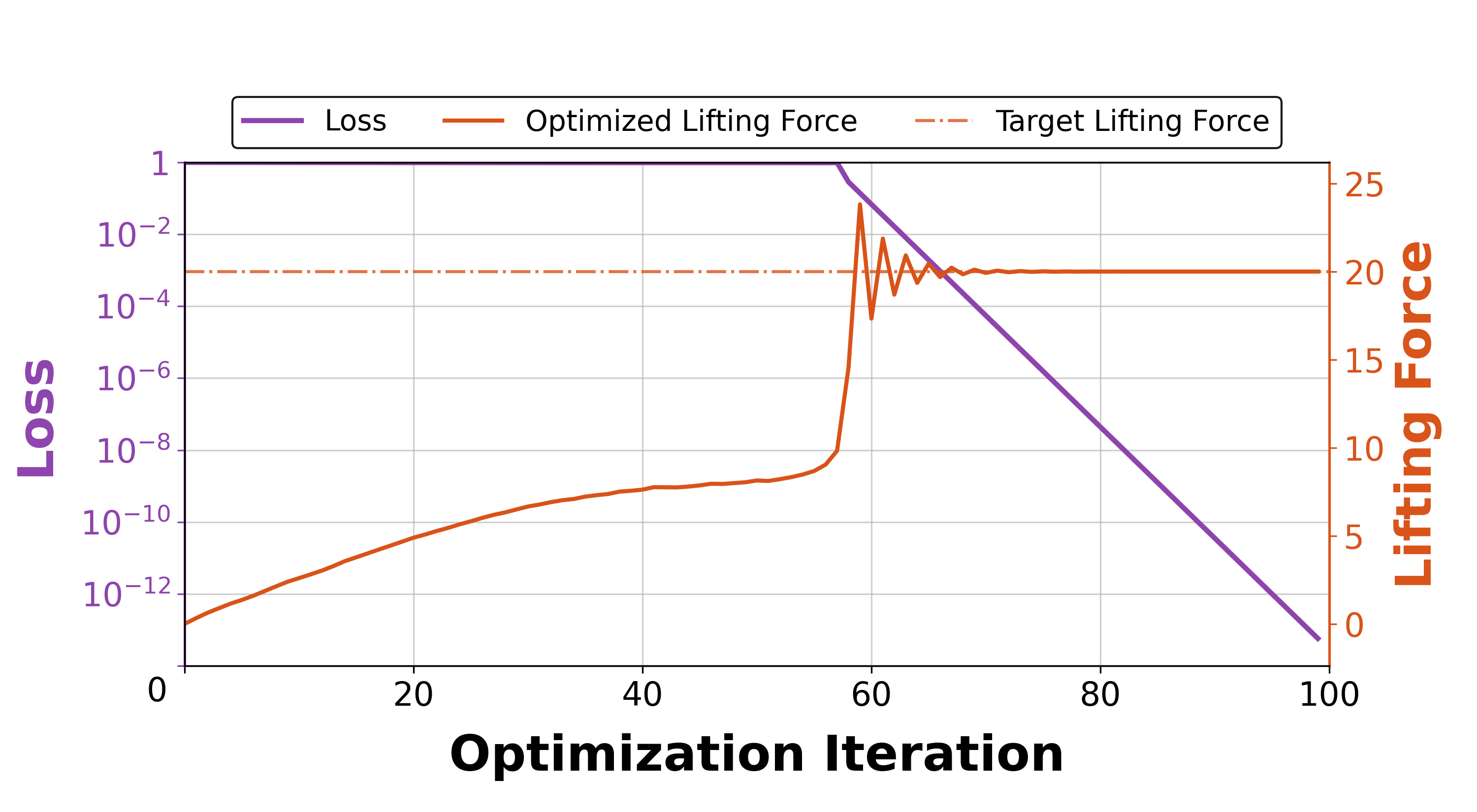}
        \caption{Loss and trajectory.}
    \end{subfigure}
    \caption{\textbf{Lifting force identification (Cow).} Identifying $F_\mathrm{lift}$ from $0$ toward $20$; the two-phase loss shape reflects the contact-to-separation crossover (Section~\ref{sec:ncp}).}
    \Description{Cow lifting-force identification: top row shows the cow at rest on the table and the optimized lifted configuration against the target; bottom row plots loss decay and the lifting-force magnitude trajectory across optimization iterations, with a long contact-regime plateau followed by a rapid post-separation drop.}
    \label{fig:demo_lifting}
\end{figure}

\paragraph*{Contact--separation.}
We place an elastic Cow on a table and optimize the upward lifting force from $F_{\mathrm{lift}}=0$ toward $\tilde{F}_{\mathrm{lift}}=20$ (Figure~\ref{fig:demo_lifting}).
A branch-wise formulation would yield zero gradients while $F_{\mathrm{lift}}<G$ (gravity), because small perturbations are absorbed by the normal reaction.
The two-phase behavior of the loss curve directly reflects the cross-regime gradient mechanism described in Section~\ref{sec:ncp}: the initial plateau corresponds to the contact regime, where the smoothed FB provides a small but correctly directed gradient signal ($\lambda_n \delta_n = \epsilon^2 \Rightarrow \partial \delta_n / \partial F \neq 0$); once the optimizer drives $F_{\mathrm{lift}}$ past the contact--separation boundary, the loss decreases rapidly as the gradient magnitude grows with the object's displacement from the target.

\begin{figure}[!t]
    \centering
    \begin{subfigure}{\linewidth}
        \centering
        \includegraphics[width=\linewidth]{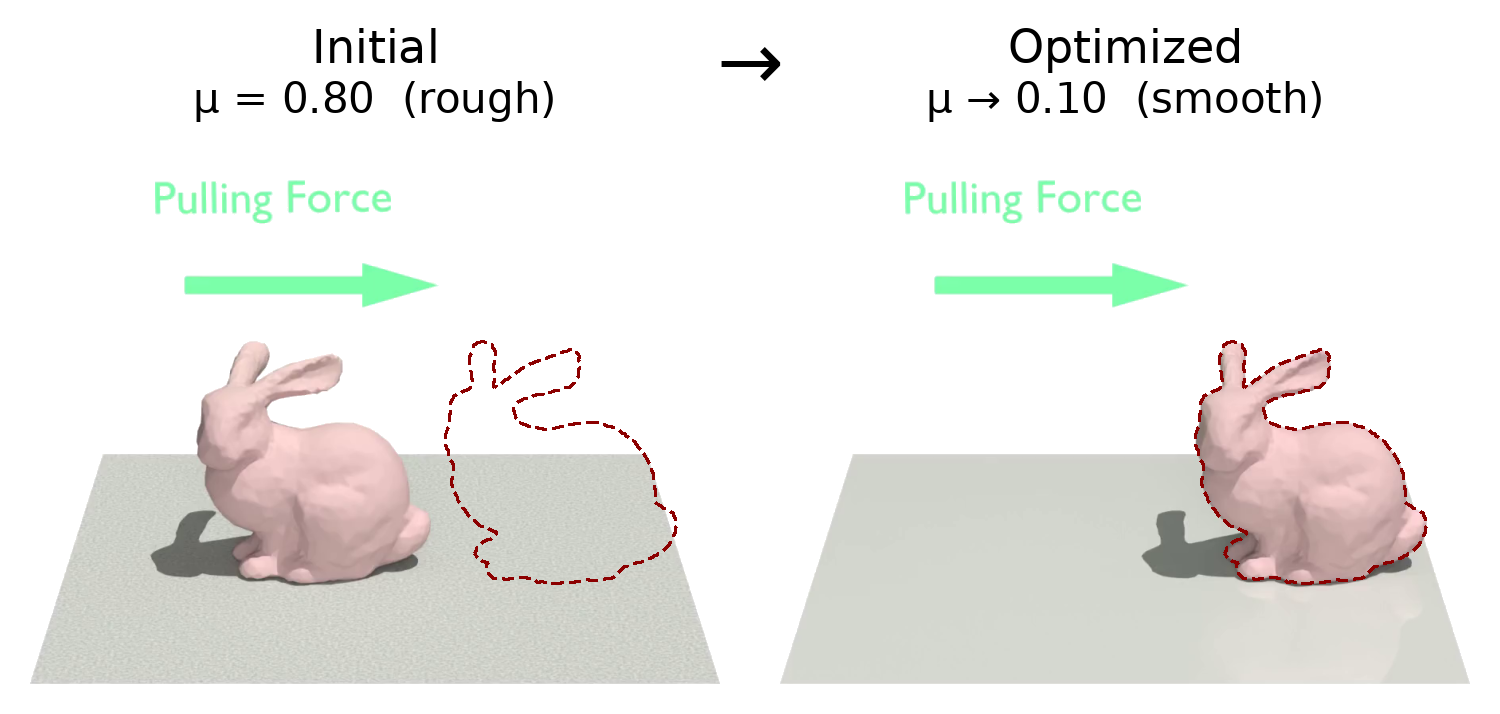}
        \caption{Configurations.}
    \end{subfigure}\\[2pt]
    \begin{subfigure}{\linewidth}
        \centering
        \includegraphics[width=\linewidth]{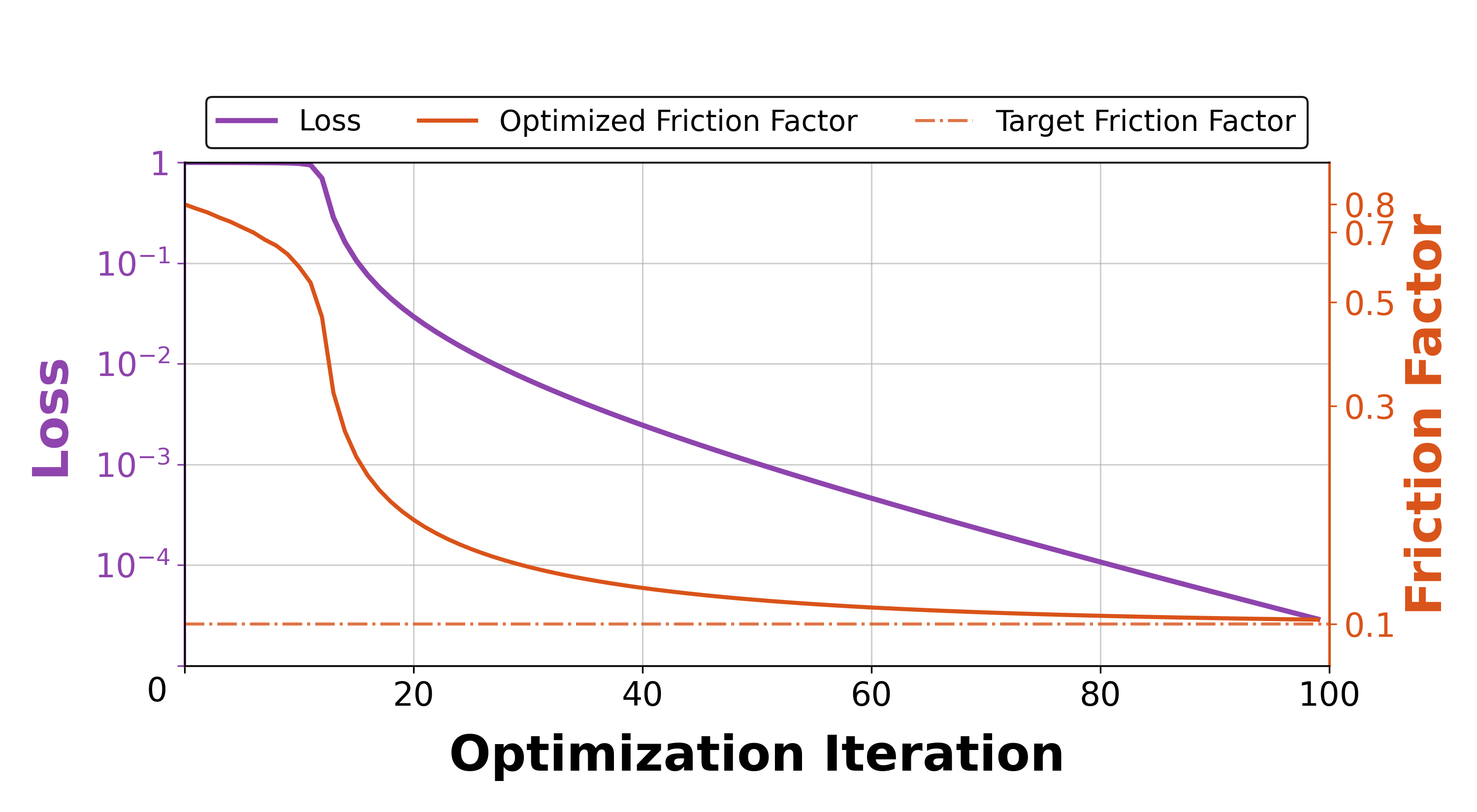}
        \caption{Loss and trajectory.}
    \end{subfigure}
    \caption{\textbf{Floor friction identification (Bunny).} Identifying $\mu$ from $0.8$ toward $0.1$; the staircase loss shape reflects the static-to-kinetic friction crossover (Section~\ref{sec:ncp}).}
    \Description{Friction-coefficient identification on a sliding rubber bunny: top row shows the bunny under the initial and optimized friction coefficients against the pink target silhouette; bottom row plots loss decay on a log scale and the friction-coefficient trajectory across optimization iterations.}
    \label{fig:demo_bunny}
\end{figure}

\begin{figure}[!t]
    \centering
    \begin{subfigure}{\linewidth}
        \centering
        \includegraphics[width=\linewidth]{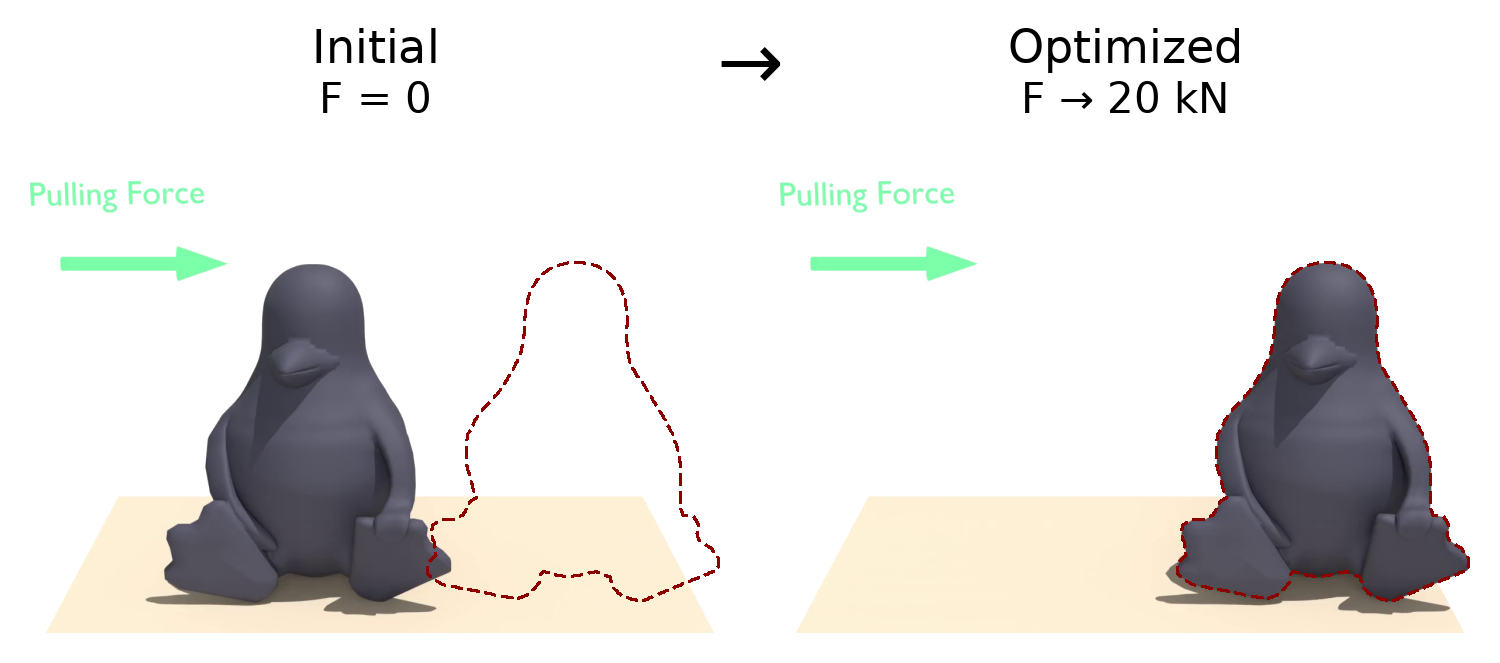}
        \caption{Configurations.}
    \end{subfigure}\\[2pt]
    \begin{subfigure}{\linewidth}
        \centering
        \includegraphics[width=\linewidth]{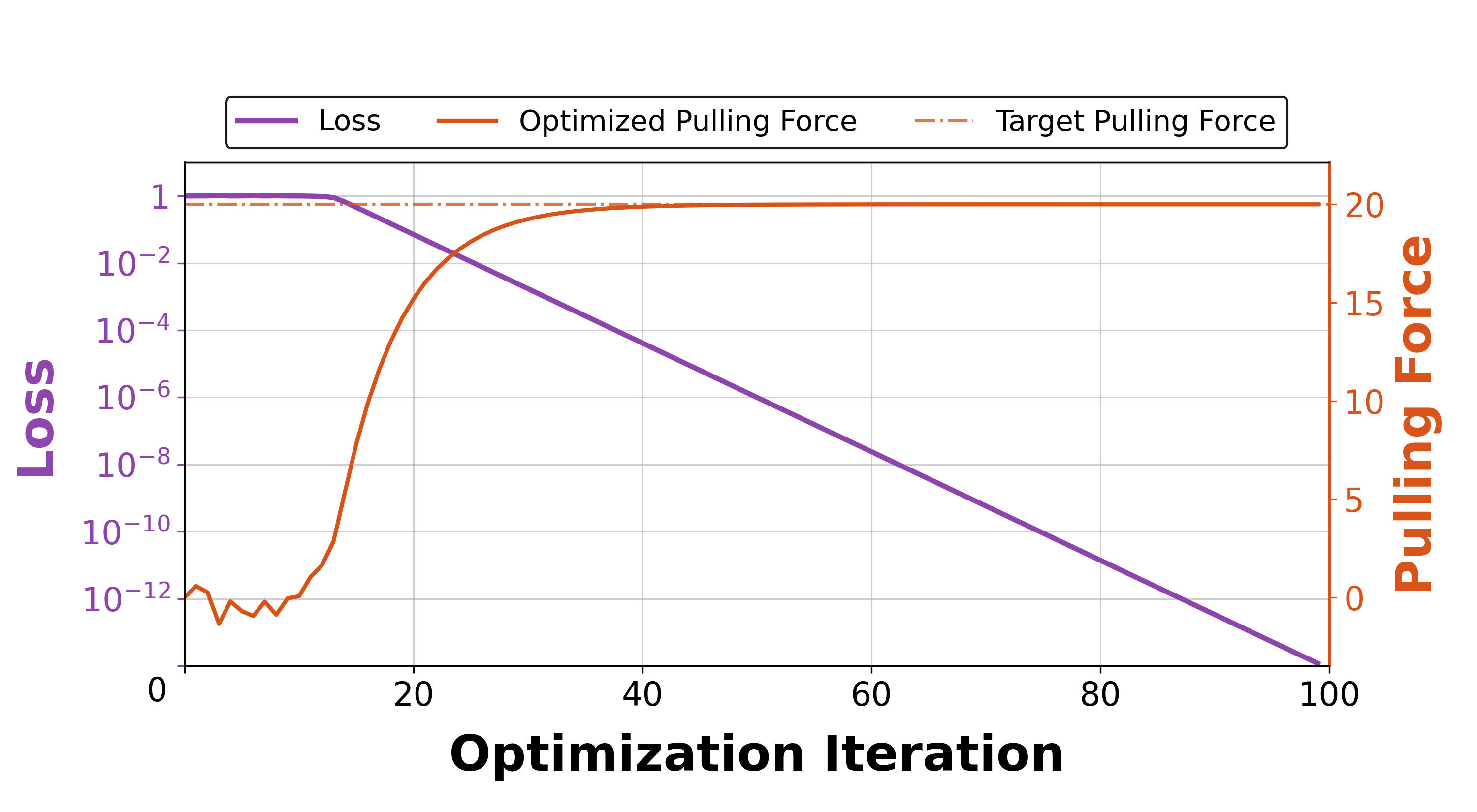}
        \caption{Loss and trajectory.}
    \end{subfigure}
    \caption{\textbf{Pulling force identification (Tux).} Identifying $F_\mathrm{pull}$ from $0$ toward $20$; the two-phase loss shape reflects the static-to-kinetic friction crossover (Section~\ref{sec:ncp}).}
    \Description{Tux pulling-force identification: top row shows the elastic Tux at rest and at the optimized horizontal force against the target position; bottom row plots loss decay and the pulling-force trajectory across optimization iterations, with a static-friction plateau followed by a sharp drop once kinetic sliding starts.}
    \label{fig:demo_tux}
\end{figure}

\paragraph*{Static--kinetic friction.}
We consider two distinct setups.
The Bunny task identifies the friction coefficient $\mu$ on a rubber bunny sliding on a slope, optimizing $\mu$ from $0.8$ toward $\tilde{\mu}=0.1$ (Figure~\ref{fig:demo_bunny}); the Tux task applies a horizontal pulling force to drag an elastic Tux across a rough table, optimizing $F_{\mathrm{pull}}$ from $0$ toward $\tilde{F}_{\mathrm{pull}}=20$ (Figure~\ref{fig:demo_tux}).
Both initializations sit deep in the static-friction regime, and both tasks exhibit the same plateau-then-drop pattern as Cow Lifting: a static-friction plateau followed by a sharp transition into kinetic sliding once the parameter sweep crosses the breakaway threshold. In Bunny ($t_{0.5}=13\%$, $t_{0.9}=16\%$ in Table~\ref{tab:analysis}), the bunny remains stuck under the applied horizontal force while $\mu\gtrsim 0.5$; once the optimizer drives $\mu$ across the static--kinetic boundary near $\mu\approx 0.47$, the gradient magnitude spikes and the loss falls two orders of magnitude within ${\sim}10$ iterations of breakaway, with the steady kinetic-regime descent carrying it another two orders by iteration $100$. Tux ($t_{0.5}=15\%$, $t_{0.9}=20\%$) shows the dual transition driven instead by the pulling force exceeding the breakaway threshold.

\begin{figure}[!t]
    \centering
    \begin{subfigure}{\linewidth}
        \centering
        \includegraphics[width=\linewidth]{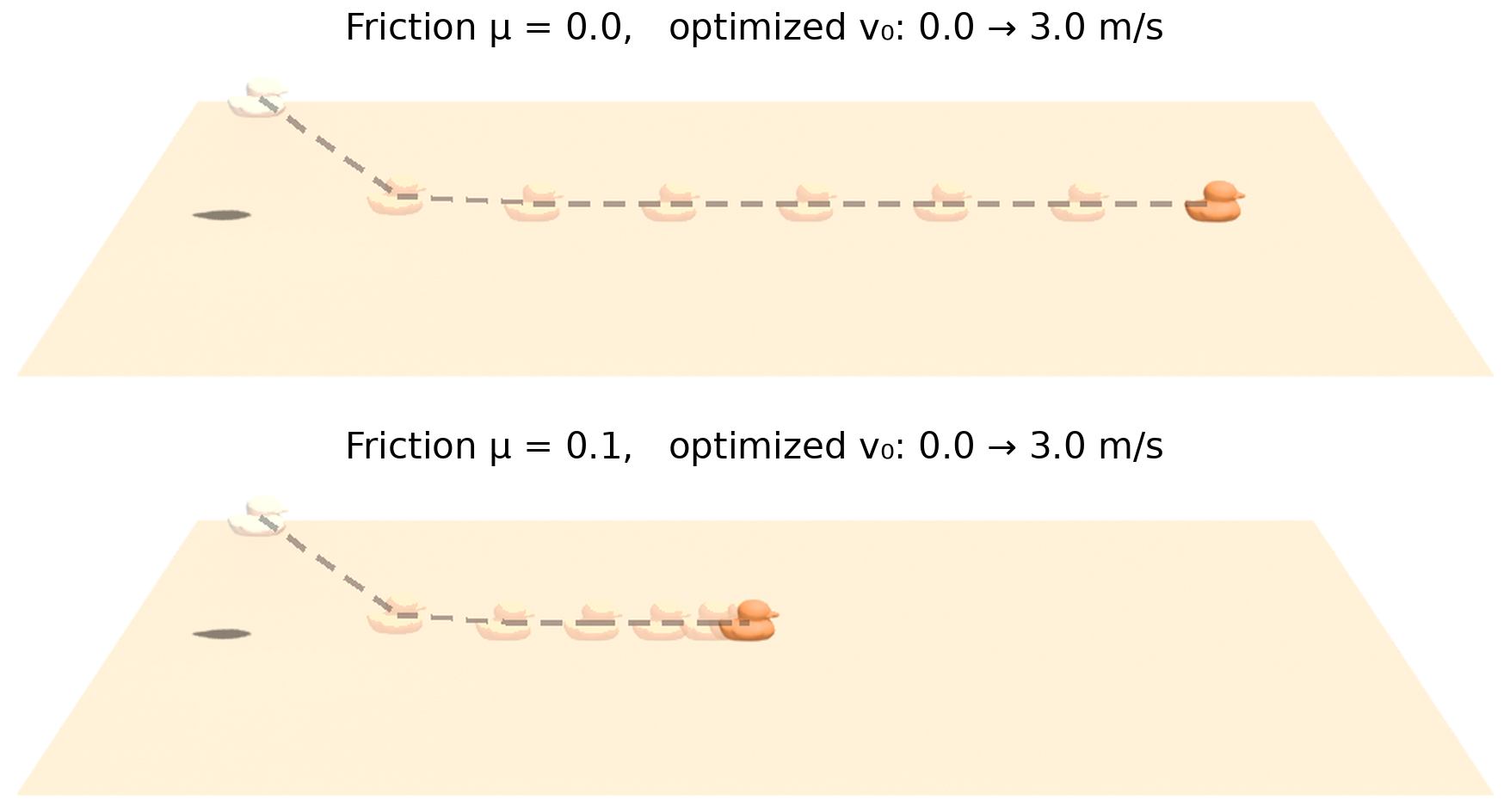}
        \caption{Configurations.}
    \end{subfigure}\\[2pt]
    \begin{subfigure}{\linewidth}
        \centering
        \includegraphics[width=\linewidth]{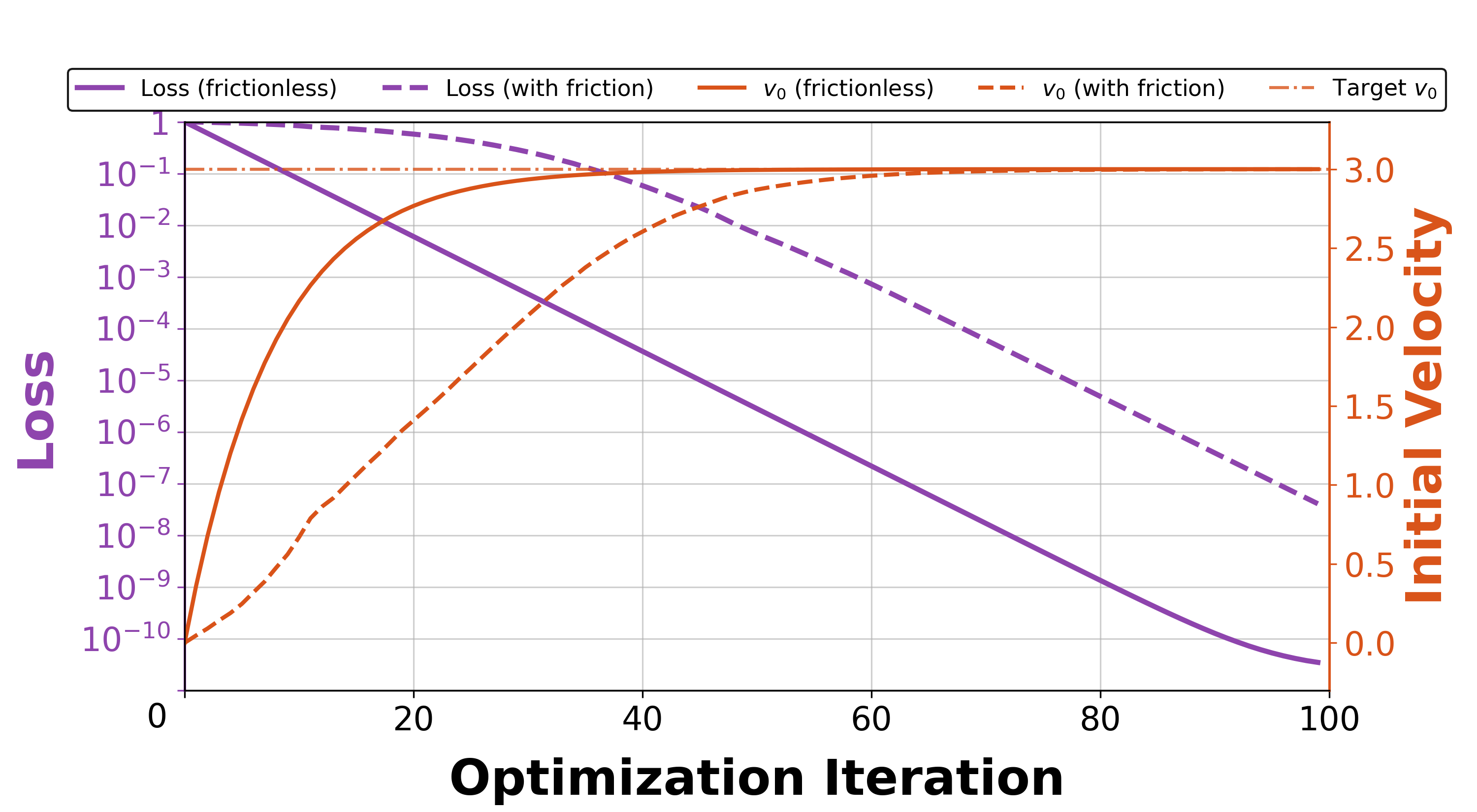}
        \caption{Loss and trajectory.}
    \end{subfigure}
    \caption{\textbf{Initial-velocity identification (Rubber Duck).} Identifying $v_0$ from $0$ toward $3$; frictionless (solid) and with-friction (dashed) variants overlaid.}
    \Description{Rubber-duck initial-velocity identification across two contact variants: top row shows the duck at rest and at the optimized launched configuration against the target; bottom row plots loss decay on a log scale and the initial-velocity trajectory across optimization iterations, with separate solid (frictionless) and dashed (with-friction) curves.}
    \label{fig:demo_duck}
\end{figure}

\paragraph*{Multi-mode contact.}
We optimize the initial velocity of an elastic rubber duck launched horizontally in mid-air to match a target landing configuration (Figure~\ref{fig:demo_duck}).
The trajectory involves free flight, ground impact with rebound, and frictional sliding, inducing repeated switching in both normal and tangential contact modes.
We include a frictionless variant for comparison.
In the frictionless case, analytical and finite-difference gradients agree to numerical precision (MRE\textbar E$=0$ in Table~\ref{tab:analysis}). The friction variant introduces repeated mode switching that injects transient noise into the forward solver near each contact event, raising MRE\textbar E to $0.48$; both gradient methods inherit this transient noise, but the average descent direction remains correct and the optimizer reaches the target $v_0 = 3$ in both variants.

\begin{table*}[ht!]
\centering
\caption{\textbf{Statistics of gradient accuracy and optimization behavior across elasticity and contact tasks.}
We evaluate a suite of identification problems spanning elasticity (left) and contact (right), and report both stage-wise optimization behavior and gradient agreement.
Stage-wise behavior is characterized by $t_{0.5}$ and $t_{0.9}$, the iteration percentages required to achieve 50\% and 90\% of the total loss decrease, and by the normalized loss area-under-curve over early/mid/late iterations (AUC|E, AUC|M, AUC|L).
Gradient agreement is summarized by the mean relative error between analytical and finite-difference gradients within each stage (MRE|E, MRE|M, MRE|L; lower is better).
For multi-parameter tasks, MRE is the mean across parameters.
For completeness, the table also lists the key task specifications for each column (e.g., shape and constitutive model), with checkmarks/crosses indicating the presence/absence of friction; learning rate is defined after reparameterization (Figure~\ref{fig:grad_grid}).}

\label{tab:analysis}
\resizebox{\textwidth}{!}{%
\begin{tabular}{ccccccccccccc}
\toprule
Target & \multicolumn{7}{c}{Elasticity} & \multicolumn{5}{c}{Contact} \\
\cmidrule(lr){1-1} \cmidrule(lr){2-8} \cmidrule(lr){9-13}
Shape & \multicolumn{4}{c}{Armadillo} & Cloth & GB-Man & Curtain
      & Bunny & Cow & Tux & \multicolumn{2}{c}{YR-Duck} \\
Model & ARAP & Co-rot & NH & ARAP & ARAP\,+\,B. & NH & NH\,+\,B.
      & NH & NH & NH & \multicolumn{2}{c}{NH} \\
Friction & \multicolumn{4}{c}{\NA} & \NA & \NA & \NA
      & \cmark & \xmark & \cmark & \xmark & \cmark \\
\cmidrule(lr){1-1} \cmidrule(lr){2-5} \cmidrule(lr){6-6} \cmidrule(lr){7-7} \cmidrule(lr){8-8}
\cmidrule(lr){1-1} \cmidrule(lr){9-9} \cmidrule(lr){10-10} \cmidrule(lr){11-11} \cmidrule(lr){12-13}
Opt. Variable & $k_s$ & $k_s$ & $k_s$ & $F_{\mathrm{wind}}$ & $k_b$ & $\nu$ & $(E,\nu)$ & $\mu$ & $F_{\mathrm{lift}}$ & $F_{\mathrm{pull}}$ & $v_0$ & $v_0$ \\
Learn. Rate & 0.1 & 0.1 & 0.1 & 10.0 & 500.0 & 500.0 & (1.0,\,2.0) & 0.05 & 10.0 & 1.0 & 0.0002 & 0.0002 \\
Init. Value & $2\!\times\!10^{5}$ & $2\!\times\!10^{5}$ & $2\!\times\!10^{5}$ & $0$ & $5\!\times\!10^{2}$ & $0.1$ & $(10^{5}, 0.2)$ & $0.8$ & $0$ & $0$ & $0$ & $0$ \\
Target Value & $10^{4}$ & $10^{4}$ & $10^{4}$ & $50$ & $10^{1}$ & $0.4$ & $(10^{4}, 0.3)$ & $0.1$ & $20$ & $20$ & $3$ & $3$ \\
$t_{0.5}$ & $68\%$ & $21\%$ & $20\%$ & $2\%$ & $10\%$ & $1\%$ & $38\%$ & $13\%$ & $59\%$ & $15\%$ & $3\%$ & $23\%$ \\
$t_{0.9}$ & $80\%$ & $29\%$ & $29\%$ & $8\%$ & $37\%$ & $4\%$ & $49\%$ & $16\%$ & $61\%$ & $20\%$ & $10\%$ & $37\%$ \\
AUC|E & $31.25$ & $19.52$ & $19.34$ & $2.96$ & $14.88$ & $1.47$ & $28.19$ & $13.15$ & $31.98$ & $15.71$ & $3.97$ & $21.32$ \\
AUC|M & $24.98$ & $0.01$ & $0.0$ & $0.0$ & $0.76$ & $0.01$ & $6.02$ & $0.05$ & $24.85$ & $0.0$ & $0.0$ & $1.06$ \\
AUC|L & $4.52$ & $0.0$ & $0.0$ & $0.0$ & $0.0$ & $0.0$ & $0.0$ & $0.0$ & $0.0$ & $0.0$ & $0.0$ & $0.0$ \\
\metric{MRE|E}{$\downarrow$} & $2.49\!\times\!10^{-3}$ & $2.10\!\times\!10^{-2}$ & $2.12\!\times\!10^{-2}$ & $5.54\!\times\!10^{-2}$ & $8.63\!\times\!10^{-5}$ & $2.11\!\times\!10^{-3}$ & $3.55\!\times\!10^{-6}$ & $6.97\!\times\!10^{-1}$ & $7.54\!\times\!10^{-3}$ & $1.53\!\times\!10^{0}$ & $0$ & $4.84\!\times\!10^{-1}$ \\
\metric{MRE|M}{$\downarrow$} & $8.37\!\times\!10^{-3}$ & $2.06\!\times\!10^{-2}$ & $2.14\!\times\!10^{-2}$ & $4.65\!\times\!10^{-2}$ & $8.67\!\times\!10^{-4}$ & $2.14\!\times\!10^{-3}$ & $6.66\!\times\!10^{-3}$ & $3.16\!\times\!10^{-3}$ & $1.55\!\times\!10^{-1}$ & $6.89\!\times\!10^{-5}$ & $4.66\!\times\!10^{-8}$ & $4.55\!\times\!10^{-2}$ \\
\metric{MRE|L}{$\downarrow$} & $1.12\!\times\!10^{-2}$ & $1.79\!\times\!10^{-2}$ & $1.77\!\times\!10^{-2}$ & $4.11\!\times\!10^{-2}$ & $2.16\!\times\!10^{-3}$ & $2.09\!\times\!10^{-3}$ & $8.48\!\times\!10^{-2}$ & $3.36\!\times\!10^{-3}$ & $0$ & $9.47\!\times\!10^{-5}$ & $2.10\!\times\!10^{-6}$ & $6.85\!\times\!10^{-3}$ \\
\bottomrule
\end{tabular}%
}
\end{table*}

\subsubsection{Quantitative Summary}
To quantify gradient quality and its impact on optimization, we report the following metrics.
Let $\loss^{(i)}$ denote the loss at iteration $i\in\{0,\dots,T\}$.
We define $t_{0.5}$ and $t_{0.9}$ as the iteration percentages at which 50\% and 90\% of the total loss decrease is achieved:
\begin{equation}
t_p \coloneqq \frac{1}{T}\min\big\{i:~\loss^{(0)}-\loss^{(i)} \ge p(\loss^{(0)}-\loss^{(T)})\big\},\quad p\in\{0.5,0.9\}.
\end{equation}
We partition the optimization into early/mid/late (E/M/L) stages separated by $t_{0.5}$/$t_{0.9}$, and report the normalized area-under-curve within each stage:
\begin{equation}
\mathrm{AUC}|S \coloneqq \frac{1}{|S|}\sum_{i\in S}\frac{\loss^{(i)}-\loss^{(T)}}{\loss^{(0)}-\loss^{(T)}},\quad S\in\{E,M,L\}.
\end{equation}
Gradient agreement is measured by the mean relative error (MRE) between analytical and finite-difference gradients:
\begin{equation}
\mathrm{MRE}|S \coloneqq \frac{1}{|S|}\sum_{i\in S}\frac{\big\|\grad^{(i)}_{\mathrm{ana}}-\grad^{(i)}_{\mathrm{fd}}\big\|}{\big\|\grad^{(i)}_{\mathrm{fd}}\big\|+\epsilon_g},\quad S\in\{E,M,L\}.
\end{equation}
Results are summarized in Table~\ref{tab:analysis}.
The largest MRE values concentrate in contact-regime transitions---Tux $F_{\mathrm{pull}}$ (MRE\textbar E$=1.53$), Bunny $\mu$ (MRE\textbar E$=0.70$), Duck-with-friction (MRE\textbar E$=0.48$), and Cow Lifting (MRE\textbar M$=0.16$)---where the forward simulation is itself transient between contact modes and both analytical and finite-difference gradients inherit short bursts of sign-flipping noise (Appendix Figure~\ref{fig:grad_grid}).
Outside these transition windows, MRE stays within $10^{-2}$ across every task and stage and reaches $10^{-4}$--$10^{-5}$ on tasks that operate in a steady regime throughout; every benchmark converges to its target.

These transition-window peaks reflect a property of the smoothing parameter $\epsilon$ rather than analytical inaccuracy: smaller $\epsilon$ drives the smoothed FB residual toward the hard-NCP limit, sharpening contact-mode transitions and amplifying transient forward-solver noise that both gradient methods inherit. The working default $2\epsilon^2 = 10^{-12}$, justified independently in Section~\ref{sec:epsilon_sweep} from $\mu$-precision considerations, balances transition smoothness against forward fidelity.

\subsection{Comparison with Backward Solvers}
\label{sec:backward_comparison}

The gradient accuracy validated above relies on the backward solve converging to sufficient precision.
We now evaluate the convergence behavior and runtime of our adjoint linear solver (Equation~\eqref{eq:adjoint}) under three regimes: contact-free (SPD), frictionless contact (ill-conditioned SPD), and frictional contact (ill-conditioned asymmetric).
We reuse the unit-test scenarios as solver benchmarks to isolate the behavior of the backward solver from the physical setup.

We benchmark Krylov solvers (CG for SPD, GMRES for asymmetric) with the preconditioners proposed in Section~\ref{sec:solver}: Sparse-Inverse, Jacobi, and Woodbury.
As a baseline, we include the semi-implicit fixed-point iteration of DiffPD~\cite{du2021diffpd} and DiffCloth~\cite{li2022diffcloth}:
$\mathbf{x}^{k+1} = \sys^{-1}(\mathbf{b} - (\lhs - \sys)\mathbf{x}^k)$.

\begin{figure*}[!t]
    \centering
    \begin{subfigure}[b]{\linewidth}
        \centering
        \includegraphics[width=0.7\linewidth]{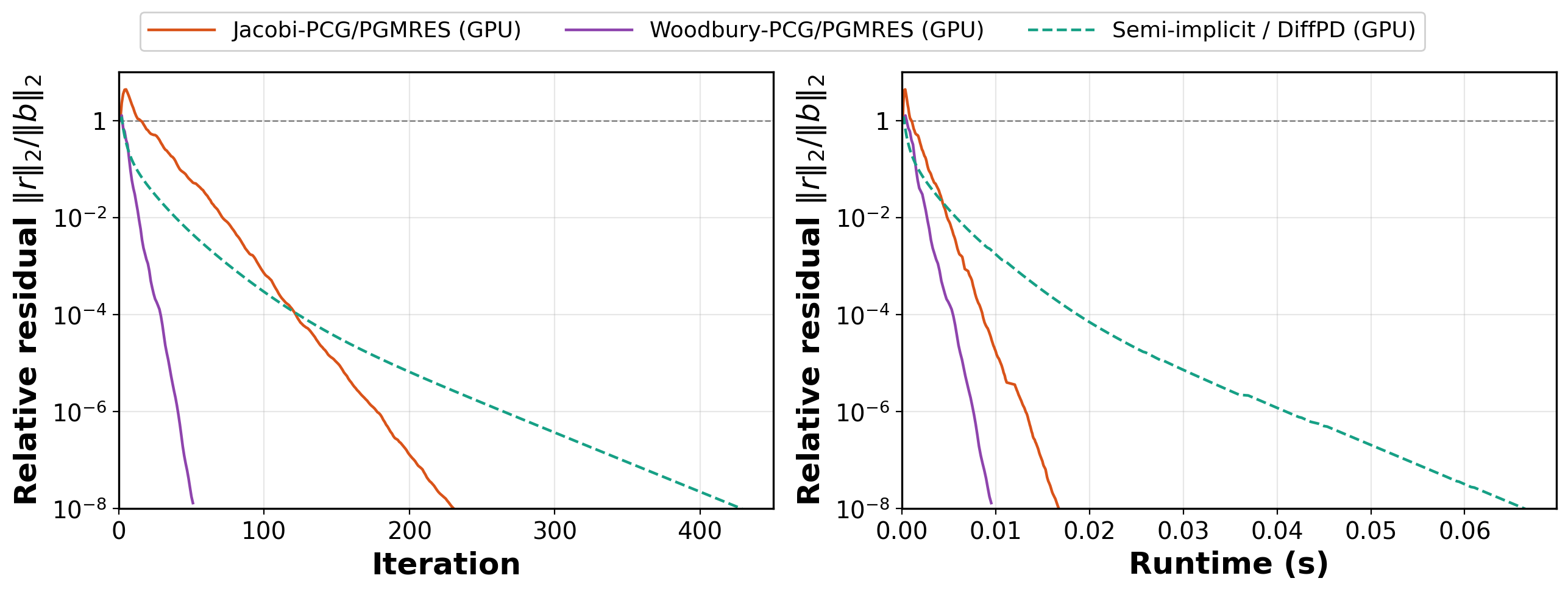}
        \caption{Contact-free.}
        \label{fig:performance1}
    \end{subfigure}
    \begin{subfigure}[b]{\linewidth}
        \centering
        \includegraphics[width=0.7\linewidth]{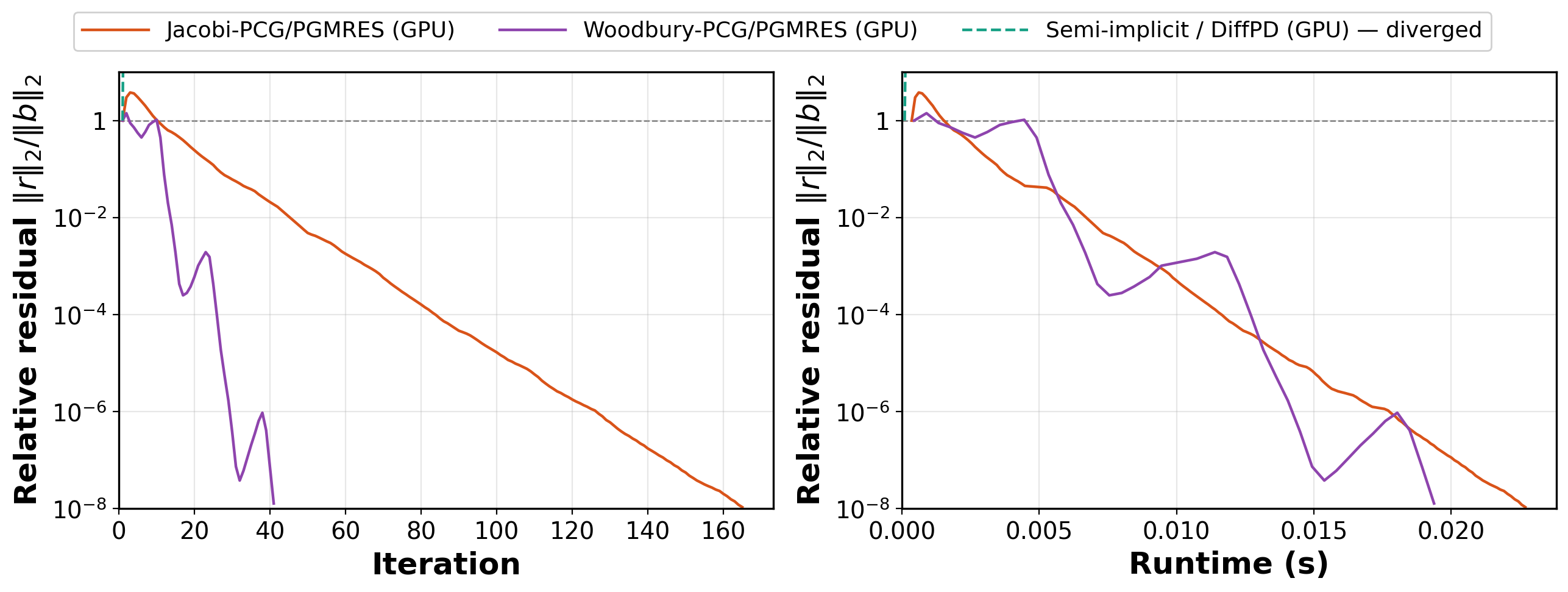}
        \caption{Frictionless contact (semi-implicit diverges).}
        \label{fig:performance2}
    \end{subfigure}
    \begin{subfigure}[b]{\linewidth}
        \centering
        \includegraphics[width=0.7\linewidth]{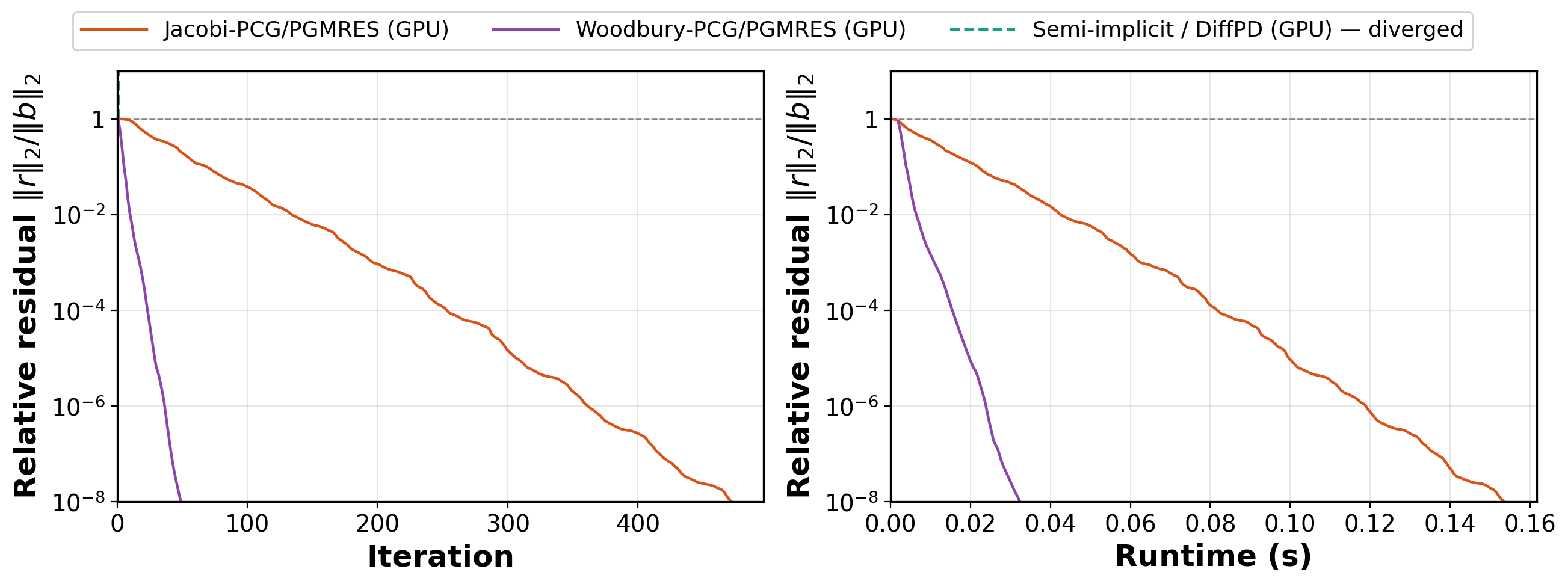}
        \caption{Frictional contact (semi-implicit diverges).}
        \label{fig:performance3}
    \end{subfigure}
    \caption{\textbf{Adjoint solver convergence across contact regimes.}
    Each row plots relative residual versus iteration count (left) and wall-clock runtime (right) for the linearized adjoint system under (a) contact-free, (b) frictionless contact, and (c) frictional contact regimes.
    Curves compare Krylov solvers (CG for SPD, GMRES for asymmetric) paired with Sparse-Inverse, Jacobi, and our Woodbury preconditioner, alongside the semi-implicit fixed-point iteration of DiffPD~\cite{du2021diffpd} and DiffCloth~\cite{li2022diffcloth}.
    Our Woodbury preconditioner uniformly dominates across all three regimes; quantitative speedups and analysis are reported in the text.}
    \Description{Three-row, two-column grid of convergence plots: relative residual error vs iteration count (left column) and vs wall-clock runtime (right column), under contact-free, frictionless contact, and frictional contact regimes; each panel overlays curves for several solver and preconditioner combinations.}
    \label{fig:performance}
\end{figure*}

\paragraph*{Results.}
Figure~\ref{fig:performance} summarizes convergence across all regimes.
In the frictionless and frictional contact regimes, the semi-implicit iteration (Equation~\eqref{eq:semi-implicit}) diverges within a few iterations.
As discussed in Section~\ref{sec:solver}, this iteration is a stationary splitting with convergence governed by $\rho(\sys^{-1}(\sys - \lhs))$; the contact contribution $\rhs$ pushes this spectral radius above unity, violating the convergence condition.

Our Woodbury preconditioner, by contrast, uniformly outperforms the alternatives across all three regimes, in both iteration count and wall-clock time.
The advantage is most pronounced in the frictional case, where Woodbury reaches a $10^{-8}$ relative residual in ${\sim}50$ iterations and ${\sim}25$\,ms while Jacobi-preconditioned GMRES requires ${\sim}450$ iterations and ${\sim}150$\,ms---a ${\sim}9{\times}$ iteration speedup and ${\sim}6{\times}$ wall-clock speedup---and the semi-implicit baseline diverges outright.
In the frictionless regime Woodbury still converges in ${\sim}4{\times}$ fewer iterations than Jacobi, while in the contact-free regime---where Woodbury degenerates to the sparse-inverse preconditioner of \cite{zeng2025fast} (since $\rhs = 0$)---it converges in ${\sim}5{\times}$ fewer iterations than Jacobi and an order of magnitude faster than the semi-implicit baseline.
The frictionless Woodbury curve also shows a mildly non-monotone descent (visible in Figure~\ref{fig:performance}(b)): the preconditioner is constructed from $\sys + \rhs$ rather than the full $\sys - \Delta\sys + \rhs$, so when the projection-Jacobian correction $\Delta\sys$ contributes a non-negligible fraction of the elastic block, the preconditioned Krylov direction misses the directional information from $\Delta\sys$ and the residual can transiently rebound.
Overall convergence nonetheless remains effective and efficient, reaching the $10^{-8}$ tolerance well below the iteration and wall-clock budgets of all alternatives.

The mechanism is the block-structure match laid out in Section~\ref{sec:solver}: the Woodbury form (Equation~\eqref{eq:woodbury_formula}) addresses both the sparse elastic and the low-rank contact part of $\lhs$ in a single pass, which is why its margin persists across all three regimes and grows with contact complexity.

These results validate the design choice of treating the adjoint as a linear system rather than reusing the forward solver's nonlinear iteration: even when the stationary splitting diverges, the proposed Krylov-with-Woodbury solver converges reliably.

\paragraph*{Inner-solver threshold.}
The $n_c = 1000$ crossover between direct and iterative inner solve (Section~\ref{sec:solver}) was determined empirically on the RTX 5090: below this size, dense LLT/LU factorization fits in cache and completes faster than several Krylov iterations; above it, direct-factorization fill-in dominates and the iterative solver wins. The crossover is mildly hardware-dependent but stable across our experiments. We use the same residual tolerance of $10^{-8}$ for the inner iterative solve as for the outer Krylov.

\subsection{Smoothing Parameter Sensitivity}
\label{sec:epsilon_sweep}

The smoothing parameter $\epsilon$ trades two competing properties: larger $\epsilon$ strengthens the cross-regime gradient signal at the cost of forward-physics fidelity, while smaller $\epsilon$ improves fidelity but weakens the signal toward the hard-NCP limit (Section~\ref{sec:ncp}).
Appendix~\ref{sec:smoothness_consistency} establishes that the matched-$\epsilon$ requirement for K-correctness holds at any magnitude; the question that remains is the largest $\epsilon$ that still resolves a stringent forward-fidelity test.
We use that value to set the default for all subsequent experiments.

\begin{figure}[!ht]
    \centering
    \begin{subfigure}{\columnwidth}
        \centering
        \includegraphics[width=0.85\columnwidth, trim=0 110 0 180, clip]{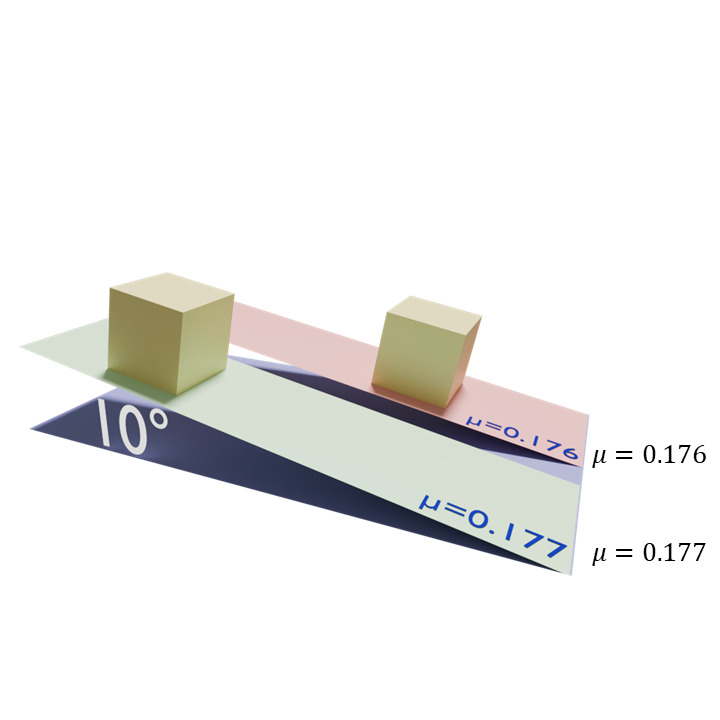}
        \caption{\textbf{Slip/stick discrimination test.} A high-stiffness rubber cube on an inclined slope with two friction strips: pink slides, green holds.}
        \label{fig:accfric_demo}
    \end{subfigure}\\[6pt]
    \begin{subfigure}{\columnwidth}
        \centering
        \includegraphics[width=\columnwidth]{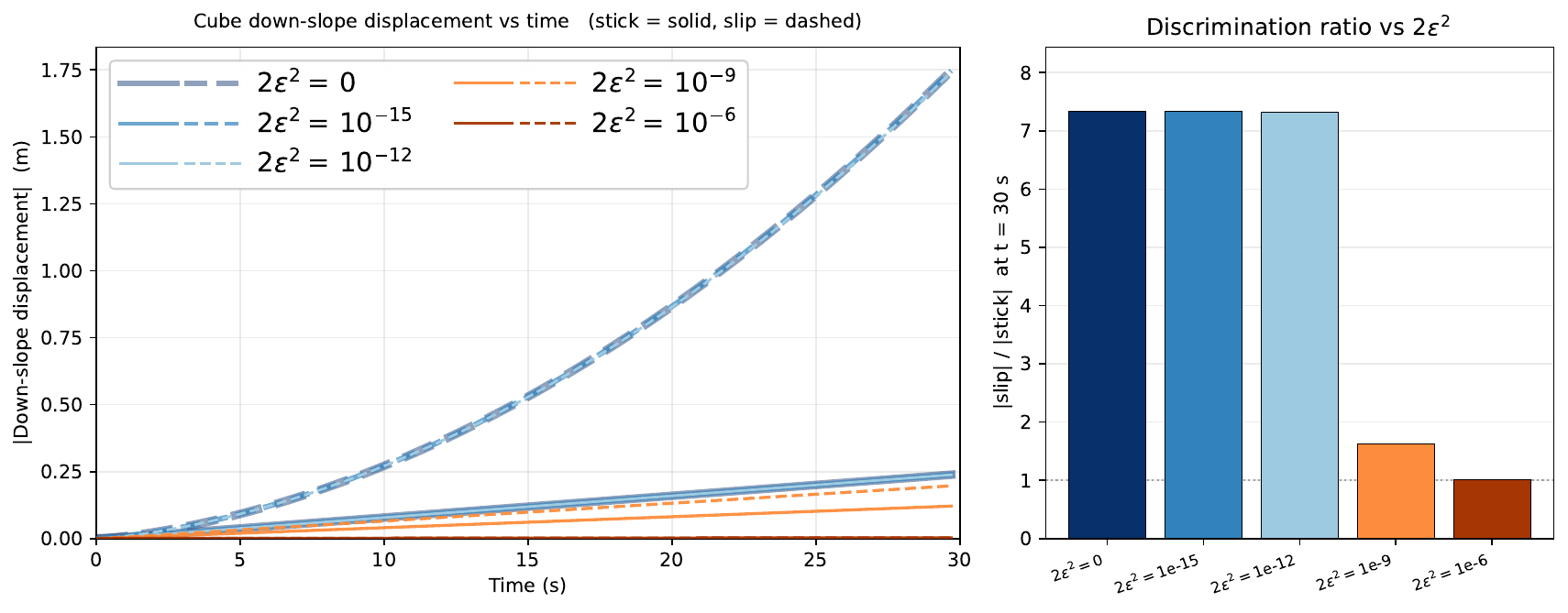}
        \caption{\textbf{Trajectories under $\epsilon$ ablation.} \emph{(Left)} Down-slope cube displacement vs.\ time, $5\,(2\epsilon^2) \times \{\mathrm{stick}, \mathrm{slip}\}$ trajectories on a linear axis. \emph{(Right)} Slip/stick discrimination ratio at $t = 30$\,s for each $2\epsilon^2$.}
        \label{fig:accfric_traj}
    \end{subfigure}
    \caption{\textbf{$\mu$-precision under $\epsilon$ ablation.} (a) The friction sliding/stick test that defines the $\Delta\mu = 10^{-3}$ benchmark, and (b) the resulting trajectories that quantify how forward fidelity collapses between $2\epsilon^2 = 10^{-12}$ and $2\epsilon^2 = 10^{-9}$.}
    \Description{Render of two cubes on a 10-degree slope with friction strips bracketing the analytical slip threshold (a), and trajectory plots of cube down-slope displacement and slip/stick discrimination ratio across five smoothing parameter values (b).}
    \label{fig:accfric}
\end{figure}

\paragraph{Setup}
We use a high-stiffness rubber-like cube resting on an inclined slope as a slip/stick discrimination benchmark (Figure~\ref{fig:accfric_demo}).
Two friction coefficients straddle the analytical slip threshold by $\pm 0.4\,\text{\textperthousand}$, giving a $\Delta\mu = 10^{-3}$ \emph{$\mu$-precision} probe.
Each $(2\epsilon^2, \mu)$ pair runs a $30$\,s forward simulation with matched $\epsilon_{\mathrm{fwd}} = \epsilon_{\mathrm{bwd}}$ so the comparison isolates $\epsilon$ magnitude.

\paragraph{Results}
At $2\epsilon^2 \le 10^{-12}$ the simulator faithfully resolves the $1\,\text{\textperthousand}$ friction difference: the slip cube reaches ${\sim}1.74$\,m while the stick cube creeps only ${\sim}0.24$\,m at $t = 30$\,s, a ${\sim}7.3\times$ separation, and the three trajectories at $2\epsilon^2 \in \{0,\,10^{-15},\,10^{-12}\}$ agree to within $0.3\%$ of each other (blue family in Figure~\ref{fig:accfric_traj}).
At $2\epsilon^2 = 10^{-9}$ (orange) this discrimination collapses to a $1.62\times$ ratio: both cubes drift at comparable rates ($\sim 0.12$\,m stick vs.\ $\sim 0.20$\,m slip), well below the threshold of useful $\mu$-precision.
At $2\epsilon^2 = 10^{-6}$ (dark red) the ratio reaches the indistinguishable line ($1\times$, gray dotted in the right panel): both cubes are pinned at the $\sim 2.6$\,mm scale, dominated by $\epsilon$-induced bias rather than the underlying $\mu$ difference.
This sharp collapse between $10^{-12}$ and $10^{-9}$ is one decade tighter than the boundary one would obtain from a softer (1\,MPa) cube, where the elasticity buffer masks the $\epsilon$-induced bias.

\paragraph{Stick-drift interpretation}
The non-zero stick displacement at small $\epsilon$ ($\sim 0.24$\,m at $2\epsilon^2 \le 10^{-12}$) is a \emph{physical} artifact of threshold proximity rather than a numerical solver floor: the stick coefficient lies only $0.4\,\text{\textperthousand}$ above the critical threshold, so elasticity-induced internal stress oscillations intermittently exceed the slip limit and produce slow creep.
The discrimination claim is therefore robust: it asserts that slip exceeds stick by $\sim 7.3\times$ at small $\epsilon$ (regardless of the absolute stick magnitude) and that this ratio collapses toward $1$ as $\epsilon$ grows past $2\epsilon^2 = 10^{-12}$.

\paragraph{Default choice}
Based on this sensitivity analysis we adopt $2\epsilon^2 = 10^{-12}$ as the working default for all subsequent experiments.
This is the largest $\epsilon$ that preserves $1\,\text{\textperthousand}$ $\mu$-precision under realistic stiffness, while remaining well above the hard-NCP limit and providing ample backward-pass conditioning (Section~\ref{sec:ncp}).
The smoothness-consistency requirement (Appendix~\ref{sec:smoothness_consistency}) holds at any matched magnitude; the choice between $10^{-15}$ and $10^{-12}$ is therefore guided purely by backward-pass numerics, where $10^{-12}$ provides more headroom against ill-conditioning while remaining indistinguishable from $10^{-15}$ in forward fidelity.

\subsection{Comparison with Baseline Frameworks}
\label{sec:baseline_comparison}

We benchmark our framework against three representative differentiable simulators that span the major design routes for deformable contact:
\textbf{DiffPD}~\cite{du2021diffpd} (CPU) uses projective-dynamics adjoint with a semi-implicit fixed-point backward iteration; its dynamics handle only frictionless contact and a static-friction approximation.
\textbf{Polyfem}~\cite{huang2024differentiable} (CPU) uses IPC barrier contact with analytical adjoint differentiation, representing the IPC route.
\textbf{Newton}~\cite{newton2025} (formerly Warp \texttt{warp.sim}~\cite{macklin2024warp}, GPU) uses semi-implicit FEM with tape-based automatic differentiation and penalty contact, representing the AD route.
We exclude DiffTaichi~\cite{hu2019difftaichi} and PlasticineLab~\cite{huang2021plasticinelab} (MPM with grid-based contact, ill-defined for direct comparison against mesh-based FEM), JAX-FEM~\cite{xue2023jaxfem} (quasi-static only), and tatva~\cite{pundir2026tatva} (concurrent preprint whose differentiable contact has not been independently validated).

\paragraph*{Tasks.}
We design three task categories of increasing contact difficulty:
(i)~\emph{Elasticity}: Poisson's ratio identification on a stretched Neo-Hookean bar with no contact ($\nu_0 = 0.1$, $\nu^* = 0.4$);
(ii)~\emph{Frictionless contact}: Young's modulus identification on an Armadillo dropping onto a frictionless ground ($E_0 = 10^4$~Pa, $E^* = 3\times 10^5$~Pa);
(iii)~\emph{Frictional contact}: friction coefficient identification on a bunny sliding on a slope ($\mu_0 = 0.5$, $\mu^* = 0.1$).
All frameworks operate on identical scene geometries with matched loss functions and optimizer settings.
DiffPD and Polyfem are CPU-only by construction; Newton and our framework run on GPU.
Reported wall times reflect each framework's native execution mode; we do not run our framework in CPU mode.

\paragraph*{Structural differences in contact and friction.}
The three baselines correspond to fundamentally different design choices for differentiable contact, each with characteristic gradient properties.
DiffPD's semi-implicit backward iteration, well-suited to the contact-free PD adjoint, struggles on coupled NCP-based adjoint systems with friction (Section~\ref{sec:backward_comparison}); we nonetheless report DiffPD on all three tasks for completeness, with the frictional results expected to suffer from this structural limitation.
Polyfem treats friction semi-implicitly: the tangential contact frame and relative velocity are evaluated at the previous time step, decoupling normal and tangential forces across steps~\cite{li2020incremental}.
Its IPC forward solver further requires continuous collision detection in line searches and adaptive barrier stiffness updates, both inherently sequential and difficult to parallelize on GPU.
Newton replaces complementarity with smooth penalty forces, which yield zero gradient when the object rests in static equilibrium (no penetration $\Rightarrow$ no penalty force $\Rightarrow$ no gradient signal); its tape-based AD also records all intermediate states, scaling memory linearly with simulation horizon $T$.
By contrast, our smoothed NCP formulation solves normal and frictional forces simultaneously within a single coupled residual, the adjoint inherits this full coupling, and the local-global structure maps naturally to GPU in both forward and backward passes.

\begin{table*}[htbp]
\centering
\caption{\textbf{Comparison against baseline differentiable frameworks across three task categories.}
DiffPD~\cite{du2021diffpd} and Polyfem~\cite{huang2024differentiable} run on CPU; Newton~\cite{newton2025} and our framework run on GPU.
For each method we report the problem size (DOFs), the simulated horizon (Sim time and time step $\Delta t$), and timing on both per-optimization-iteration and per-time-step granularity for the forward and backward passes, plus peak GPU memory and the end-to-end optimization wall time taken to reach a loss threshold of $10^{-10}$.
Newton uses 10--200$\times$ smaller substeps than the other frameworks for semi-implicit stability; its per-step costs are reported at this substep granularity.}
\label{tab:baseline_comparison}
\footnotesize
\setlength{\tabcolsep}{4pt}
\begin{tabular*}{\textwidth}{@{\extracolsep{\fill}}llccccccccc@{}}
\toprule
\multirow{2}{*}{Task} & \multirow{2}{*}{Method} & \multirow{2}{*}{DOFs} & \multirow{2}{*}{Sim time (s)} & \multirow{2}{*}{$\Delta t$ (s)}
     & \multicolumn{2}{c}{Per opt iter (s)} & \multicolumn{2}{c}{Per step (ms)}
     & \multirow{2}{*}{\makecell{Memory\\(MB)}} & \multirow{2}{*}{\makecell{Total\\(s)}} \\
\cmidrule(lr){6-7} \cmidrule(lr){8-9}
& & & & & Fwd & Bwd & Fwd & Bwd & & \\
\midrule
\multirow{4}{*}{Elasticity}    & DiffPD (CPU)   & 12675 & 3.0 & 0.05 & 3.38 & 4.91 & 56.3 & 81.9 & 859 & 140.9 \\
                                & Polyfem (CPU)  & 12675 & 3.0 & 0.05 & 22.56 & 7.08 & 376.0 & 118.0 & 1905 & 770.7 \\
                                & Newton (GPU)   & 12675 & 3.0 & 0.00025 & 1.53 & 1.40 & 0.128 & 0.117 & 3501 & 29.3 \\
                                & Ours (GPU)     & 12675 & 3.0 & 0.05 & 0.372 & 0.161 & 6.2 & 2.7 & 1112 & 9.6 \\
\midrule
\multirow{4}{*}{Frictionless}  & DiffPD (CPU)   & 38274 & 0.6 & 0.01 & 5.95 & 12.95 & 99.1 & 215.9 & 1291 & 302.4 \\
                                & Polyfem (CPU)  & 38274 & 0.6 & 0.01 & 291.34 & 31.30 & 4855.7 & 521.6 & 7265 & 11292.2 \\
                                & Newton (GPU)   & 38274 & 0.6 & 0.001 & 0.08 & 0.11 & 0.140 & 0.184 & 554 & 3.3 \\
                                & Ours (GPU)     & 38274 & 0.6 & 0.01 & 1.66 & 1.62 & 27.6 & 27.1 & 2454 & 93.8 \\
\midrule
\multirow{4}{*}{Frictional}    & DiffPD (CPU)   & 29250 & 1.5 & 0.01 & 99.35 & 98.29 & 662.3 & 655.3 & 1220 & 1976.4 \\
                                & Polyfem (CPU)  & 29250 & 1.5 & 0.01 & 703.35 & 10.95 & 4689.0 & 73.0 & 8941 & 39286.4 \\
                                & Newton (GPU)   & 29250 & 1.5 & 0.0001 & 2.34 & N/A & 0.156 & N/A & 10068 & N/A \\
                                & Ours (GPU)     & 29250 & 1.5 & 0.01 & 4.67 & 10.78 & 31.1 & 71.9 & 3682 & 325.2 \\
\bottomrule
\end{tabular*}
\end{table*}

\begin{figure*}[!t]
    \centering
    \begin{subfigure}{\linewidth}
        \centering
        \includegraphics[width=0.95\linewidth]{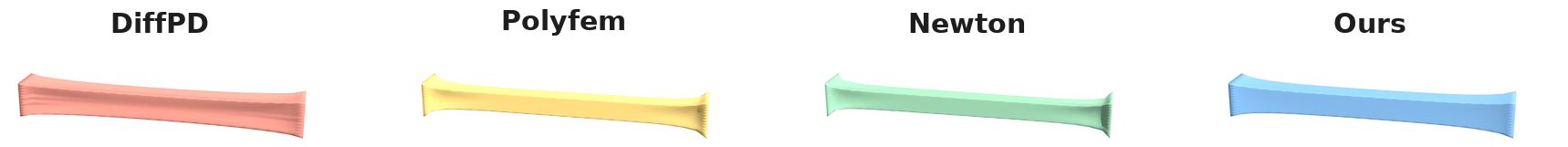}
        \caption{Elasticity (bar).}
        \label{fig:baseline_bar}
    \end{subfigure}\\[4pt]
    \begin{subfigure}{\linewidth}
        \centering
        \includegraphics[width=0.95\linewidth]{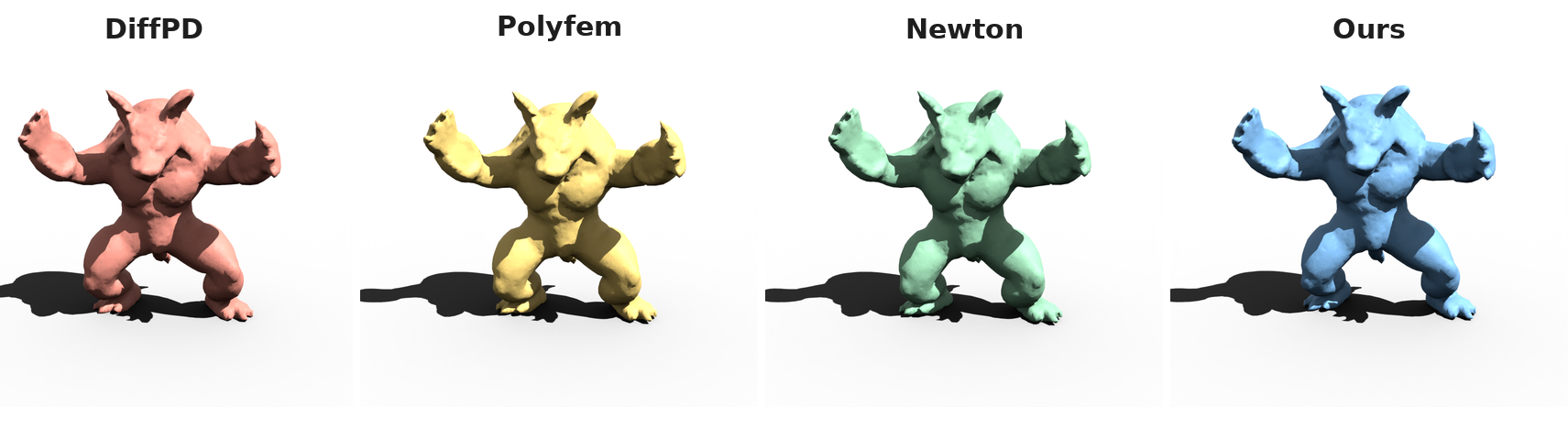}
        \caption{Frictionless contact (Armadillo).}
        \label{fig:baseline_armadillo}
    \end{subfigure}\\[4pt]
    \begin{subfigure}{\linewidth}
        \centering
        \includegraphics[width=0.95\linewidth]{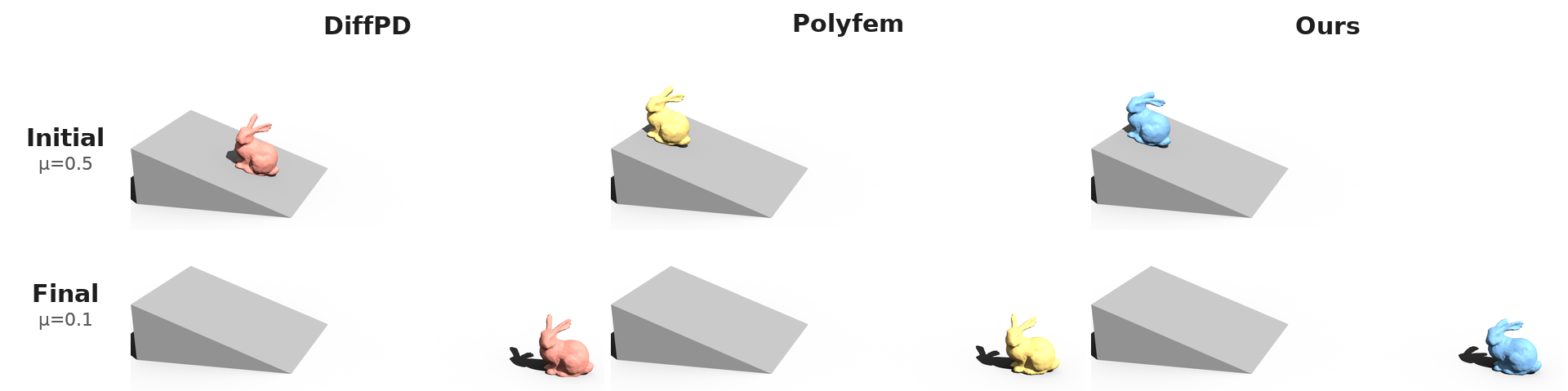}
        \caption{Frictional contact (Bunny on slope): initial (top, $\mu = 0.5$) and final (bottom, $\mu \to 0.1$) configurations; Newton omitted (NaN gradients).}
        \label{fig:baseline_bunny}
    \end{subfigure}
    \caption{\textbf{Visual comparison on three baseline tasks.} All four frameworks recover the target on (a) elasticity and (b) frictionless contact; on (c) frictional contact only our framework reaches a fast, correct solution (DiffPD: lagged friction, $\sim 33$\,min; Polyfem: $\sim 11$\,hr; Newton: NaN gradients). Quantitative timing in Table~\ref{tab:baseline_comparison}.}
    \Description{Three-row visual comparison: (a) a stretched elastic bar shown in four colors corresponding to the four frameworks (DiffPD pink, Polyfem yellow, Newton green, Ours blue), all in similar deformed shapes; (b) four armadillos in the same colors, all in similar dropped-and-deformed poses; (c) a 3-by-2 grid of bunnies on a sloped ramp comparing DiffPD (pink), Polyfem (yellow), and our framework (blue) at the initial high-friction configuration ($\mu=0.5$, top row, bunny resting on slope) and the final low-friction configuration ($\mu=0.1$, bottom row, bunny slid down off the slope).}
    \label{fig:baseline_comparison}
\end{figure*}

\paragraph*{Results.}
Table~\ref{tab:baseline_comparison} reports end-to-end optimization wall time as the headline metric (rightmost column); the per-iteration and per-step Fwd/Bwd costs decomposed in the middle columns indicate where the time is spent. Figure~\ref{fig:baseline_comparison} shows the corresponding final configurations.

On the elasticity bar task, all four frameworks converge to the target Poisson's ratio, confirming basic correctness of each adjoint route in the absence of contact.
Polyfem, Newton, and our framework use Neo-Hookean hyperelasticity for the forward solve, while DiffPD uses a hyperelastic approximation supported by its projective-dynamics solver.

On the frictionless Armadillo task, DiffPD and Polyfem converge to the target Young's modulus, while Newton requires a 4\% offset of the initial Young's modulus to $E_0 = 1.04\times 10^4$\,Pa since it produces a spurious local maximum at $E_0 = 10^4$\,Pa that traps the optimization.
With this workaround Newton is the fastest end-to-end at $3.3$\,s, reflecting the cheap per-step cost of penalty contact; our framework completes the same task in $93.8$\,s with NCP-based contact and no parameter offset.

On the frictional bunny task, DiffPD completes in $1{,}976.4$\,s ($\sim$\,33\,min) via its lagged static-friction approximation that decouples normal and tangential forces across time steps: the regime predicted to suffer in Section~\ref{sec:backward_comparison}.
Polyfem reaches the target after $39{,}286.4$\,s ($\sim$\,11\,hr)---the highest wall time among the four frameworks at this DOF count---reflecting IPC's CCD line search and adaptive barrier-stiffness updates, design choices favoring robustness over parallel throughput.
Newton produces NaN gradients, likely from a normalization singularity in its friction adjoint.
Our framework finishes in $325.2$\,s ($6.1\times$ faster than DiffPD), preserving full normal--frictional coupling in both passes; it is the only framework of the four to combine completion, structural correctness, and end-to-end wall time within minutes on this benchmark.

\subsection{Applications}
\label{sec:applications}

We demonstrate the practical value of our differentiable simulator on four contact-rich tasks: soft-body locomotion with high-dimensional muscle control, inverse-control trajectory matching through wall contact, batched differentiable simulation at the million-DOF scale, and sim-to-real material identification on real-world observations.
These applications exercise the framework's three core strengths: continuous cross-regime gradients (smoothed FB), accurate friction coupling (NCP), and efficient long-horizon backpropagation through the analytical adjoint.

\subsubsection{Seal Locomotion}

\begin{figure}[ht]
    \centering
    \begin{subfigure}{0.7\columnwidth}
        \centering
        \includegraphics[width=\linewidth]{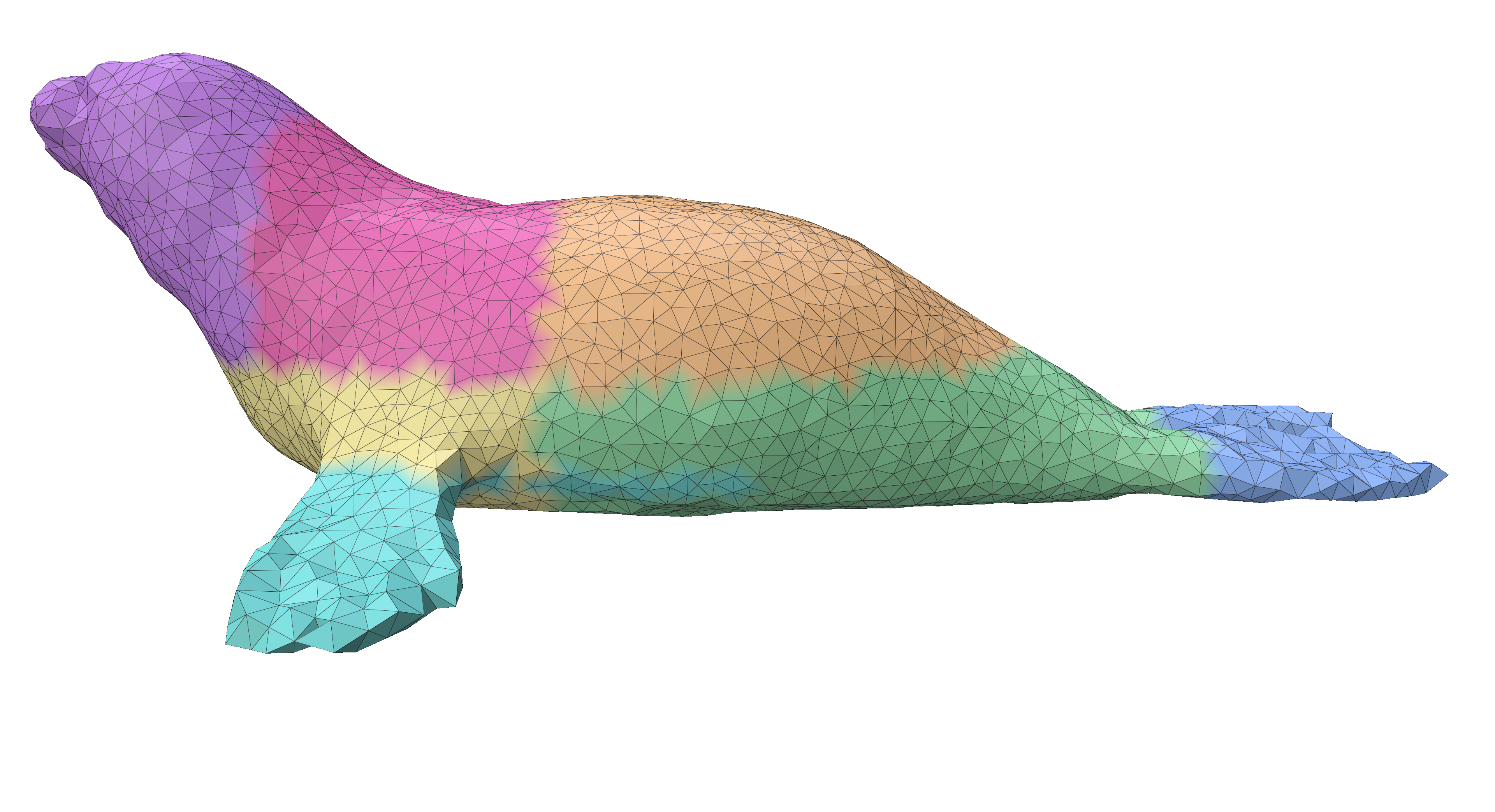}
        \caption{Side view.}
    \end{subfigure}\hfill
    \begin{subfigure}{0.3\columnwidth}
        \centering
        \includegraphics[width=\linewidth]{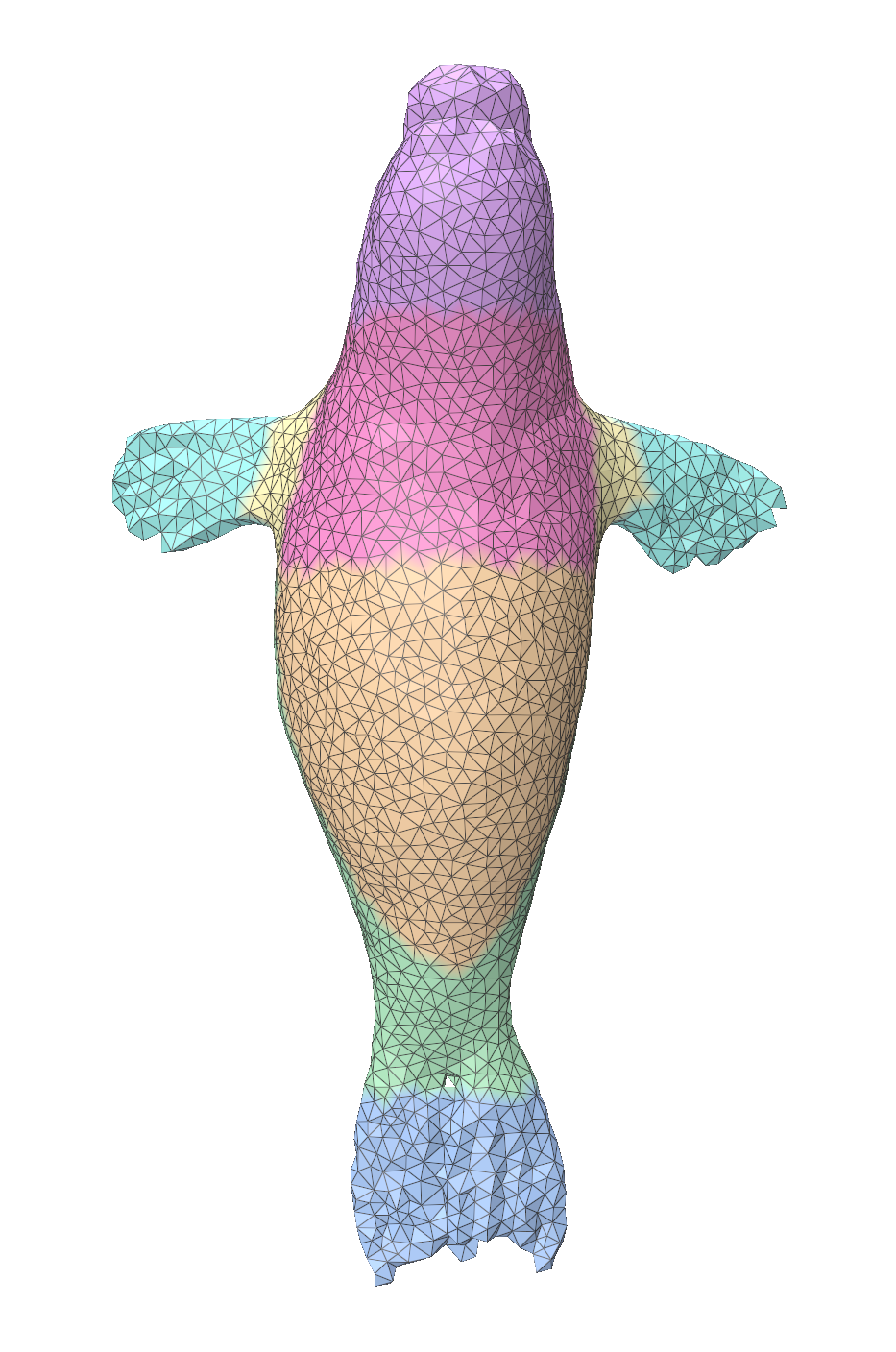}
        \caption{Top view.}
    \end{subfigure}
    \caption{\textbf{Muscle-group layout on the elastic seal.}
    The volumetric tetrahedral body is partitioned into $14$ contractile-fiber muscle groups (dorsal/ventral pairs at fore-left/-right and rear-left/-right, plus flipper, tail, and neck on each side), color-coded above.
    Each group contains $3$ fibers along an anatomically prescribed direction; per-group activations $a_i(t)$ scale the resulting anisotropic contractile force.
    The top view shows the left/right symmetry of the flipper, neck, and tail groups.}
    \Description{Two views of an elastic tetrahedral seal mesh with surface colored by 14 muscle-group regions: a side view showing dorsal versus ventral organization along the body, and a top view showing left/right paired groups at the flippers, neck, and tail.}
    \label{fig:seal_muscles}
\end{figure}

We optimize the muscle-like actuation of an elastic seal on a frictional ground to maximize its forward crawl distance (Figure~\ref{fig:seal_stride}).
The seal is driven by $14$ contractile-fiber muscle groups laid out as dorsal/ventral pairs at fore-left/-right and rear-left/-right positions, plus flipper, tail, and neck groups on each side (Figure~\ref{fig:seal_muscles}); each group carries $3$ fibers along an anatomically prescribed direction, and a per-group activation $a_i(t) \in [0.2, 3.0]$ scales the resulting anisotropic contractile force.
Activations evolve over $500$ time steps, yielding $7{,}000$ control parameters in total, well beyond the reach of gradient-free or sample-based methods at any reasonable budget.
The objective combines forward displacement with a shape-preservation term, a front-lift bonus, and temporal smoothness on $a_i(t)$.

Forward motion is contact-driven: a hand-tuned sinusoidal gait that ignores contact actually drifts \emph{backward} by $0.21$\,m.
We use this sinusoidal gait as the optimizer's initialization, from which our adjoint pipeline must flip the trajectory direction by learning the right muscle--friction coupling; it drives the seal to $6.13$\,m ($\sim 4.4$ body lengths) of forward travel at $\sim 1.1$\,m/s, within the published range for real terrestrial seal crawl.
Optimization completes in $69$ iterations across a two-stage curriculum: stage 1 discovers a $100$-step stride from rest ($59$ iterations), and stage 2 tiles this short stride $5\times$ as a warm start and refines the full $500$-step horizon ($10$ iterations), for $\sim 48$ minutes of total wall time on a single RTX 5090.

\paragraph*{Long-horizon optimization.}
We extend the seal optimization to longer horizons to verify that the analytical adjoint produces stable, informative gradients across deep chain rules.
Starting from the converged $100$-step gait tiled to $T \in \{1500, 2000, 2500, 3000\}$ active steps ($15$--$30$\,s of simulated locomotion), we apply a single Adam step at each horizon and measure the gradient norm, the displacement update, and per-step cost (Table~\ref{tab:long_horizon}).

Two findings support the stability of the adjoint pipeline at long horizons.
First, the gradient norm $\|\partial L / \partial a\|$ remains bounded at $\le 5$ across all tested $T$ up to $3000$, growing roughly linearly rather than exponentially with $T$; this is direct evidence that backpropagation through $3000$ implicit-Euler steps does not destabilize the signal.
Second, the single optimization step improves forward displacement at every tested horizon ($\Delta \in [0.012, 0.096]$\,m, $+0.06\%$ to $+0.22\%$), confirming that the gradient direction remains correct at all tested horizons.

Per-iteration costs scale linearly with $T$ at approximately $22$\,ms/step (forward) and $17$\,ms/step (backward) on a single RTX 5090.
The forward-rolled-out gait further sustains indefinitely without policy collapse: tiling the converged $100$-step actuation to a $2000$-step rollout produces a linear forward trajectory at $\sim 1.43$\,m/s---approximately one body length per second, consistent with adult harbor-seal terrestrial crawl speed---over $20$\,s of simulated time.

\begin{table}[ht]
\centering
\caption{\textbf{Long-horizon optimization on the seal task.}
Starting from the converged $100$-step gait tiled to the target horizon, we apply a single Adam step at each $T$ and report the gradient norm and displacement before / after the step.}
\label{tab:long_horizon}
\footnotesize
\begin{tabular*}{\columnwidth}{@{\extracolsep{\fill}}rrrrr@{}}
\toprule
$T$ & $\|\partial L/\partial a\|$ & \makecell{Disp init\\(m)} & \makecell{Disp upd\\(m)} & $\Delta\%$ \\
\midrule
$1500$ & $0.74$ & $21.38$ & $21.39$ & $+0.06\%$ \\
$2000$ & $1.57$ & $28.63$ & $28.67$ & $+0.15\%$ \\
$2500$ & $2.86$ & $35.87$ & $35.94$ & $+0.20\%$ \\
$3000$ & $4.84$ & $43.11$ & $43.21$ & $+0.22\%$ \\
\bottomrule
\end{tabular*}
\end{table}


\subsubsection{Elephant Trunk Writing}
\label{sec:elephant_trunk_writing}

\begin{figure*}[!t]
    \centering
    \begin{subfigure}{\linewidth}
        \centering
        \includegraphics[width=\linewidth]{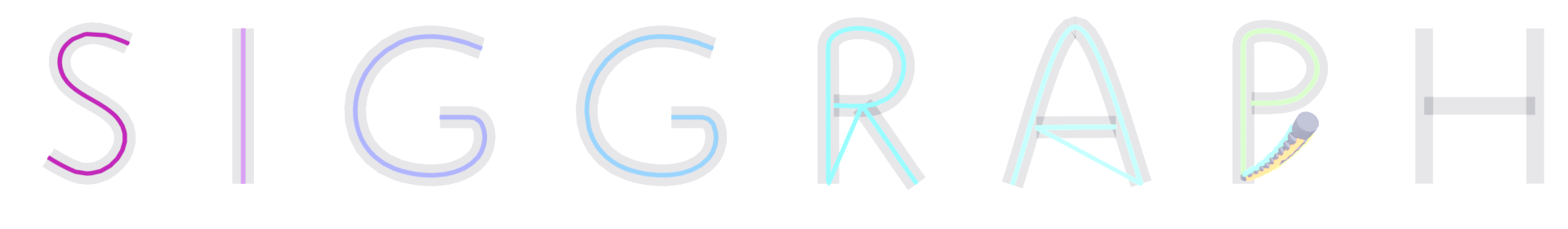}
        \caption{Tip trajectories (one color per letter) overlaid on target characters (thick gray); 3D trunk visible mid-stroke on ``P''.}
    \end{subfigure}\\[2pt]
    \begin{subfigure}{\linewidth}
        \centering
        \includegraphics[width=\linewidth]{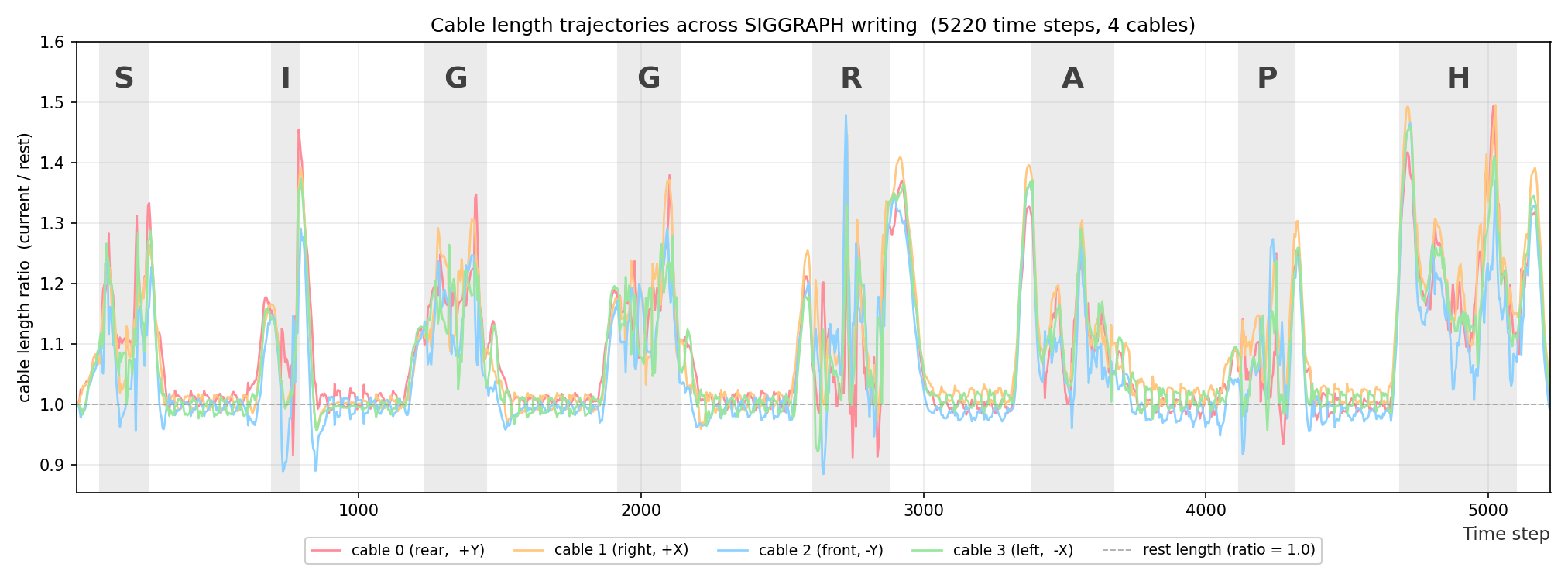}
        \caption{Cable-length traces (rear/right/front/left) across the $5{,}220$-step rollout; eight gray bands mark per-letter windows, dashed line is rest length.}
    \end{subfigure}
    \caption{\textbf{Writing ``SIGGRAPH''.} Optimized cable-driven trunk tracing the eight target letters.}
    \Description{Top: the word SIGGRAPH rendered as eight thick gray letter outlines arranged horizontally; most letters are overlaid with a thinner colored polyline tracing the trunk tip's actual trajectory --- pink for S, purple for I, light blue for the first G, slightly darker blue for the second G, cyan for R and A, and green for P; the final H carries no overlay yet, and the 3D elephant trunk model is visible at the lower-right portion of the P mid-stroke. Bottom: a wide line plot of the four cables' rest-length ratios (rear, right, front, left) over the full $5{,}220$-step rollout, with the eight SIGGRAPH letter intervals highlighted as labeled gray bands; the curves stay near a horizontal dashed line at ratio one for most of the rollout and then make sharp, paired counter-movements inside each letter band, peaks reaching roughly $1.5\times$ rest and troughs dropping to roughly $0.9\times$ rest.}
    \label{fig:trunk_writing}
\end{figure*}

While our core contribution focuses on frictional contact, robustly differentiating through complex, coupled hyperelasticity and hard bilateral constraints over long horizons is a prerequisite for contact-rich tasks. To stress-test this elastic foundation in isolation, we model a soft elephant trunk as a tetrahedral elastic body and optimize its cable actuation to trace the letters of "SIGGRAPH" along a planar target trajectory (Fig.~\ref{fig:trunk_writing}).

\begin{figure}[htbp]
    \centering
    \includegraphics[width=\linewidth]{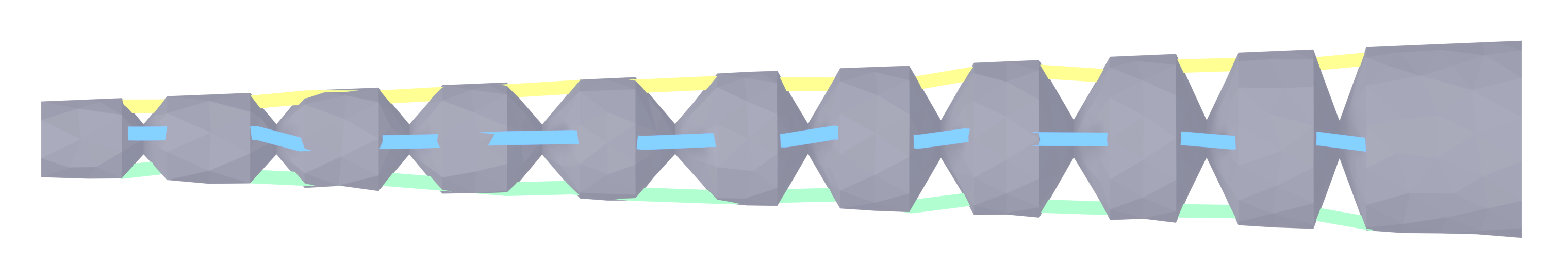}
    \caption{\textbf{Cable-driven actuation of the elephant trunk.} The body (gray) is discretized as a tetrahedral mesh, tapering from a thicker root on the right to a thinner tip on the left. The four actuator cables are visible as yellow along the top, green along the bottom, and blue along the central axis, with a fourth cable on the far side that is occluded in this view; each cable is drawn as the polyline connecting the mesh anchors it threads through.}
    \Description{Side view of a soft elephant trunk discretized as a gray tetrahedral mesh, tapering from a thicker root on the right to a thinner tip on the left. Three colored polylines run longitudinally along the trunk body --- yellow along the top, blue along the central axis, and pale green along the bottom --- representing three of the four embedded actuator cables; the fourth cable is on the far side of the trunk and not visible in this view.}
    \label{fig:trunk_actuation}
\end{figure}

The body is actuated by four cables routed longitudinally through the mesh: each cable threads through a chain of anchor points and enforces a distance constraint on that chain, so shortening or lengthening a cable contracts or extends the corresponding portion of the body (Fig.~\ref{fig:trunk_actuation}).
The control variables exposed to the optimizer are the four cables' target lengths over time.
The core difficulty is the body's elastic response: every cable adjustment propagates through the compliant material as a coupled deformation, so a small length change perturbs the entire tip trajectory across a long horizon, and the optimizer must plan through these nonlinear dynamics to trace each target stroke.

\paragraph*{Setup.}
The trunk is discretized into $709$ vertices and $1{,}972$ tetrahedra ($2{,}127$ DOF) and modeled with a Neo-Hookean energy at density $\rho = 1000$\,kg/m$^{3}$, Young's modulus $E = 10^{5}$\,Pa, and Poisson ratio $\nu = 0.2$, plus a small anti-twist soft regularizer that suppresses unphysical torsion of the body.
The four cables each thread through $20$ mesh-vertex anchors ($10$ segments per cable), with each segment's length imposed as a hard bilateral Lagrange constraint (compliance $10^{-10}$).
Forward dynamics use an implicit Projective-Dynamics-style local--global solver ($10$ local--global rounds per step, with a Fischer--Burmeister inner Newton for the bilateral constraints) at $\Delta t = 0.01$\,s and zero gravity, over a $5{,}220$-step rollout that interleaves the eight letter-writing phases ($1{,}884$ steps in total) with inter-letter lift / lower / sidestep connectors ($3{,}340$ steps).
Each letter's target stroke is sampled at stride $5$, yielding $377$ write-phase keypoints in total.
The targets occupy an aggregate $\approx\!2.9\,\text{m} \times 1.0\,\text{m}$ rectangle on the wall, comparable to the trunk's own rest length of $2.617$\,m, so writing each glyph requires the tip to traverse much of the body's full reach (per-letter principal-axis lengths range from $0.55$ to $2.16$\,m, mean $1.37$\,m).
We minimize the time-indexed squared distance between the trunk tip (barycentric-interpolated from a fixed tet) and its target keypoint, and report the segment-averaged RMS tip-target distance in millimetres.
The four cables' per-step target lengths are optimized by a windowed-shooting curriculum on $5$-step segments. This sliding-window approach leverages the efficiency of our analytical adjoint to maintain high-quality gradient signals over long total durations without the vanishing or exploding issues of full-horizon unrolling: each segment is locally optimized with plain gradient descent (learning rate $3.0$, at most $20$ iters per segment) and advances once its loss drops below $10^{-4}$ or the segment-end tip displacement is within $12$\,mm of its target.

\paragraph*{Results.}
Figure~\ref{fig:trunk_writing}(a) overlays the optimized tip trajectories on the eight target glyphs, one color per letter: the cable-driven body recovers the stroke-level shape of each character despite the long-horizon coupled elastic dynamics, and the per-letter mean RMS tracking errors sit in a tight $14.75$--$15.75$\,mm band ($15.36$\,mm averaged across all eight letters, or $\approx\!1.4\%$ of each letter's principal-axis length given that the writing scene is at trunk-scale; Table~\ref{tab:elephant_timing}).
The cable-length traces in Fig.~\ref{fig:trunk_writing}(b) reveal how the optimizer organizes this writing: each cable sits near its rest length during the inter-letter connectors and concentrates sharp, coordinated excursions inside the eight letter windows, with antagonistic pairs (left/right and front/rear) extending and contracting in opposition to drive the lateral and elevation swings of the tip.
Peak stretch reaches $\approx\!1.5\times$ rest and slack episodes dip to $\approx\!0.9\times$ rest, well within the physical range of a tendon-driven soft actuator.
The full $5{,}220$-step demo completes end-to-end in $\approx\!18$\,min of wall time --- $\approx\!9.5$\,min on the eight write phases and $\approx\!8.8$\,min on the inter-letter connectors.

\begin{table}[htbp]
\centering
\caption{\textbf{Elephant trunk writing: per-letter timing and tracking accuracy.} Write-phase step count, target keypoint count, mean forward and adjoint wall time per optimization iteration (a single $5$-step segment forward + adjoint), total optimization iterations summed across the letter's segments, converged RMS tip-target distance, and total wall time spent on the letter's write phases (excludes inter-letter lift / lower / sidestep connectors).}
\label{tab:elephant_timing}
\footnotesize
\setlength{\tabcolsep}{4pt}
\begin{tabular*}{\columnwidth}{@{\extracolsep{\fill}}lrrrrrrr@{}}
\toprule
Char & Steps & Keypts & \makecell{Fwd\\(ms/iter)} & \makecell{Bwd\\(ms/iter)} & Iters & \makecell{Error\\(mm)} & \makecell{Wall\\(min)} \\
\midrule
S     & $173$     & $34$  & $110.9$ & $132.2$ & $224$     & $15.28$ & $0.9$ \\
I     & $101$     & $21$  & $107.0$ & $131.4$ & $144$     & $15.53$ & $0.6$ \\
G     & $221$     & $44$  & $104.0$ & $130.8$ & $242$     & $14.75$ & $1.0$ \\
G     & $221$     & $44$  & $104.3$ & $128.7$ & $252$     & $14.94$ & $1.0$ \\
R     & $269$     & $54$  & $104.0$ & $132.1$ & $409$     & $15.75$ & $1.6$ \\
A     & $289$     & $57$  & $112.5$ & $130.0$ & $315$     & $15.28$ & $1.3$ \\
P     & $197$     & $40$  & $105.5$ & $131.7$ & $251$     & $15.61$ & $1.0$ \\
H     & $413$     & $83$  & $101.8$ & $130.9$ & $518$     & $15.59$ & $2.1$ \\
\midrule
Total & $1{,}884$ & $377$ & $105.7$ & $131.0$ & $2{,}355$ & $15.36$ & $9.5$ \\
\bottomrule
\end{tabular*}
\end{table}

\subsubsection{Million-DOF Batched Differentiable Simulation}
\label{sec:parallel_sysid}

The applications above each operate on a single deformable scene; we now demonstrate that the same analytical adjoint pipeline scales to approximately $1.56$\,M degrees of freedom in a single batched forward and backward pass, aggregating up to $512$ independent simulations on one GPU.
Typical differentiable simulators in graphics operate at $10^{3}$--$10^{5}$ DOFs (cf.\ the baselines in Section~\ref{sec:baseline_comparison}); the experiment below shows that batched aggregation is a practical route to an order-of-magnitude larger scale.

\paragraph*{Setup.}
Each environment instantiates an identical scene: a $1{,}013$-vertex cloth patch ($1{,}936$ triangles, $3{,}039$ DOF per env) draping under gravity onto a frictionless sphere collider, with ARAP elasticity and isometric bending.
Per-environment variation comes from the Young's modulus $E_e$, drawn log-uniformly from $[10^{4}, 10^{6}]$\,Pa; the per-env target is the final-state cloth configuration produced by a forward simulation at the same $E_e^{\text{true}}$.
Each optimization iteration performs one batched forward rollout (50 steps, $\Delta t = 0.01$\,s) followed by one batched adjoint sweep returning the aggregate gradient with respect to $\{E_e\}_{e=1}^{N}$.

\paragraph*{Scaling protocol.}
We sweep the number of parallel environments $N \in \{4, 8, 16, 32, 64, 128, 256, 512\}$, reporting aggregate vertex count and DOF, per-step forward and backward times, total opt-iter wall time, and peak GPU memory.
At $N=512$ the simulation aggregates $518{,}656$ vertices and approximately $1.56$\,M DOF in a single differentiable rollout, with peak VRAM of $29.2$\,GB ($91\%$ of the 32\,GB capacity of one RTX 5090); $N=1024$ exceeds the device's memory.

\begin{figure}[!t]
    \centering
    \includegraphics[width=\linewidth]{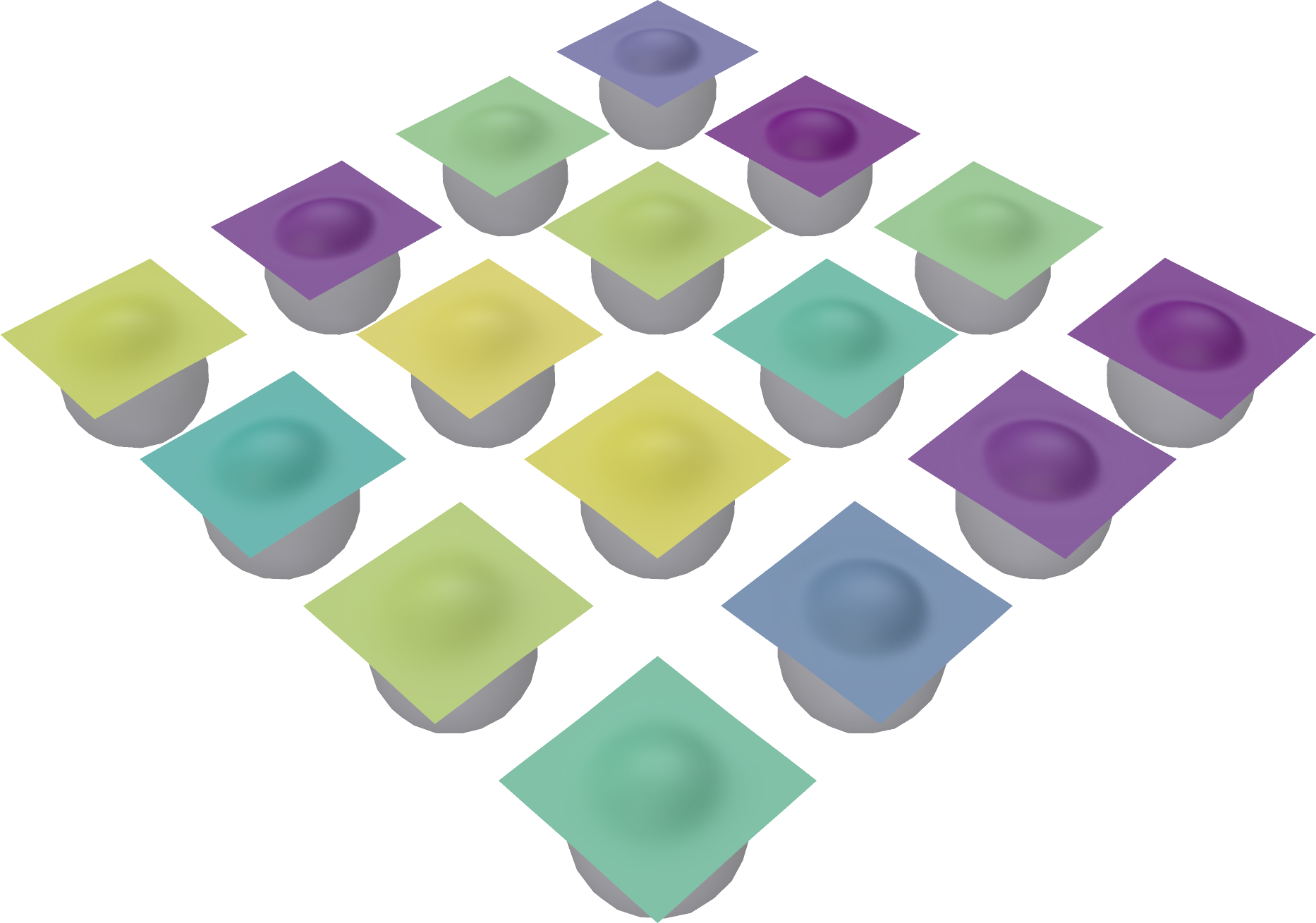}
    \caption{\textbf{Parallel-environment differentiable rollout ($N=16$).}
    A $4\times4$ grid of cloth-on-frictionless-sphere environments at the end of the same $50$-step batched rollout ($t = 0.5$\,s).
    Cloth color encodes per-environment Young's modulus $E_e$ on a $\log_{10}$ viridis scale: purple ($E_e \approx 10^{4}$\,Pa, softest, deepest drape) to yellow ($E_e \approx 7.4\times 10^{5}$\,Pa, stiffest, only the contact patch dips).
    All $16$ rollouts share a single fused forward and adjoint pass and return one gradient per $E_e$.}
    \Description{Sixteen cloth patches arranged on a 4-by-4 grid, each draping over a sphere; the cloth colors range from purple (soft, deeply draped) to yellow (stiff, minimally deformed), showing the per-environment Young's-modulus diversity in a single batched rollout.}
    \label{fig:parallel_env_render}
\end{figure}

\paragraph*{Results.}
Table~\ref{tab:parallel_env_scaling} reports the scaling sweep, and two distinct cost regimes emerge as $N$ grows.
At small $N$ ($\le 32$) the per-step backward PCG dominates ($\sim 1.2$--$5\times$ the forward cost): Krylov dot products and SpMVs on a small aggregated system cannot saturate the 5090's bandwidth.
Around $N = 32$--$64$ the two phases reach parity, and above $N = 64$ the forward sparse solve dominates with per-step forward cost growing super-linearly with DOF (roughly $3\times$ per $2\times$ envs at $N=512$) as factor-graph fill-in increases; the backward PCG, by contrast, scales sub-linearly from $32$ envs onward as the GPU pipeline saturates.
Per-environment per-step cost is therefore minimized at $N=64$--$128$ ($\sim 2$\,ms per env per step combined), forming the practical sweet spot of the pipeline.

\begin{table}[!t]
\centering
\caption{\textbf{Scaling of the parallel-environment differentiable rollout.}
Each environment is a $1{,}013$-vertex cloth patch on a frictionless sphere (see Section text).
The table reports aggregate vertex count and DOF, per-step forward and backward wall times, total optimization-iteration wall time (one forward rollout plus one adjoint sweep over $50$ time steps), and peak GPU memory.
The bottleneck migrates with scale: backward-bound at small $N$, forward-bound above $N=64$.}
\label{tab:parallel_env_scaling}
\footnotesize
\setlength{\tabcolsep}{4pt}
\begin{tabular*}{\columnwidth}{@{\extracolsep{\fill}}rrrrrrr@{}}
\toprule
$N$ & Verts & DOF & \makecell{Fwd\\(ms/step)} & \makecell{Bwd\\(ms/step)} & \makecell{Opt iter\\(s)} & \makecell{VRAM\\(GB)} \\
\midrule
4    & $4{,}052$     & $12{,}156$      & $3.4$    & $11.9$  & $0.82$  & $3.1$  \\
8    & $8{,}104$     & $24{,}312$      & $6.1$    & $29.9$  & $1.91$  & $3.8$  \\
16   & $16{,}208$    & $48{,}624$      & $12.3$   & $33.3$  & $2.55$  & $3.7$  \\
32   & $32{,}416$    & $97{,}248$      & $29.2$   & $34.3$  & $3.78$  & $5.0$  \\
64   & $64{,}832$    & $194{,}496$     & $64.5$   & $61.0$  & $7.51$  & $6.1$  \\
128  & $129{,}664$   & $388{,}992$     & $159.3$  & $94.1$  & $15.1$  & $9.8$  \\
256  & $259{,}328$   & $777{,}984$     & $436.0$  & $158.6$ & $34.6$  & $15.8$ \\
512  & $518{,}656$   & $1{,}555{,}968$ & $1317.7$ & $294.3$ & $90.3$  & $29.2$ \\
\bottomrule
\end{tabular*}
\end{table}

This places batched differentiable simulation at the million-DOF scale within reach of a single consumer GPU, enabling batched system identification and reinforcement-learning rollouts with gradient propagation across hundreds of parallel environments: scales at which tape-based AD memory growth becomes prohibitive.

\subsubsection{System Identification from Real-World Observation}

\begin{figure}[!t]
    \centering
    \begin{subfigure}{0.485\linewidth}\centering
        \includegraphics[width=\linewidth]{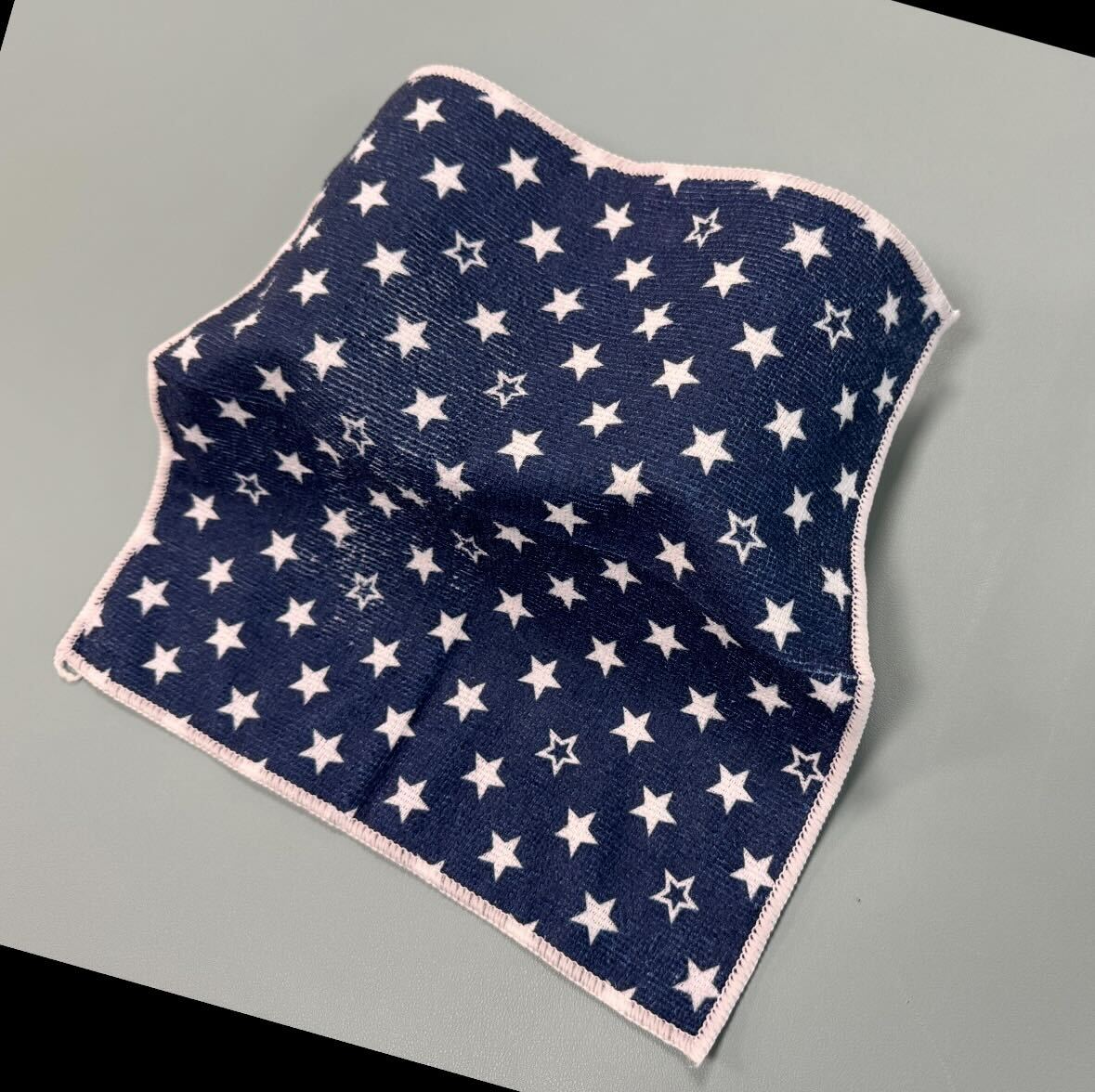}
        \caption{Real photo}
    \end{subfigure}\hfill
    \begin{subfigure}{0.485\linewidth}\centering
        \includegraphics[width=\linewidth]{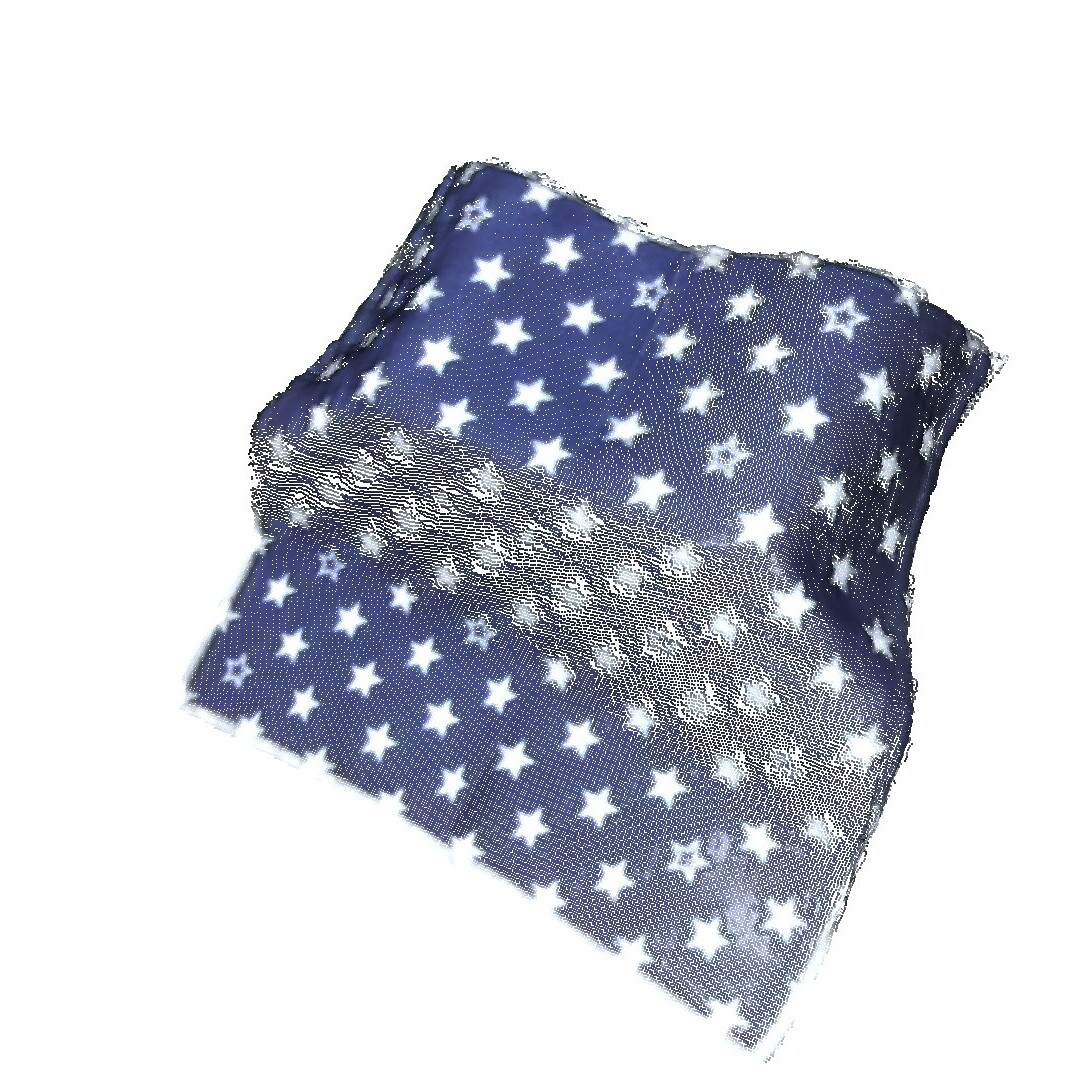}
        \caption{Point cloud}
    \end{subfigure}\\[3pt]
    \begin{subfigure}{0.485\linewidth}\centering
        \includegraphics[width=\linewidth]{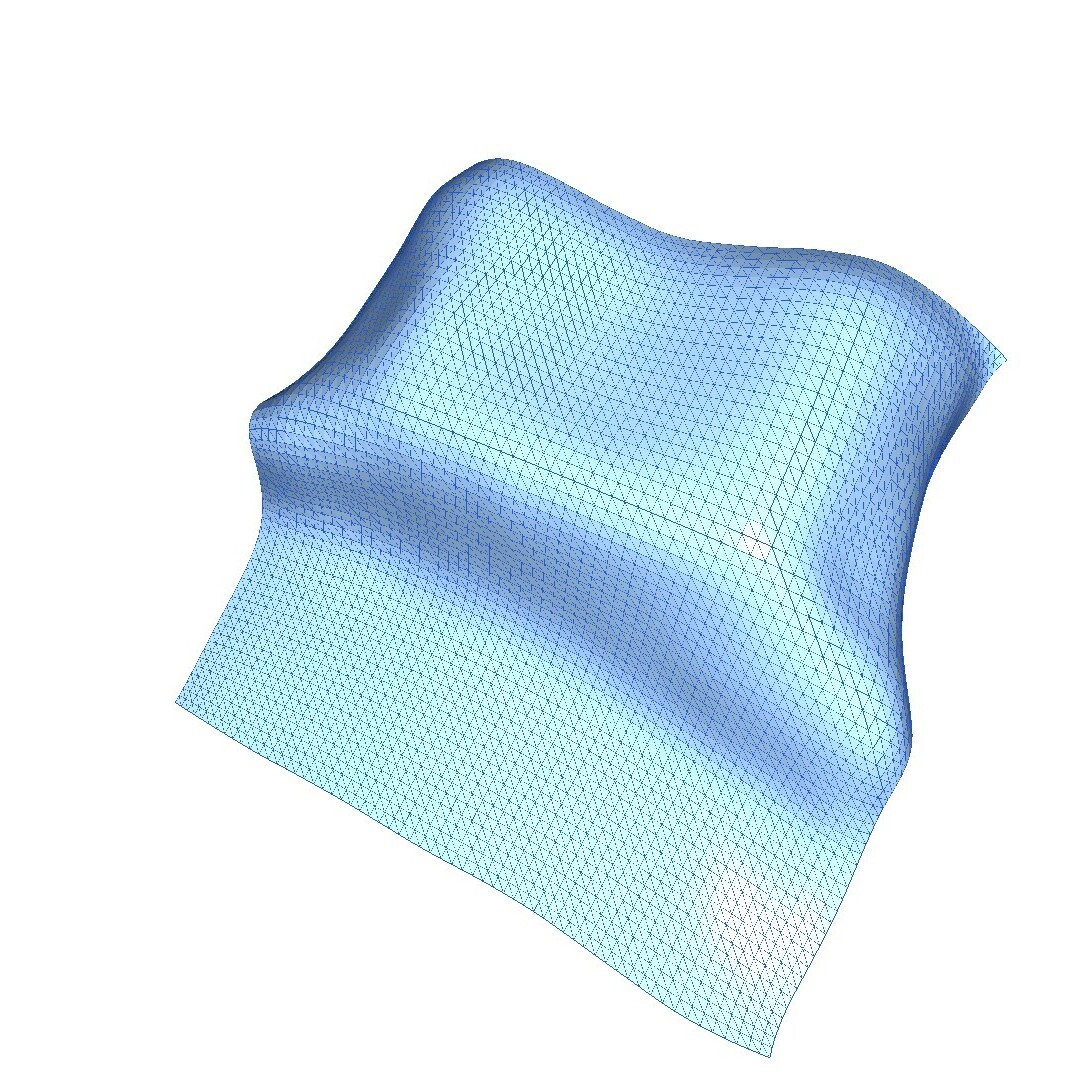}
        \caption{Reconstructed mesh}
    \end{subfigure}\hfill
    \begin{subfigure}{0.485\linewidth}\centering
        \includegraphics[width=\linewidth]{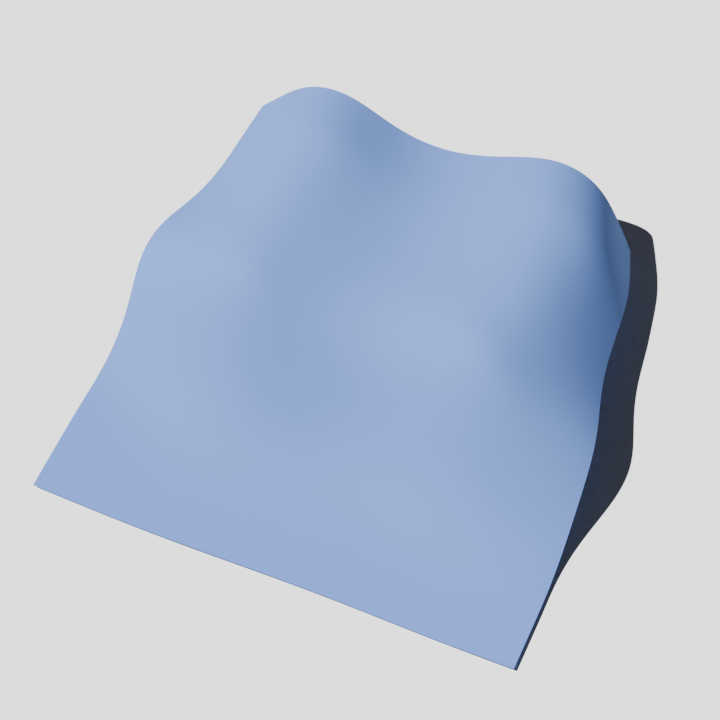}
        \caption{Simulation at identified $(E, k_b)$}
    \end{subfigure}
    \caption{\textbf{Real-world system identification on a cotton towel.}
    A real cloth draped over two cylindrical cups (a) is captured into a depth-camera point cloud (b), reconstructed into a $3{,}721$-vertex mesh (c), and ICP-aligned to the simulation frame; our differentiable simulator then identifies $E \approx 1.78\times 10^{5}$\,Pa and $k_b \approx 1.59$ by minimizing the per-vertex residual against the target mesh (d).}
    \Description{Four-panel figure: a real photo placeholder, a captured point cloud of a cloth draped over two cups, a reconstructed mesh of the same cloth, and a simulated cloth drape at the identified material parameters.}
    \label{fig:realworld}
\end{figure}

We demonstrate end-to-end transfer of our differentiable simulator to real-world material identification on a cotton towel draped over two cylindrical cups (Figure~\ref{fig:realworld}).
A depth-camera capture is reconstructed into a $3{,}721$-vertex mesh and ICP-aligned to the simulation frame ($6$\,mm RMSE); we then identify the Young's modulus $E$ and the bending coefficient $k_b$ by minimizing the mean per-vertex residual between simulation and target.

Optimization recovers $E \approx 1.78 \times 10^{5}$\,Pa (within the order of magnitude expected for soft textiles) and $k_b \approx 1.59$, reducing the mean per-vertex residual from $21.4$\,mm to $19.4$\,mm ($\sim 6.5\%$ of the cloth dimension).
Without ground-truth material parameters this is a qualitative demonstration of pipeline feasibility; the residual floor reflects sim noise, ICP alignment error, and constitutive mismatch rather than optimization failure.

\section{Conclusion}

We have presented an analytical adjoint framework for differentiable deformable-body simulation with NCP-based Coulomb friction.
By implicitly differentiating the converged coupled residual of the forward system, our approach addresses three challenges in a unified manner: non-smooth contact transitions via a smoothed Fischer--Burmeister formulation, isotropic hyperelastic constitutive models (e.g., Neo-Hookean) via SVD-based projection differentiation, and large-scale adjoint solves via a GPU-parallel Krylov solver with a Woodbury preconditioner that jointly exploits the elastic and contact structure of the linearized adjoint matrix and uniformly outperforms simpler choices across all contact regimes.

\paragraph*{Limitations.}
The validity of our gradients depends on the smoothing parameter $\epsilon$ in the NCP formulation.
While small $\epsilon$ values induce negligible physical error, extremely stiff contact may require careful tuning to balance conditioning with accuracy.
The global coupling inherent in implicit integration means cost scales super-linearly with degrees of freedom; for very high-resolution meshes, the memory footprint of sparse factorizations remains a potential bottleneck.
Our formulation also assumes fixed topology: differentiating through fracture or tearing involves discontinuous state-space jumps that our continuous relaxation does not address.
Unlike barrier-based formulations such as IPC~\cite{li2020incremental, huang2024differentiable}, our smoothed NCP does not provide intersection-free guarantees by construction; we rely on time-step size and contact tolerance to control penetration in practice.

\paragraph*{Future work.}
The combination of a high-fidelity forward solver and reliable analytical gradients provides a foundation for sim-to-real transfer; validating this pipeline against real-world measurements is a natural next step.
The framework's differentiability and performance also suggest potential for model predictive control (MPC) in soft robotics, where analytical gradients through contact events could enable real-time nonlinear control.
Adapting the adjoint formulation to multi-GPU architectures via domain decomposition would further extend scalability to large-scale simulations.

\begin{acks}
This work was supported in part by the NUS Presidential Young Professorship grant (A-0009982-00-00) and the MOE AcRF Tier 1 24-1234-P0001 Funding.
The authors would like to thank the PranaLab for their generous support and for providing the collaborative environment necessary to conduct this research.
We also thanks the support from Swiss AI Initiative Small Grant, NVIDIA Academic Grant, and Google Research Funding.
\end{acks} 

\bibliographystyle{ACM-Reference-Format}
\bibliography{content/graphics}

\appendix

\section{Frictional Contact}

\subsection{Construct Local Transformation}
\label{sec:apx_A1}

When both $\|\pene_f\|$ and $\|\force_f\|$ are nonzero, we define
\begin{equation}
\begin{aligned}
    \ortho &= \frac{1}{\|\pene_f\|}
    \begin{bmatrix} 
        \|\pene_f\| & 0 & 0 \\ 
        0 & \delta_{f_1} & \delta_{f_2} \\
        0 & -\delta_{f_2} & \delta_{f_1}
    \end{bmatrix} \\
    &= \frac{1}{\|\force_f\|}
    \begin{bmatrix} 
        \|\force_f\| & 0 & 0 \\ 
        0 & -\lambda_{f_1} & -\lambda_{f_2} \\
        0 & \lambda_{f_2} & -\lambda_{f_1}
    \end{bmatrix},
\end{aligned}
\end{equation}
which naturally aligns $\pene$, $\force$, and their differentials in the contact-aligned basis in Equation~\eqref{eq:ortho}. 
However, this nonzero assumption does not always eliminate numerical singularities. Note that $|\pene_f|$ can be near zero under static friction, and $|\force_f|$ can be near zero under separation. Although our smooth formulation keeps these quantities nominally nonzero, finite-precision round-off can still drive them effectively to zero, leading to ill-conditioned normalization and unstable matrix assembly.

So in practice, we define a threshold $\tau\ll 1$.
As long as at least one of $\|\pene_f\|$ and $\|\force_f\|$ is larger than $\tau$, we just use it to assemble $\ortho$ (i.e., determine the main friction direction).
If both $\|\pene_f\|$ and $\|\force_f\|$ are less than $\tau$, it means current contact is nearly symmetric under rotation along normal direction; so we just choose $\ortho=\ids$.

\subsection{Forward-Backward Smoothing Consistency}
\label{sec:smoothness_consistency}

Section~\ref{sec:ncp} derives the FB subgradient under the assumption that the forward solve and the backward subgradient evaluation use the same smoothing parameter $\epsilon$.
We test this requirement empirically on the Pulling Tux task (low resolution), which simultaneously exercises sliding, sticking, and separating contact regimes within a single optimization run, and is therefore the most sensitive to violations of the consistency assumption.
The three configurations in Table~\ref{tab:smoothness_consistency} differ only in the choice of forward and backward $\epsilon$; the piecewise/root branch in the implementation is forced to root form throughout ($\eta=+\infty$) so the comparison isolates the consistency variable from any branch-switching effect.

\begin{table}[htbp]
\centering
\caption{\textbf{Forward-backward smoothing consistency.} Optimization on Pulling Tux (low res, 100 iterations, root-only backward).  Both matched configurations succeed regardless of $\epsilon$ magnitude; the mismatched configuration fails to make any progress despite identical formulation in every other respect.}
\label{tab:smoothness_consistency}
\begin{tabular*}{\columnwidth}{@{\extracolsep{\fill}}lccccc@{}}
\toprule
Config & $2\epsilon_{\mathrm{fwd}}^2$ & $2\epsilon_{\mathrm{bwd}}^2$ & $L_0$ & $L_{\mathrm{final}}$ & Reduction \\
\midrule
Matched, tight   & $10^{-12}$ & $10^{-12}$ & 7.001 & 2.19$\times$10$^{-3}$ & 99.97\% \\
Mismatched       & $10^{-12}$ & $10^{-6}$  & 7.002 & 6.989                  & 0.18\%  \\
Matched, loose   & $10^{-6}$  & $10^{-6}$  & 7.001 & 6.84$\times$10$^{-4}$ & 99.99\% \\
\bottomrule
\end{tabular*}
\end{table}

\paragraph*{Why mismatched $\epsilon$ fails.}
At a forward-converged active normal contact row, the relations $x_n y_n=\epsilon_{\mathrm{fwd}}^2$ and $y_n=\mathcal O(1)$ give $x_n=\mathcal O(\epsilon_{\mathrm{fwd}}^2)$.
Under matched smoothing, $r=\sqrt{x_n^2+y_n^2+2\epsilon_{\mathrm{fwd}}^2}=x_n+y_n$ on the manifold, and the K diagonal entry $\omega/E=(1-x_n/r)/(1-y_n/r)=y_n/x_n$ scales as $1/\epsilon_{\mathrm{fwd}}^2$.
With backward $\epsilon_{\mathrm{bwd}}\!\neq\!\epsilon_{\mathrm{fwd}}$, the same row is evaluated as $r'=\sqrt{x_n^2+y_n^2+2\epsilon_{\mathrm{bwd}}^2}\approx y_n\sqrt{1+2\epsilon_{\mathrm{bwd}}^2/y_n^2}$, $E\approx \epsilon_{\mathrm{bwd}}^2/y_n^2$, and $\omega/E\sim 1/\epsilon_{\mathrm{bwd}}^2$, a multiplicative error of $\epsilon_{\mathrm{bwd}}^2/\epsilon_{\mathrm{fwd}}^2$.
By the analogous analysis on inactive rows ($x_n=\mathcal O(1)$, $y_n=\mathcal O(\epsilon_{\mathrm{fwd}}^2)$), the corresponding error is reciprocal: $\epsilon_{\mathrm{fwd}}^2/\epsilon_{\mathrm{bwd}}^2$.
At our mismatched configuration with $2\epsilon_{\mathrm{fwd}}^2=10^{-12}$ and $2\epsilon_{\mathrm{bwd}}^2=10^{-6}$, this amounts to a $10^{6}$-fold under\-estimation of K on active rows and a $10^{6}$-fold over\-estimation on inactive rows, compressing both regimes toward a common $1/\epsilon_{\mathrm{bwd}}^2$ scale and erasing the contrast that distinguishes constraining from non-constraining contacts.

\FloatBarrier

\section{Local Projection Differentiation}

\subsection{Full SVD Differential Derivation}
\label{sec:apx_B0}

This appendix provides the complete derivation of the vectorized Jacobian $\pp{\vect{\pdeform}}{\vect{\deform}}$ summarized in the main text.
We denote $\vect{\cdot}$ as the column-stacking vectorization of a matrix, $\diag{\cdot}$ as the diagonal vector of a matrix and $\Diag{\cdot}$ as the diagonal matrix embedded with a vector.
We also use auxiliary matrices $\mathbf{D}\mathbf{x} = \vect{\Diag{\mathbf{x}}}$ and $\mathbf{T}\vect{\mathbf{X}} = \vect{\trans{\mathbf{X}}}$ (exact values in Appendix~\ref{sec:apx_B1}), with $\mathbf{D}^\top\vect{\mathbf{X}} = \diag{\mathbf{X}}$.
We define $\svds=\diag{\svdS}$, $\svdss=\diag{\svdSS}$, and denote $\svdW \coloneqq \pp{\svdss}{\svds}$.

In 3D, $\deform$ and $\pdeform$ are full-rank square matrices with orthogonal singular vectors $\svdU$ and $\svdV$.
Thus $\dd\svdU=\svdU\rot^U$ and $\dd\svdV=\svdV\rot^V$, where $\rot^U = -\trans{(\rot^U)}$ and $\rot^V = -\trans{(\rot^V)}$ are skew-symmetric.
Matching the diagonal and off-diagonal elements in the differential of $\deform = \svdU\svdS\trans{\svdV}$:
\begin{subequations}
\begin{align}
    \trans{\svdU}\dd\deform\svdV &= \dd\svdS + \rot^U \svdS - \svdS \rot^V \\
    \dd\svdS_{ii} &= (\trans{\svdU}\dd\deform\svdV)_{ii} \\
    \rot^U_{ij} &= \frac{\sigma_j(\trans{\svdU}\dd\deform\svdV)_{ij} + \sigma_i(\trans{\svdU}\dd\deform\svdV)_{ji}}{\sigma_j^2-\sigma_i^2} \\
    \rot^V_{ij} &= \frac{\sigma_i(\trans{\svdU}\dd\deform\svdV)_{ij} + \sigma_j(\trans{\svdU}\dd\deform\svdV)_{ji}}{\sigma_j^2-\sigma_i^2}.
\end{align}
\end{subequations}
Similarly, for the projected deformation $\pdeform = \svdU\svdSS\trans{\svdV}$:
\begin{subequations}
\begin{align}
    &\trans{\svdU}\dd\pdeform\svdV = \dd\svdSS + \rot^U \svdSS - \svdSS \rot^V \\
    &\begin{aligned}
    (\trans{\svdU}\dd\pdeform\svdV)_{ij} &= \sum_k \svdW_{ik}(\trans{\svdU}\dd\deform\svdV)_{kk} \delta_{ij} \\
    &\quad + \frac{1}{\sigma_i^2-\sigma_j^2} \Big[ (\sigma_i\theta_i-\sigma_j\theta_j)(\trans{\svdU}\dd\deform\svdV)_{ij} \\
    &\quad + (\sigma_j\theta_i-\sigma_i\theta_j)(\trans{\svdU}\dd\deform\svdV)_{ji} \Big],
    \end{aligned}
\end{align}
\end{subequations}
where $\delta_{ij}$ is the Kronecker delta.

To convert this element-wise relation into a vectorized Jacobian, we use the identity $\mathrm{vec}(\mathbf{A}\mathbf{X}\mathbf{B}) = (\mathbf{B}^\top\otimes\mathbf{A}) \mathrm{vec}(\mathbf{X})$ and define
\begin{subequations}
\begin{align}
    \mathbf{M}&: \{ M_{ii}=0,\, M_{ij} = \frac{\sigma_i\theta_i-\sigma_j\theta_j}{\sigma_i^2-\sigma_j^2} = \frac{1}{2} \big( \frac{\theta_i-\theta_j}{\sigma_i-\sigma_j} + \frac{\theta_i+\theta_j}{\sigma_i+\sigma_j} \big) \} \\
    \mathbf{N}&: \{ N_{ii}=0,\, N_{ij} = \frac{\sigma_j\theta_i-\sigma_i\theta_j}{\sigma_i^2-\sigma_j^2} = \frac{1}{2} \big( \frac{\theta_i-\theta_j}{\sigma_i-\sigma_j} - \frac{\theta_i+\theta_j}{\sigma_i+\sigma_j} \big) \}.
\end{align}
\end{subequations}
Moving to the column-stacking vector basis yields the final expression:
\begin{multline}
    \pp{\vect{\pdeform}}{\vect{\deform}} = (\svdV\otimes\svdU) \big[ \mathbf{D}\svdW\trans{\mathbf{D}} \\
    + \Diag{\vect{\mathbf{M}}} + \Diag{\vect{\mathbf{M}}} \mathbf{T} \big] (\trans{\svdV}\otimes\trans{\svdU}).
\end{multline}

\subsection{Construction of \texorpdfstring{$\mathbf{D}$}{D} and \texorpdfstring{$\mathbf{T}$}{T}, 2D formulation}
\label{sec:apx_B1}

In the main text, we have defined auxiliary matrices: $\mathbf{D}\mathbf{x} = \vect{\Diag{\mathbf{x}}}$ and $\mathbf{T}\vect{\mathbf{X}} = \vect{\trans{\mathbf{X}}}$, whose exact values are
\begin{equation}
    \mathbf{D} = 
    \begin{bmatrix}
        1&0&0\\
        0&0&0\\
        0&0&0\\
        0&0&0\\
        0&1&0\\
        0&0&0\\
        0&0&0\\
        0&0&0\\
        0&0&1
    \end{bmatrix},\,
    \mathbf{T} =
    \begin{bmatrix}
        1&0&0&0&0&0&0&0&0\\
        0&0&0&1&0&0&0&0&0\\
        0&0&0&0&0&0&1&0&0\\
        0&1&0&0&0&0&0&0&0\\
        0&0&0&0&1&0&0&0&0\\
        0&0&0&0&0&0&0&1&0\\
        0&0&1&0&0&0&0&0&0\\
        0&0&0&0&0&1&0&0&0\\
        0&0&0&0&0&0&0&0&1
    \end{bmatrix},
\end{equation}
for 3D formulation and 
\begin{equation}
    \mathbf{D} = 
    \begin{bmatrix}
        1&0\\
        0&0\\
        0&0\\
        0&1
    \end{bmatrix},\,
    \mathbf{T} =
    \begin{bmatrix}
        1&0&0&0\\
        0&0&1&0\\
        0&1&0&0\\
        0&0&0&1
    \end{bmatrix},
\end{equation}
for 2D formulation.

For triangular elements, the preceding 3D derivation carries over with only a change in matrix dimensions. The displacement matrix $\mathbf{D}_s = [\mathbf{x}_1-\mathbf{x}_0, \mathbf{x}_2-\mathbf{x}_0]$ is no longer square, and the deformation gradient $\mathbf{F} = \mathbf{D}_s \mathbf{D}_m^{-1}$ has rank 2. To maintain positive singular values, we adopt the thin SVD with $\mathbf{U}\in\mathbb{R}^{3\times 2}$, $\mathbf{V}\in\mathbb{R}^{2\times 2}$, and $\boldsymbol{\sigma}\in\mathbb{R}^{2}$.

\subsection{Construction of \texorpdfstring{$\mathbf{W} = \pp{\theta}{\sigma}$}{W = dθ/dσ}}
\label{sec:apx_B2}
In the main text, we defined the derivative matrix of singular values $\mathbf{W} = \pp{\theta}{\sigma}$, which encodes the constitutive-law dependence.
We consider three models: As-Rigid-As-Possible (ARAP), co-rotational, and Neo-Hookean.

\paragraph*{ARAP}
This model only keeps rotation information in SVD by setting $\boldsymbol{\theta}(\boldsymbol{\sigma}) = [1,1,1]^\top$, so we directly have $\mathbf{W}=\mathbf{0}$ and $M_{ij}=N_{ij}=\frac{1}{\sigma_i+\sigma_j}$.
For invariant-based reformulations of the ARAP energy, see~\cite{lin_isotropic_2022}.

\paragraph*{Co-rotational}
The co-rotational energy density is $\zeta(\boldsymbol{\theta}) = \mu\|\boldsymbol{\theta}-\mathbf{1}\|^2 + \frac{\lambda}{2}\big(\sum_i(\theta_i-1)\big)^2$, with $w=2\mu$. The optimality condition is
\begin{equation}
    \mathbf{g} = 4\mu(\boldsymbol{\theta}-\mathbf{1}) + \lambda\,\mathrm{tr}(\boldsymbol{\theta}-\mathbf{1})\,\mathbf{1} - 2\mu(\boldsymbol{\sigma}-\mathbf{1}) = 0,
\end{equation}
yielding the constant Hessian $\mathbf{H}=\frac{\partial\mathbf{g}}{\partial\boldsymbol{\theta}} = 4\mu\mathbf{I} + \lambda\mathbf{1}\mathbf{1}^\top$. Since $\frac{\partial\mathbf{g}}{\partial\boldsymbol{\sigma}}=-2\mu\mathbf{I}$, by the Sherman--Morrison formula:
\begin{equation}
    \mathbf{W} = 2\mu\mathbf{H}^{-1} = \frac{1}{2}\mathbf{I} - \frac{\lambda}{2(4\mu+n\lambda)}\mathbf{1}\mathbf{1}^\top,
\end{equation}
where $n$ is the spatial dimension. Unlike ARAP and Neo-Hookean, $\mathbf{W}$ is independent of $\boldsymbol{\sigma}$ and $\boldsymbol{\theta}$---a distinctive property of the co-rotational model.

\paragraph*{Neo-Hookean}
In this model, $\boldsymbol{\theta}$ is the minimizer of $\Psi(\boldsymbol{\theta}, \boldsymbol{\sigma}) = \frac{w}{2}\|\boldsymbol{\theta}-\boldsymbol{\sigma}\|_F^2 + \zeta(\boldsymbol{\theta})$: 
\begin{equation}
    \zeta(\boldsymbol{\theta}) = \frac{1}{2}\mu(I_1-\log{I_3}-3) + \frac{1}{8}\lambda(\log{I_3})^2
\end{equation}
where $I_1=\theta_1^2+\theta_2^2+\theta_3^2$, $I_3=J^2$, $J=\theta_1\theta_2\theta_3$, $\mu$ denotes Shear modulus and $\lambda$ denotes Lamé's first parameter. 
By definition, we have $w = 2 \mu$.
Since $\boldsymbol{\theta}$ minimizes $\Psi(\boldsymbol{\theta}, \boldsymbol{\sigma})$, we have:

\begin{equation}
    \mathbf{g} \equiv \frac{\partial\Psi}{\partial\boldsymbol{\theta}} = w(\boldsymbol{\theta}-\boldsymbol{\sigma}) + \mu(\boldsymbol{\theta}-\boldsymbol{\theta}^{-1}) + \lambda \log{J}~\boldsymbol{\theta}^{-1} = 0 
\end{equation}
\begin{equation}
    \frac{\partial\mathbf{g}}{\partial\boldsymbol{\sigma}} = -2\mu \mathbf{I},\quad \frac{\partial\mathbf{g}}{\partial\boldsymbol{\theta}}=\underbrace{(3\mu)\mathbf{I} + (\mu-\lambda\log{J})\boldsymbol{\Theta}^{-2}}_{\mathcal{D}} + \lambda\boldsymbol{\theta}^{-1}\boldsymbol{\theta}^{-\top}
\end{equation}
\begin{equation}
    \mathbf{W} = \frac{\partial\boldsymbol{\theta}}{\partial\boldsymbol{\sigma}} = -(\frac{\partial\mathbf{g}}{\partial\boldsymbol{\theta}})^{-1}\frac{\partial\mathbf{g}}{\partial\boldsymbol{\sigma}} = 2\mu\mathcal{D}^{-1} \Big( \mathbf{I} - \frac{\lambda\boldsymbol{\theta}^{-1}\boldsymbol{\theta}^{-\top}}{1+\lambda\boldsymbol{\theta}^{-\top}\mathcal{D}^{-1}\boldsymbol{\theta}^{-1}}\mathcal{D}^{-\top} \Big)
\end{equation}

\subsection{Differentiation of Elastic Parameters}
\label{sec:apx_B3}

Before calculating the gradient with respect to Young's modulus $E$ and Poisson's ratio $\nu$, we first consider Shear modulus $\mu$ and Lamé's first parameter $\lambda$ as the bridge:
\begin{equation}
    \mu = \frac{E}{2(1+\nu)},\quad \lambda=\frac{E\nu}{(1+\nu)(1-2\nu)}.
\end{equation}

For hyperelastic materials, it is necessary to calculate $\frac{\partial\mathbf{p}_i}{\partial \mu}$ and $\frac{\partial\mathbf{p}_i}{\partial \lambda}$ when local projection is influenced by $\mu$ and $\lambda$.

In the co-rotational model (with $w=2\mu$), $\mathbf{g} = 4\mu(\boldsymbol{\theta}-\mathbf{1}) + \lambda\,\mathrm{tr}(\boldsymbol{\theta}-\mathbf{1})\,\mathbf{1} - 2\mu(\boldsymbol{\sigma}-\mathbf{1}) = 0$, giving
\begin{equation}
    \frac{\partial\mathbf{g}}{\partial\mu} = 2(2\boldsymbol{\theta}-\boldsymbol{\sigma}-\mathbf{1}),\quad \frac{\partial\mathbf{g}}{\partial\lambda} = \mathrm{tr}(\boldsymbol{\theta}-\mathbf{1})\,\mathbf{1},
\end{equation}
and, since $\mathbf{H}^{-1}$ is constant (Section~\ref{sec:apx_B2}),
\begin{equation}
    \frac{\partial\boldsymbol{\theta}}{\partial\mu} = -\mathbf{H}^{-1}\frac{\partial\mathbf{g}}{\partial\mu},\quad
    \frac{\partial\boldsymbol{\theta}}{\partial\lambda} = -\mathbf{H}^{-1}\frac{\partial\mathbf{g}}{\partial\lambda}.
\end{equation}

In the Neo-Hookean model (with $w=2\mu$):
\begin{equation}
    \mathbf{g} = \mu(3\boldsymbol{\theta}-2\boldsymbol{\sigma}-\boldsymbol{\theta}^{-1}) + \lambda \log{J}~\boldsymbol{\theta}^{-1} = 0
\end{equation}
\begin{equation}
    \frac{\partial\mathbf{g}}{\partial\mu} = 3\boldsymbol{\theta}-2\boldsymbol{\sigma}-\boldsymbol{\theta}^{-1},\quad \frac{\partial\mathbf{g}}{\partial\lambda} = \log{J}~\boldsymbol{\theta}^{-1}
\end{equation}
\begin{equation}
    \begin{aligned}
        \frac{\partial\boldsymbol{\theta}}{\partial\mu} &= -(\frac{\partial\mathbf{g}}{\partial\boldsymbol{\theta}})^{-1}\frac{\partial\mathbf{g}}{\partial\mu} \\
        &= -\mathcal{D}^{-1} \Big( \mathbf{I} - \frac{\lambda\boldsymbol{\theta}^{-1}\boldsymbol{\theta}^{-\top}}{1+\lambda\boldsymbol{\theta}^{-\top}\mathcal{D}^{-1}\boldsymbol{\theta}^{-1}}\mathcal{D}^{-\top} \Big)(3\boldsymbol{\theta}-2\boldsymbol{\sigma}-\boldsymbol{\theta}^{-1}) \\
        \frac{\partial\boldsymbol{\theta}}{\partial\lambda} &= -(\frac{\partial\mathbf{g}}{\partial\boldsymbol{\theta}})^{-1}\frac{\partial\mathbf{g}}{\partial\lambda} \\
        &= -\mathcal{D}^{-1} \Big( \mathbf{I} - \frac{\lambda\boldsymbol{\theta}^{-1}\boldsymbol{\theta}^{-\top}}{1+\lambda\boldsymbol{\theta}^{-\top}\mathcal{D}^{-1}\boldsymbol{\theta}^{-1}}\mathcal{D}^{-\top} \Big)(\log{J}~\boldsymbol{\theta}^{-1}).
    \end{aligned}
\end{equation}
For both models, then
\begin{equation}
    \begin{aligned}
        \frac{\partial \mathrm{vec}(\mathbf{P})}{\partial \mu}&=(\mathbf{V} \otimes \mathbf{U}) \mathbf{D} \frac{\partial \boldsymbol{\theta}}{\partial \mu} \\
        \frac{\partial \mathrm{vec}(\mathbf{P})}{\partial \lambda}&=(\mathbf{V} \otimes \mathbf{U}) \mathbf{D} \frac{\partial \boldsymbol{\theta}}{\partial \lambda} 
    \end{aligned}
\end{equation}
\begin{equation}
    \begin{aligned}
        \frac{\partial \mathbf{P}}{\partial \mu}&=\mathbf{U} \mathrm{Diag} (\frac{\partial \boldsymbol{\theta}}{\partial \mu}) \mathbf{V}^\top\\
        \frac{\partial \mathbf{P}}{\partial \lambda}&=\mathbf{U} \mathrm{Diag} (\frac{\partial \boldsymbol{\theta}}{\partial \lambda}) \mathbf{V}^\top.
    \end{aligned}
\end{equation}
After differentiation, we have 
\begin{equation}
\begin{aligned}
    \rhs_\mu &= h^2 \sum_i 2 \mathbf{G}_i^\top \big(\mu_i\frac{\partial\mathbf{p}_i}{\partial \mu_i} + \mathbf{p}_i - \mathbf{G}_i \mathbf{q}\big) \\
    \rhs_\lambda &= h^2 \sum_i 2 \mathbf{G}_i^\top \big(\mu_i\frac{\partial\mathbf{p}_i}{\partial \lambda_i} \big),
\end{aligned}
\end{equation}
and
\begin{equation}
    \begin{aligned}
        \frac{\partial L}{\partial \mathbf{\mu}} &= \mathbf{z}^\top\rhs_\mu, \quad \frac{\partial L}{\partial \mathbf{\lambda}} = \mathbf{z}^\top\rhs_\lambda \\
        \frac{\partial L}{\partial E} &= \frac{\partial L}{\partial \mathbf{\mu}} \frac{\partial \mu}{\partial E} + \frac{\partial L}{\partial \mathbf{\lambda}} \frac{\partial \lambda}{\partial E} \\
        \frac{\partial L}{\partial \nu} &= \frac{\partial L}{\partial \mathbf{\mu}} \frac{\partial \mu}{\partial \nu} + \frac{\partial L}{\partial \mathbf{\lambda}} \frac{\partial \lambda}{\partial \nu}.
    \end{aligned} 
\end{equation}

\subsection{Degeneracy Analysis}
\label{sec:apx_B4}

In the main text, we assumed the non-degeneracy of $\boldsymbol{\Sigma}$, which does not always hold. When degeneracy occurs, i.e.\ $\sigma_i=\sigma_j$, the expressions for $\mathrm{d}\mathbf{U}$ and $\mathrm{d}\mathbf{V}$ derived above blow up. This is because $\boldsymbol{\Sigma}$, $\mathbf{U}$, and $\mathbf{V}$ are no longer deterministic functions of $\mathbf{F}$, and their derivatives are not well-defined. Attempting to differentiate in the neighborhood requires considering the off-diagonal terms in $\mathrm{d}\boldsymbol{\Sigma}$, which are no longer zero and become symmetric. Resolving this involves a finite rotation in the degenerate eigenspace that cannot be absorbed into the infinitesimal $\mathrm{d}\mathbf{U}$ and $\mathrm{d}\mathbf{V}$. However, although the non-degeneracy assumption is not rigorous and the resulting gradient expressions contain singularities, their limits are well-defined when $\boldsymbol{\theta}(\boldsymbol{\sigma})$ satisfies certain conditions.

We first analyze the source of the singularity $\frac{\theta_i-\theta_j}{\sigma_i-\sigma_j}$, which arises from the degeneracy $\sigma_i=\sigma_j$. Performing a first-order expansion around the degenerate point, $\boldsymbol{\sigma}=\bar{\boldsymbol{\sigma}}+\mathbf{h}$ and $\boldsymbol{\theta}=\boldsymbol{\theta}(\bar{\boldsymbol{\sigma}})+\mathbf{W}\mathbf{h}$, and writing $\mathbf{h}=t\mathbf{v}$:
\begin{equation}
    \lim_{t\to 0} \frac{\theta_i-\theta_j}{\sigma_i-\sigma_j} = \frac{(W_{i:}-W_{j:})\mathbf{v}}{v_i-v_j}.
\end{equation}
This limit depends on the path direction $\mathbf{v}$ in general, rendering $\frac{\partial\mathrm{vec}(\mathbf{P})}{\partial\mathrm{vec}(\mathbf{F})}$ ill-defined. It becomes path-independent if and only if
$W_{i:}-W_{j:} = \alpha(e_i^\top-e_j^\top)$, in which case the limit reduces to $\alpha$. Let $I(\bar{\boldsymbol{\sigma}})$ denote the degenerate block at $\bar{\boldsymbol{\sigma}}$, defined by $\forall~\Pi\in S_I:\Pi\bar{\boldsymbol{\sigma}} = \bar{\boldsymbol{\sigma}}$, where $S_I$ is the symmetric group (permutation group) on $I$. The condition can be written as
\begin{equation}
    \begin{aligned}
        &\forall~\Pi\in S_I: \mathbf{W}_{II}(\bar{\boldsymbol{\sigma}})\Pi = \Pi\mathbf{W}_{II}(\bar{\boldsymbol{\sigma}}) \\
        \Leftrightarrow&~\mathbf{W}_{II}(\bar{\boldsymbol{\sigma}}) = \alpha\mathbf{I} + \beta\mathbf{1}\mathbf{1}^T,
    \end{aligned}
    \label{eq:within_block}
\end{equation}
and is called within-block commutation.

Moreover, the $\mathbf{P}(\mathbf{F})$ itself must be well-defined with degeneracy, requiring that $\boldsymbol{\theta}(\boldsymbol{\sigma})$ preserves degeneracy, $\forall~\Pi\in S_I: \boldsymbol{\theta}(\bar{\boldsymbol{\sigma}}) = \Pi\boldsymbol{\theta}(\bar{\boldsymbol{\sigma}})$. Otherwise $\mathbf{P} = \mathbf{U}\boldsymbol{\Theta}\mathbf{V}^\top$ will depend on the gauge selection (the arbitrary rotation in degenerate space). So combining Eq.~\ref{eq:within_block}, the whole sufficient and necessary condition for $\mathbf{P}(\mathbf{F})$'s Fréchet differentiability is
\begin{equation}
    \forall~\Pi\in S_I,~\exists~U\ni\bar{\boldsymbol{\sigma}}: \boldsymbol{\theta}(\Pi\boldsymbol{\sigma}) = \Pi\boldsymbol{\theta}(\boldsymbol{\sigma}),~\forall~\boldsymbol{\sigma}\in U,
    \label{eq:local_equivariance}
\end{equation}
which is called local equivariance near the degenerate point $\bar{\boldsymbol{\sigma}}$.

For the elastic models of interest, $\boldsymbol{\theta}(\boldsymbol{\sigma})$ is the \textbf{unique} minimizer of $\Psi(\boldsymbol{\theta}, \boldsymbol{\sigma}) = \frac{w}{2}\|\boldsymbol{\theta}-\boldsymbol{\sigma}\|_F^2 + \zeta(\boldsymbol{\theta})$. Since our models assume \textbf{isotropic} elasticity, $\forall~\Pi\in S_n:\zeta(\boldsymbol{\theta}) = \zeta(\Pi\boldsymbol{\theta})$ and $\|\boldsymbol{\theta}-\boldsymbol{\sigma}\|_F^2 = \|\Pi\boldsymbol{\theta}-\Pi\boldsymbol{\sigma}\|_F^2$ (the norm kernel is the identity for the isotropic $L^2$ norm), and hence $\Psi(\boldsymbol{\theta}, \boldsymbol{\sigma}) = \Psi(\Pi\boldsymbol{\theta}, \Pi\boldsymbol{\sigma})$. Since $\boldsymbol{\theta}(\Pi\boldsymbol{\sigma})$ minimizes $\Psi(\boldsymbol{\theta}, \Pi\boldsymbol{\sigma})$, it also minimizes $\Psi(\Pi^{-1}\boldsymbol{\theta}, \boldsymbol{\sigma})$ and therefore equals $\Pi\boldsymbol{\theta}(\boldsymbol{\sigma})$. Thus Eq.~\ref{eq:local_equivariance} holds globally, so $\mathbf{P}(\mathbf{F})$ is Fr\'{e}chet differentiable, and at a degenerate point $\sigma_i=\sigma_j$ we have
\begin{equation}
    \begin{aligned}
        M_{ij} = \frac{1}{2} \big( W_{ii}-W_{ij} + \frac{\theta}{\sigma} \big) \\
        N_{ij} = \frac{1}{2} \big( W_{ii}-W_{ij} - \frac{\theta}{\sigma} \big).
    \end{aligned}
\end{equation}

\section{Per-Demo Gradient Comparison}
\label{sec:apx_grad_grid}

\begin{figure*}[p]
    \centering
    \begin{subfigure}[t]{0.33\textwidth}
        \centering
        \includegraphics[width=\linewidth]{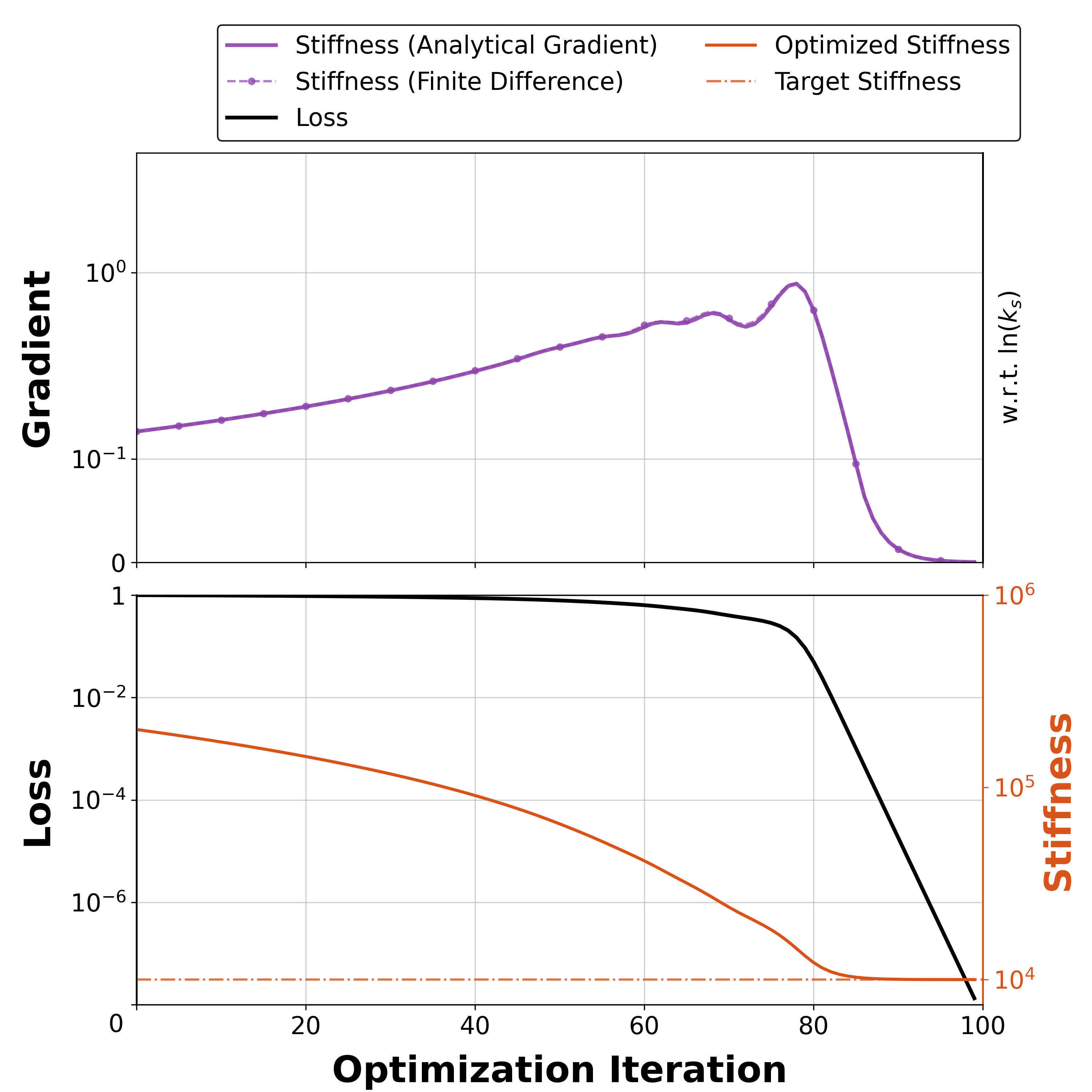}
        \caption{Stiffness}
        \label{fig:r1c1}
    \end{subfigure}\hfill
    \begin{subfigure}[t]{0.33\textwidth}
        \centering
        \includegraphics[width=\linewidth]{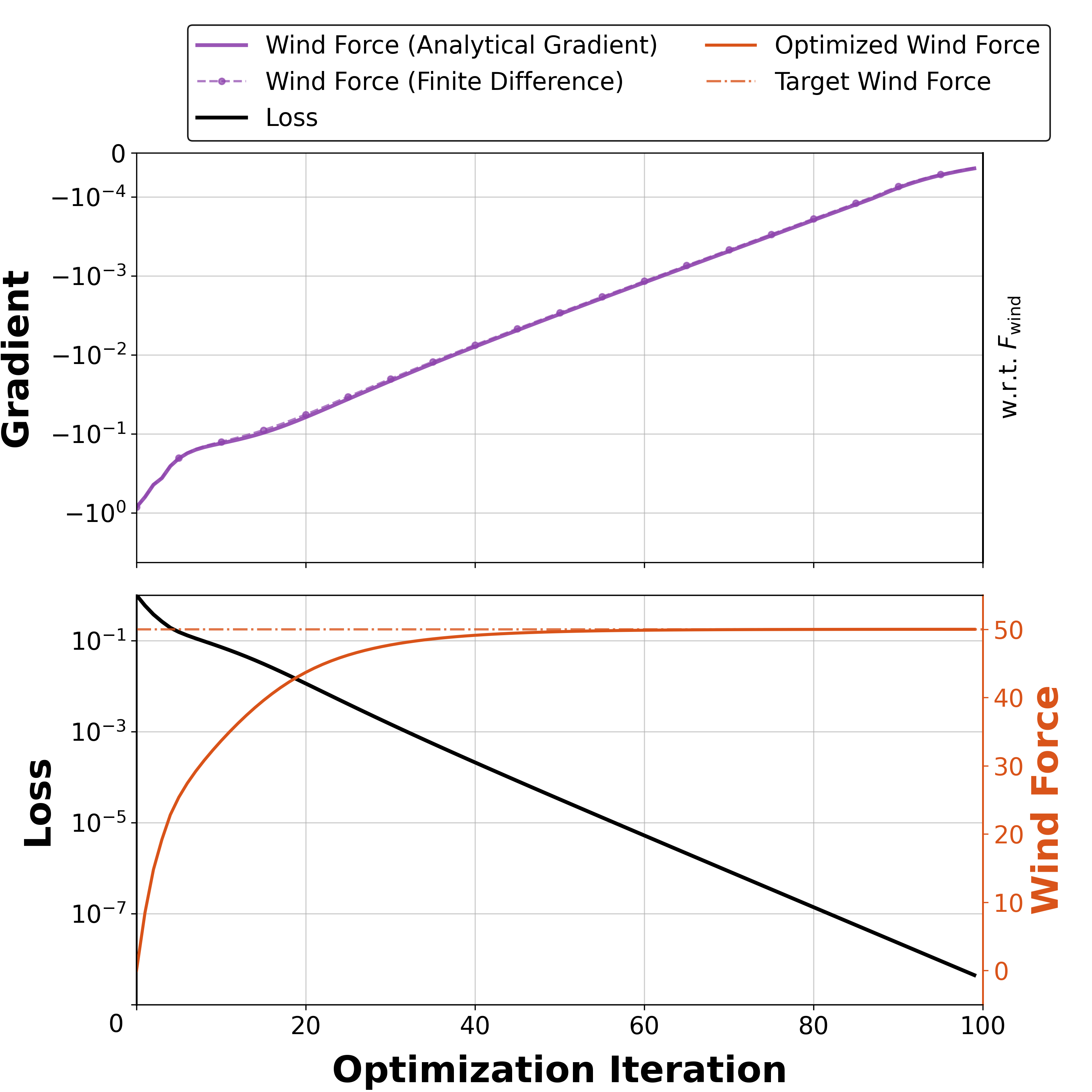}
        \caption{Wind force}
        \label{fig:r1c2}
    \end{subfigure}\hfill
    \begin{subfigure}[t]{0.33\textwidth}
        \centering
        \includegraphics[width=\linewidth]{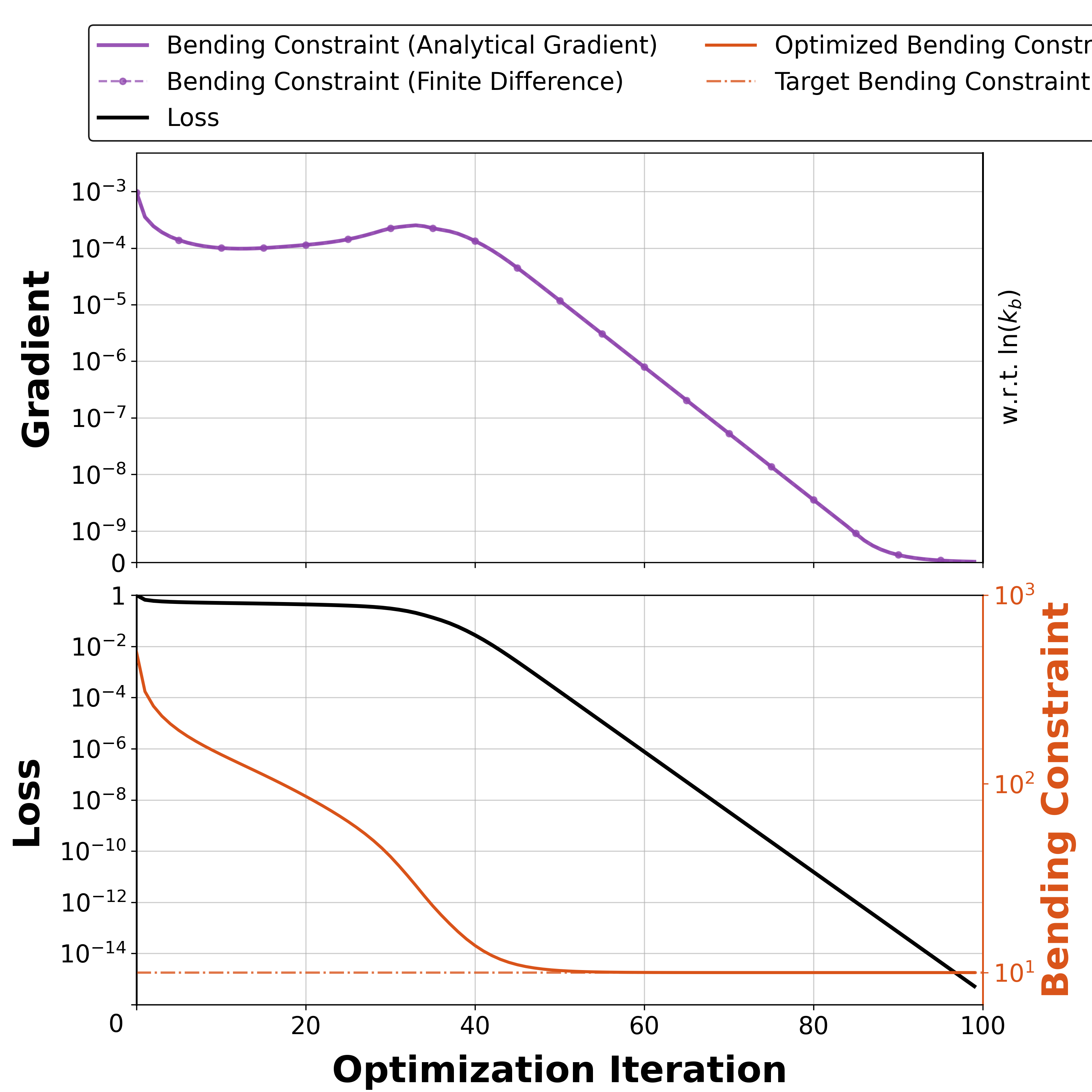}
        \caption{Bending constraint}
        \label{fig:r1c3}
    \end{subfigure}

    \vspace{2pt}
    \begin{subfigure}[t]{0.33\textwidth}
        \centering
        \includegraphics[width=\linewidth]{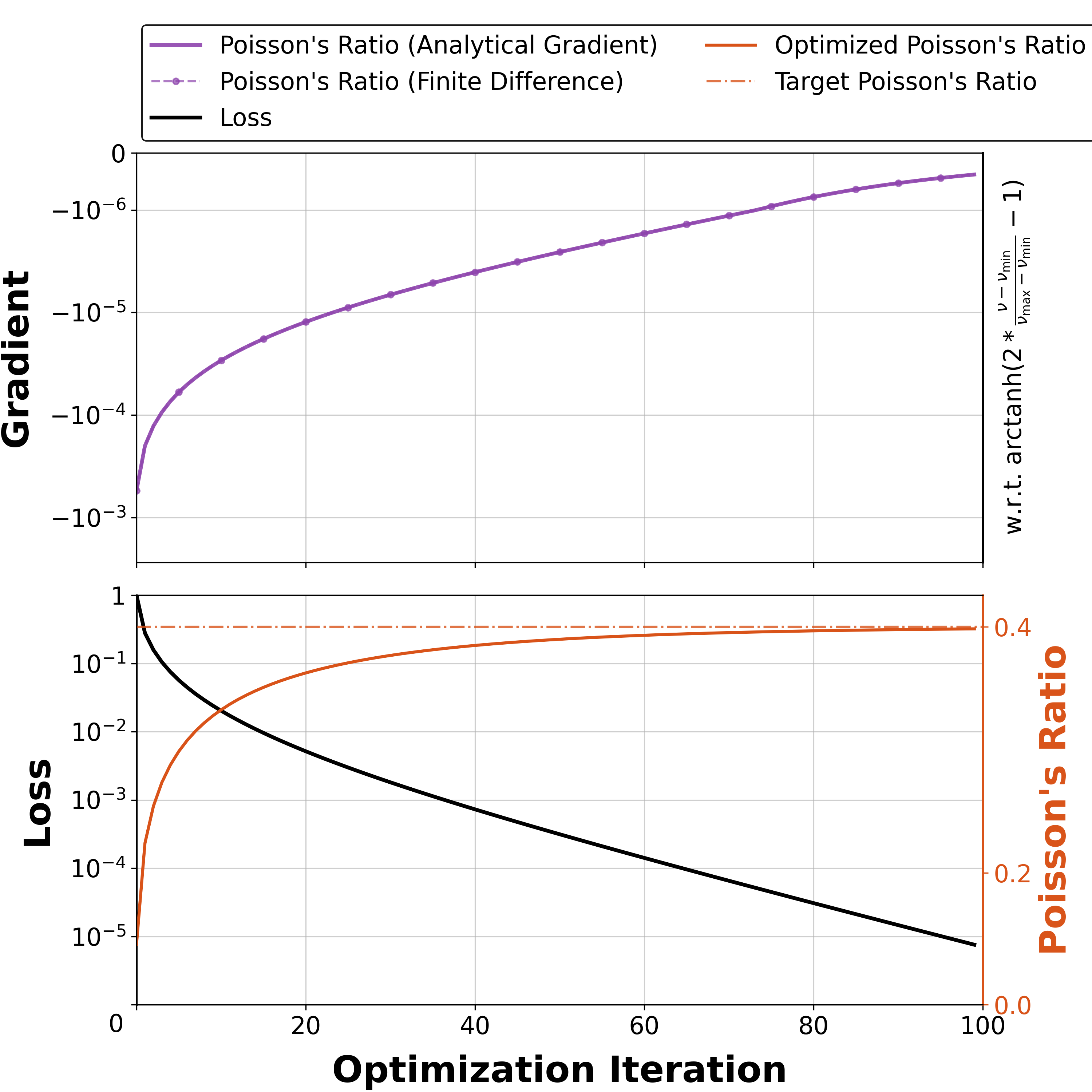}
        \caption{Poisson's ratio}
        \label{fig:r2c1}
    \end{subfigure}\hfill
    \begin{subfigure}[t]{0.33\textwidth}
        \centering
        \includegraphics[width=\linewidth]{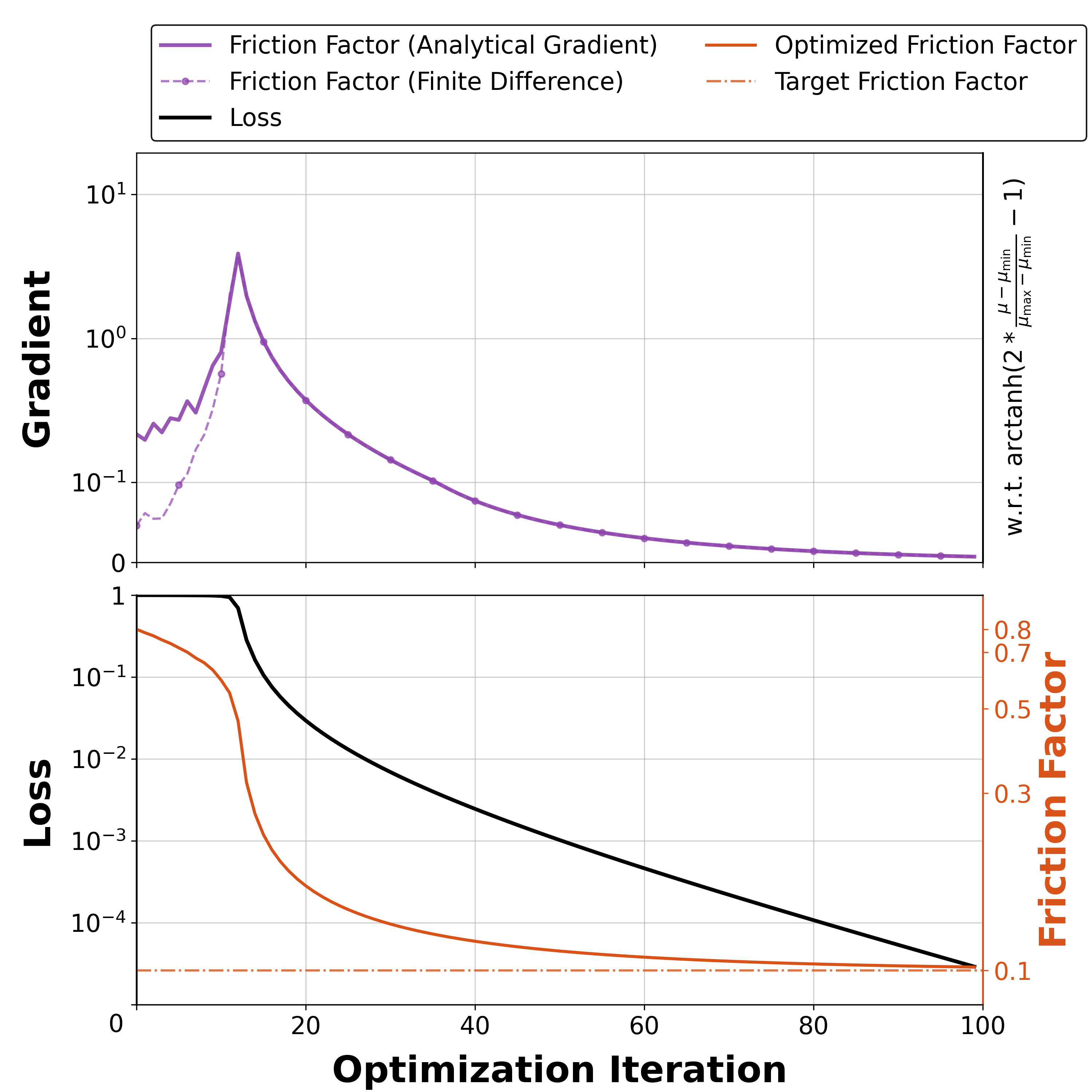}
        \caption{Friction factor}
        \label{fig:r2c2}
    \end{subfigure}\hfill
    \begin{subfigure}[t]{0.33\textwidth}
        \centering
        \includegraphics[width=\linewidth]{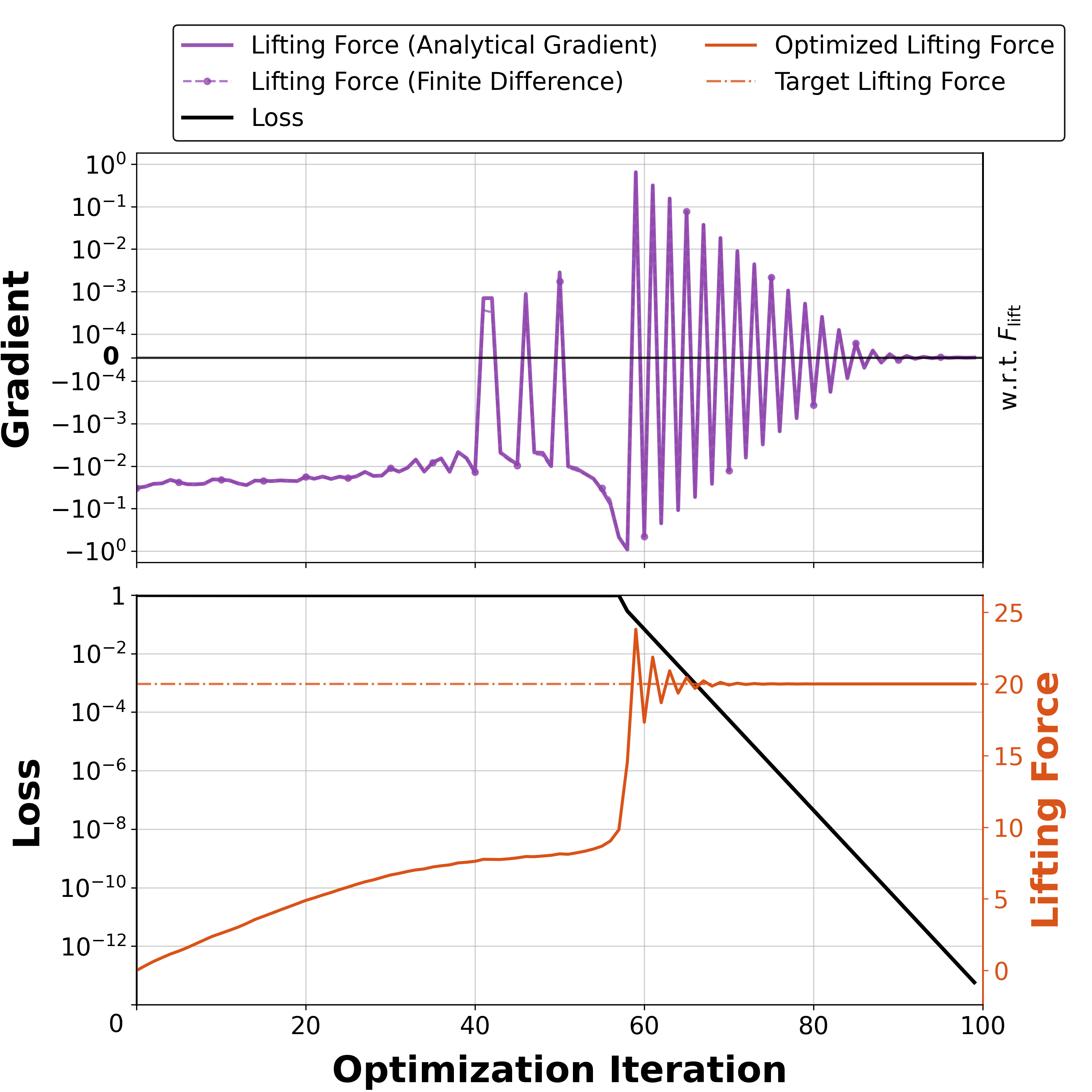}
        \caption{Lifting force}
        \label{fig:r2c3}
    \end{subfigure}

    \vspace{2pt}
    \begin{subfigure}[t]{0.33\textwidth}
        \centering
        \includegraphics[width=\linewidth]{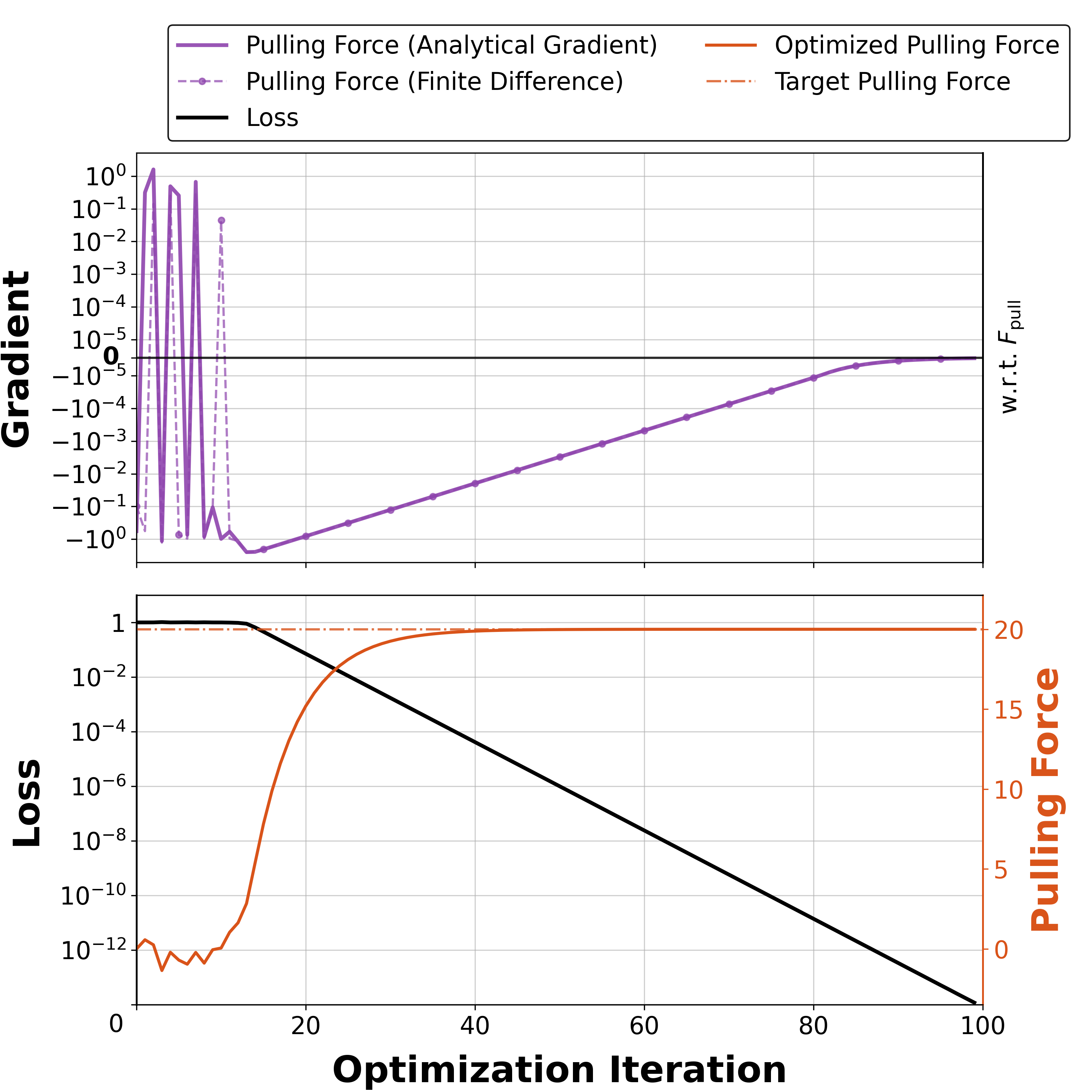}
        \caption{Pulling force}
        \label{fig:r3c1}
    \end{subfigure}\hfill
    \begin{subfigure}[t]{0.33\textwidth}
        \centering
        \includegraphics[width=\linewidth]{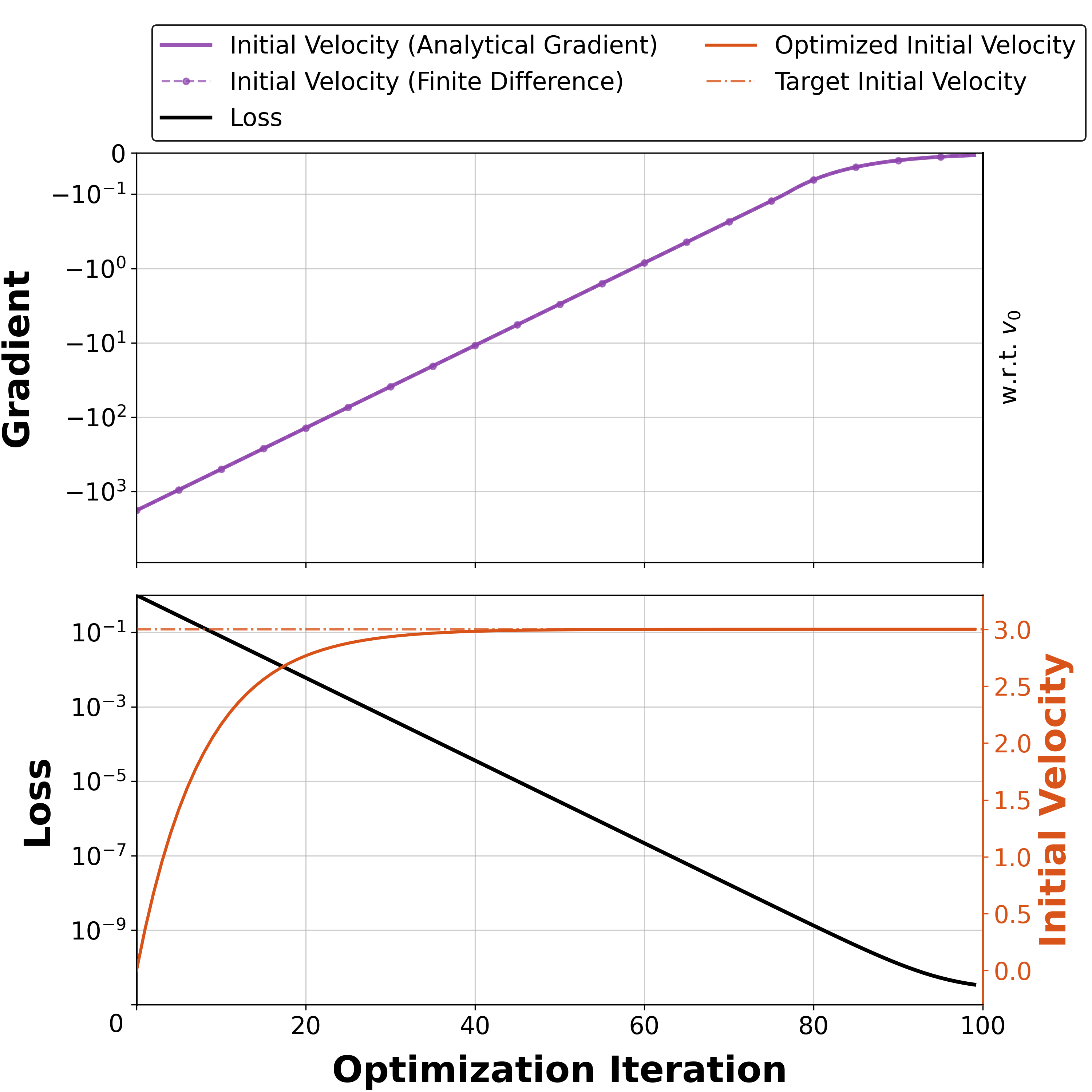}
        \caption{Initial velocity without friction}
        \label{fig:r3c2}
    \end{subfigure}\hfill
    \begin{subfigure}[t]{0.33\textwidth}
        \centering
        \includegraphics[width=\linewidth]{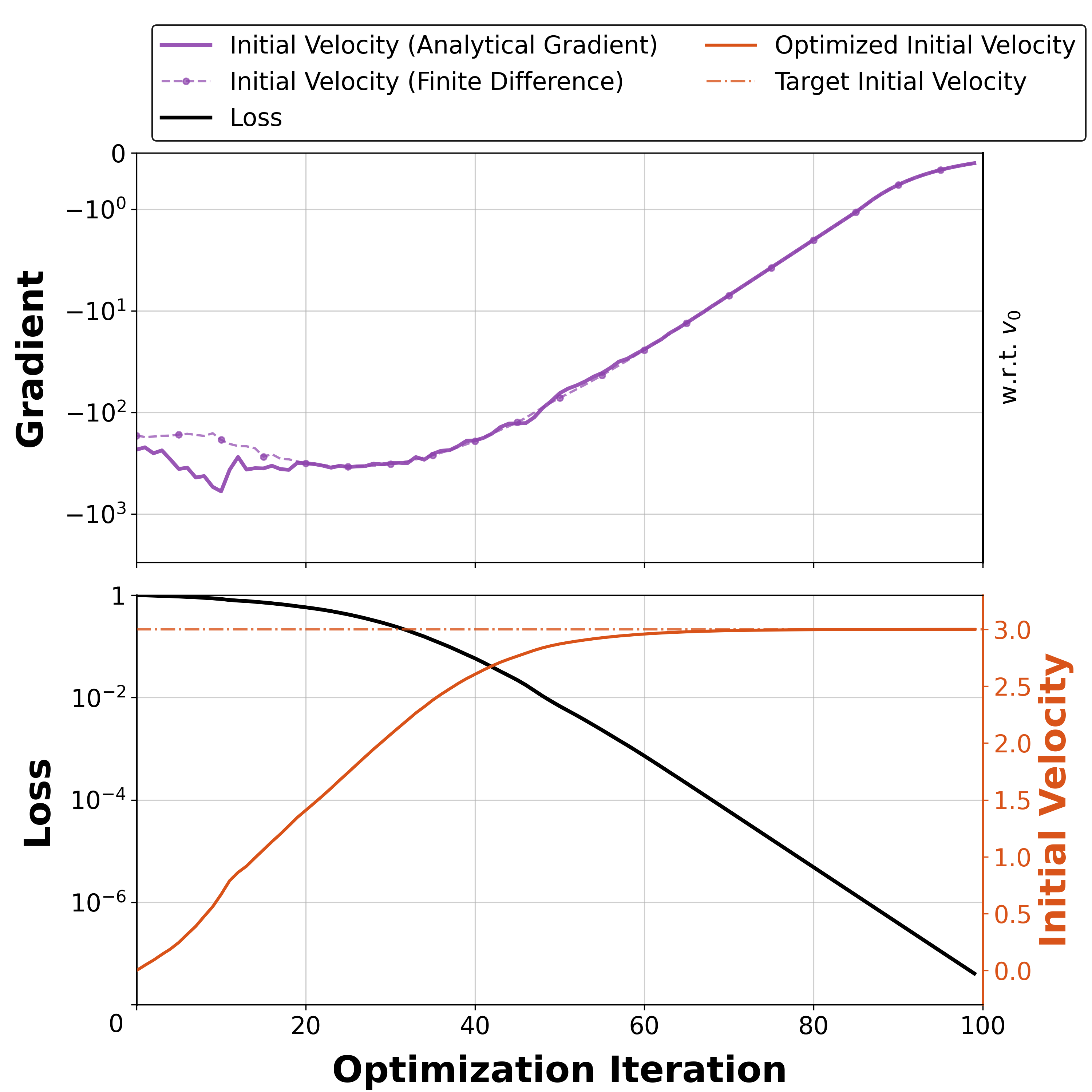}
        \caption{Initial velocity with friction}
        \label{fig:r3c3}
    \end{subfigure}
    \caption{\textbf{Gradient unit tests for elasticity and contact.} Each panel reports a gradient-based identification task: (a) stiffness, (b) wind force, (c) bending coefficient, (d) Poisson’s ratio, (e) friction factor, (f) lifting force, (g) pulling force, and initial-velocity identification (h) without friction and (i) with friction. In each panel, the \textbf{top} subplot compares analytical gradients with finite-difference references (with “w.r.t.” indicating the reparameterized optimization variable), and the \textbf{bottom} subplot shows the loss and the corresponding parameter trajectory over iterations. In contact--separation and static-friction transitions (panels (f) and (g)), both analytical and finite-difference gradients exhibit transient sign-flip excursions that reflect the marginal stability of the forward simulation in these regimes; agreement between the two methods is recovered once the system enters a steady regime (kinetic sliding, full separation, contact-pinned plateau).}
    \Description{Three-by-three grid of gradient unit tests; each cell stacks a top plot of analytical-vs-finite-difference gradients and a bottom plot of loss and parameter trajectory across optimization iterations, covering stiffness, wind force, bending, Poisson's ratio, friction factor, lifting force, pulling force, and frictionless and frictional initial-velocity identification.}
    \label{fig:grad_grid}
\end{figure*}

For each identification task in Section~\ref{sec:gradient_accuracy}, we report the analytical-vs-finite-difference gradient comparison alongside the loss / parameter trajectory in a single grid (Figure~\ref{fig:grad_grid}). This is the appendix-side detailed view of the per-demo loss curves shown in the main text.

\end{document}